\documentclass[aps,prd,superscriptaddress,showpacs,floatfix,letterpaper,preprintnumbers,nofootinbib]{revtex4}
\usepackage{epsfig}
\usepackage{bm}
\usepackage{amsmath}
\usepackage{amstext}
\usepackage{amssymb}
\usepackage{mathrsfs}
\usepackage{color}
\usepackage{subfigure}
\usepackage{graphicx}

\begin{document}

\title{Investigating the temperature dependence of the specific shear viscosity of QCD matter with dilepton radiation}
\author{Gojko Vujanovic}
\affiliation{Department of Physics, The Ohio State University, 191 West Woodruff Avenue, Columbus, Ohio 43210, USA}
\author{Gabriel S. Denicol}
\affiliation{Instituto de F\'{i}sica, Universidade Federal Fluminense, UFF, Niter\'{o}i, 24210-346, RJ, Brazil}
\author{Matthew Luzum}
\affiliation{Instituto de F\'isica - Universidade de S\~ao Paulo, Rua do Mat\~ao Travessa R, no. 187, 05508-090, Cidade Universit\'aria, S\~ao Paulo, Brazil}
\author{Sangyong Jeon}
\affiliation{Department of Physics, McGill University, 3600 University Street, Montr\'eal, Qu\'ebec, Canada H3A 2T8}
\author{Charles Gale}
\affiliation{Department of Physics, McGill University, 3600 University Street, Montr\'eal, Qu\'ebec, Canada H3A 2T8}

\begin{abstract}
This work reports on investigations of the effects on the evolution of viscous hydrodynamics and on the flow coefficients of thermal dileptons, originating from a temperature-dependent specific shear viscosity $\eta/s (T)$ at temperatures beyond 180 MeV formed at the Relativistic Heavy-Ion Collider (RHIC). We show that the elliptic flow of thermal dileptons can resolve the magnitude of $\eta/s$ at the high temperatures, where partonic degrees of freedom become relevant, whereas discriminating between different specific functional forms will likely not be possible at RHIC using this observable.
\end{abstract}

\maketitle
\date{\today }
\section{Introduction}

Recently, two laboratories---the Relativistic Heavy Ion Collider (RHIC) at the Brookhaven National Laboratory and the Large Hadron Collider at CERN---have created an exotic state of matter: the quark-gluon plasma (QGP). Since this  discovery, the characterization of the properties of the QGP has been a mainstream goal of high-energy nuclear physics. One of the striking discoveries of RHIC, also confirmed at the LHC, has been the fluid-dynamical behavior of the QGP \cite{Heinz:2013th,Gale:2013da}. The progress in hydrodynamic modeling of relativistic heavy-ion collisions has been so rapid over the last decade or so, that there now exists genuine hope to soon be able to precisely quantify the degree of departure for equilibrium of the QGP, by assessing its transport coefficients. Much of the theoretical activity has up to now concentrated on the determination of the shear viscosity to entropy density ratio, as revealed by measurements of the hadronic collectivity \cite{Romatschke:2007mq,Shen:2011zc}. In addition, it is now clear that these same flow data  also demand a non zero value of the specific bulk viscosity, $\zeta/s$ \cite{Ryu:2015vwa}.

A promising class of probes with which to investigate the QGP directly are electromagnetic (EM) signals, {\it i.e.}, photons and dileptons, as they do not participate in strong interactions and can thus escape with negligible final state interactions \cite{Gale:2009gc}. Furthermore, these probes are being emitted throughout the entire evolution of the medium, thereby giving  local information about the state of the medium, from the initial nucleon-nucleon collisions to kinetic freeze-out. The penetrating nature of EM observables makes them a particularly useful tool to study the temperature dependence of the transport coefficients of the QCD medium. A key coefficient present in all recent hydrodynamical calculations is the shear viscosity $\eta$ whose temperature dependence is often assumed to be identical with that of the entropy density $s$ of strongly interacting media, such that $\eta/s$ is left as a constant to be determined by experimental data. However, it is clear that this is an approximation \cite{Csernai:2006zz} and, in fact, calculations based on perturbative QCD \cite{Arnold:2003zc}, on hadronic degrees of freedom in the confined sector \cite{Prakash:1993bt,Gorenstein:2007mw,Itakura:2007mx,NoronhaHostler:2008ju,Greiner:2011rx}, and on functional renormalization group techniques \cite{Christiansen:2014ypa} show that $\eta/s$ changes with temperature. Calculations from first principles that address the temperature dependence of $\eta/s$ are still challenging and it is therefore imperative to investigate whether this information can be extracted from empirical data. 

After much work on the extraction of an effective value of $\eta/s$ in relativistic heavy-ion collisions \cite{Gale:2013da}, the temperature dependence of the $\eta/s$ ratio has seen increased interest recently, using hadronic observables \cite{Niemi:2015qia,Denicol:2015nhu,Denicol:2015bnf} to quantify its behavior. A recent study \cite{Niemi:2011ix} has shown that the elliptic flow of charged hadrons as a function of transverse momentum at mid-rapidity is sensitive to a temperature-dependent $\eta/s$ in the hadronic phase both at the top RHIC energy and the LHC, while only LHC data are sensitive to the value of $\eta/s (T)$ in the QGP. This investigation continues in the same spirit, but using a complementary EM probe, and focusing on the $\eta/s$ ratio in the QGP. More specifically, the goal of this paper is to explore the sensitivity of thermal dileptons to a temperature-dependent $\eta/s$ at temperatures $T$ above 180 MeV and at top RHIC energy, in order to determine whether thermal dileptons break the degeneracy in $v^{ch}_2(p_T)$, shown in Ref. \cite{Niemi:2011ix}, and further be used to extract the value of the specific shear viscosity, and even possibly  of its low order derivatives as a function of $T$. This investigation focuses on thermal dileptons, rather than photons, because dileptons have an additional degree of freedom, the center of mass energy of the lepton pair, also known as the invariant mass, that allows us to separate the hadronic from partonic emission sources. As will be shown later in this contribution, small invariant mass dileptons are radiated preferentially by hadronic sources, while intermediate to high invariant mass dileptons originate mostly from partonic interactions.    

This paper is organized as follows: the next section gives the details of the relativistic fluid-dynamical modeling of the strongly interacting medium. Section \ref{rates} contains a discussion of the lepton pair emission rates in both the QGP and non-perturbative hadronic medium sectors, together with their respective viscous corrections. Results are shown and discussed in Sec. \ref{results}, followed by a conclusion. 

\section{Modeling the evolution of the medium created at RHIC}

\subsection{Viscous hydrodynamics}\label{sec:hydro}

In this work, we assume that the medium created in relativistic heavy-ion collisions very quickly reaches a state close to thermal equilibrium, such that relativistic dissipative fluid dynamics is a valid description of its space-time behavior. This assumption is supported by the good agreement between the measured  flow coefficients of charged hadrons and the ones calculated through fluid-dynamical simulations (see, e.g., \cite{Gale:2013da} for a recent review). In fluid dynamics, the energy-momentum tensor $T^{\mu \nu}$, satisfies the continuity equation,
\begin{equation}
\partial _{\mu }T^{\mu \nu }=0,  \label{1}
\end{equation}
where $T^{\mu \nu}=T^{\mu\nu}_0+\delta T^{\mu\nu}$. Inviscid (ideal) hydrodynamics is contained within $T^{\mu\nu}_0$, which is expressed as $T^{\mu\nu}_0=\varepsilon u^{\mu }u^{\nu }-\Delta ^{\mu \nu }P$, where $\varepsilon$ is the energy density, $P$ is the thermodynamic pressure, and $\Delta ^{\mu \nu}=g^{\mu \nu}-u^{\mu }u^{\nu }$ is the projection operator orthogonal to the four-velocity $u^\mu$, and $g^{\mu \nu} = {\rm diag}\{1, -1, -1, -1\}$. Throughout this study, deviations from ideal hydrodynamics appear exclusively via the shear viscous pressure tensor, i.e., $\delta T^{\mu\nu}=\pi^{\mu \nu}$, with all other dissipative effects being neglected. Furthermore, we set the net baryon four-current to vanish for all space-time points. The equation of state, which dictates how the thermodynamic pressure changes as a function of energy density, is taken from Ref. \cite{Huovinen:2009yb} and corresponds to a parametrization of a lattice QCD calculation, at high temperatures, smoothly connected to a parametrization of a hadron resonance gas at lower temperatures, which below $T_{\rm ch}=0.16$ GeV follows a partial chemical equilibrium (PCE) prescription \cite{Bebie:1991ij,Hirano:2002ds}. 

The dynamics of the shear-stress tensor is given by Israel-Stewart theory \cite{Israel1976310,Israel:1979wp,Baier:2007ix},
\begin{equation}
\tau _{\pi }\left[\Delta _{\alpha \beta }^{\mu \nu }u^\lambda \partial_\lambda \pi ^{\alpha\beta }+\frac{4}{3}\pi^{\mu \nu }\partial _{\lambda }u^{\lambda }\right]=2\eta \sigma ^{\mu \nu }-\pi ^{\mu \nu }\text{,}
\label{eq:shear_relax}
\end{equation}
where $\sigma^{\mu\nu}=\Delta _{\alpha\beta}^{\mu\nu}\partial^{\alpha}u^{\beta}$ is the shear tensor and $\Delta_{\alpha\beta}^{\mu\nu}=\frac{1}{2}\left( \Delta_{\alpha}^{\mu}\Delta_{\beta}^{\nu}+\Delta_{\beta}^{\mu}\Delta_{\alpha}^{\nu}\right)-\frac{1}{3}\Delta_{\alpha\beta}\Delta^{\mu\nu}$ is the double, symmetric, traceless projection operator. Israel-Stewart theory introduces two transport coefficients, the shear viscosity coefficient ($\eta$), which is already present in Navier-Stokes fluid dynamics, and the shear relaxation time $\tau_{\pi}$, germane to Israel-Stewart hydrodynamics. In this study, the relaxation time is fixed at $\tau_{\pi}=5\frac{\eta}{\varepsilon+P}$ \cite{Denicol:2012cn,Denicol:2014loa}. In principle, additional nonlinear terms exist in second order dissipative fluid dynamics \cite{Denicol:2012cn,Denicol:2014loa}, however we will not be studying their effects here. 

\begin{figure}[!h]
\begin{center}
\includegraphics[width=0.497\textwidth]{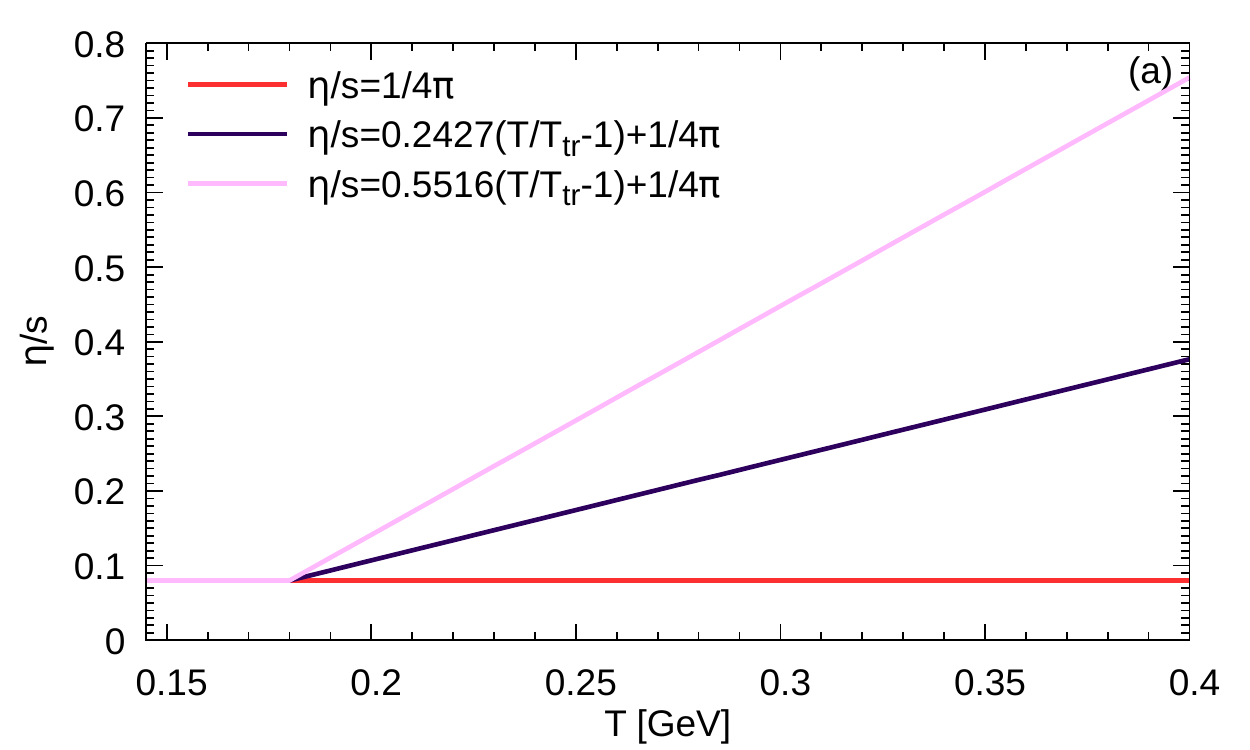}
\includegraphics[width=0.497\textwidth]{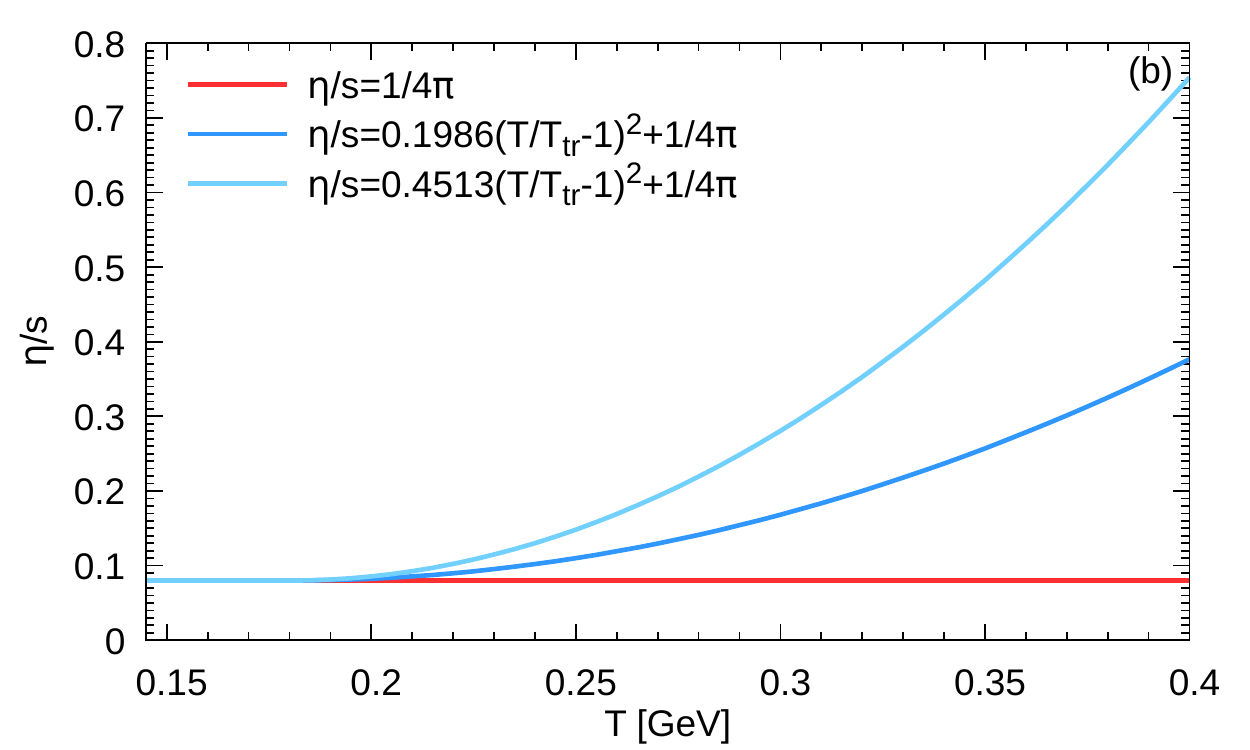}
\end{center}
\caption{(Color online) Linear (a) and quadratic (b) temperature dependence of $\eta/s$.}
\label{fig:eta_s_T}
\end{figure}

Presently, nonperturbative estimates of the aforementioned temperature-dependent transport coefficients in the strongly coupled regime are still a rare commodity \cite{Nakamura:2004sy,Meyer:2009jp,Haas:2013hpa,Christiansen:2014ypa}. Inspired by the recent Bayesian analysis within a hydrodynamical simulation \cite{Bernhard:2016tnd}, which shows an increase in $\eta/s(T)$ for temperatures above $\sim 180$ MeV, we focus on the growth of the specific shear viscosity at temperatures above the threshold $T_{tr}=180$ MeV in our hydrodynamical simulations. The growth of $\eta/s(T)$ at high temperature is also present in perturbative analysis \cite{Arnold:2003zc}. $\eta/s(T)$ is modeled by choosing two linear and two quadratic parametrizations of the temperature:
\begin{eqnarray}
\frac{\eta}{s}(T)= m\left(\frac{T}{T_{tr}}-1\right)+\frac{1}{4\pi}\text{,}\nonumber\\ 
\frac{\eta}{s}(T)= a\left(\frac{T}{T_{tr}}-1\right)^2+\frac{1}{4\pi}\text{,}
\label{eq:m_a_lin_quad}
\end{eqnarray}
where $T_{tr}=0.18$ GeV, while $m=0.5516$ and $a=0.4513$ are selected such that $\eta/s=0.755$ at $T=0.4$ GeV. Furthermore, the values $m=0.2427$ and $a=0.1986$ correspond to $\eta/s=0.3775$ at $T=0.4$ GeV. For temperatures below $T_{tr}$, $\eta/s=1/(4\pi)$. Figure \ref{fig:eta_s_T} shows all the various forms of temperature dependence used in this calculation. 
The goal of introducing different temperature-dependent $\eta/s$ is to investigate the sensitivity of thermal dileptons to this transport coefficient. The fluid-dynamical equations are solved numerically using \textsc{music}, which has recently been shown to be in very good agreement with semi-analytic solutions of Israel-Stewart theory \cite{Marrochio:2013wla}. A simulation using $\Delta \tau =0.03$ fm/$c$, a grid spacing of $\Delta x=\Delta y=1/6$ fm, and $\Delta \eta =1/5$ was precise enough to capture all the relevant physics present in the continuum limit.

\subsection{Initial conditions and hadronic particle production}\label{sec:IC}

As the initial conditions are not currently known in detail, especially those of the shear viscous pressure tensor, we assume that  $\pi^{\mu\nu}(\tau_0)\equiv0$ at $\tau_{0}=0.4$ fm/$c$ when the hydrodynamical evolution begins. Given that the incoming nuclei have a large longitudinal velocity, while their transverse velocity is assumed to be negligible, we initialize the local fluid velocity $u^\mu$ distribution to the Bjorken solution \cite{Bjorken:1982qr}. Thus, we have factorized the initial energy density profile containing a longitudinal part along the space-time rapidity ($\eta_s$) direction,\footnote{$\eta_s=\frac{1}{2}\ln\left[\frac{t+z}{t-z}\right]$.} and transverse part \cite{Hirano:2002ds} in the transverse ($x$--$y$) plane: 
\begin{equation*}
\varepsilon \left( \tau _{0},x,y,\eta_s \right) = \left\{\exp \left[ -\frac{\left(\left\vert \eta_s\right\vert -\eta _{\mathrm{flat}}/2\right) ^{2}}{2\eta _{\sigma}^{2}}\theta \left( \left\vert \eta_s\right\vert -\eta _{\mathrm{flat}}/2\right) \right]\right\}\left\{ W\left[\alpha n_{\mathrm{WN}}\left( x,y\right) + \left(1-\alpha\right) n_{\mathrm{BC}}\left(x,y\right) \right]\right\}\text{,}
\end{equation*}
where the transverse piece is being modeled according to the Monte Carlo (MC) Glauber prescription, while $n_{\mathrm{WN}}$ is the density of wounded nucleons, $n_{\mathrm{BC}}$ is the density of binary collision, $W$ is an overall normalization factor, and $\alpha$ is the proportion in which wounded nucleons and binary collisions contribute to the energy density profile in the transverse plane. The density of wounded nucleons and binary collisions is expressed as
\begin{equation*}
n_{\mathrm{BC/WN}}\left(x,y\right) =\frac{1}{2\pi\sigma^{2}}\sum_{i=1}^{N_{\mathrm{bin/part}}}\exp \left[ -\frac{\left( x-x_{i}\right)^{2}+\left( y-y_{i}\right) ^{2}}{2\sigma ^{2}}\right], 
\end{equation*}
where $N_{\mathrm{part}}$ and $N_{\mathrm{bin}}$ are the number of participants and binary collision of a given event, while ($x_{i}$,$y_{i}$) are the coordinates of the corresponding participant or binary collision on the transverse plane. In order to determine the number and coordinates of participants and binary collisions, the nucleon-nucleon inelastic cross section, $\sigma _{NN}=42.1$ mb at $\sqrt{s_{NN}}=200$ GeV, is used. Table \ref{table:IC} summarizes the parameters used by the MC Glauber model to describe the charged pion yield and charged hadron elliptic flow at RHIC in the 20--40\% centrality class (see also Ref. \cite{Vujanovic:2016anq}).
\begin{table}[!ht]
\caption{Initial state parameters.} 
\centering 
\begin{tabular}{c | c} 
Parameter & Value\\
\hline \hline 
$\eta_{\mathrm{flat}}$ & 5.9 \\
\hline 
$\eta_{\sigma}$ & 0.4 \\
\hline 
$W$ & 6.16 GeV/fm\\
\hline 
$\alpha$ & 0.25\\
\hline 
$\sigma$ & 0.4 fm\\
\hline 
\end{tabular}
\label{table:IC} 
\end{table}
Two hundred MC Glauber events were generated in this study for each of the four $\eta/s (T)$ parametrizations, along with another 200 events where $\eta/s =1/(4\pi)$. The same events in the 20--40\% centrality class are also used to compute dilepton observables.  

Hadron production proceeds through the Cooper-Frye prescription \cite{Cooper:1974mv}, where the dissipative degrees of freedom are converted to particles through the 14-moment Israel-Stewart (IS) approximation \cite{Teaney:2003kp}. The freeze-out temperature hypersurface was chosen to be $T_{FO}=145$ MeV \cite{Vujanovic:2016anq} and all two-- and three--particle decays of hadronic resonances up to 1.3 GeV are computed according to Ref. \cite{Sollfrank:1991xm}. 

\section{Thermal Dilepton Rates}
\label{rates}

Modern equations of state used to describe the medium in relativistic heavy ion collisions, such as the one used in this study, employ a continuous crossover phase transition between the partonic and the hadronic degrees of freedom. In the high temperature regime, perturbative partonic reactions are used to characterize the dilepton production rates, whereas in the low temperature sector, various hadronic interactions are responsible for dilepton radiation. The current calculation follows this prescription and describes the crossover region via a linear interpolation in temperature between the high and the low temperature regions, occurring at $0.184<T<0.22$ GeV \cite{Vujanovic:2016anq}. Specifically, the four-momentum dependent total dilepton rate density $\frac{d^4 R}{d^4 q}$ is:
\begin{equation}
\frac{d^4 R}{d^4 q} = f_{QGP} \frac{d^4 R_{QGP}}{d^4 q} + \left(1-f_{QGP}\right) \frac{d^4 R_{HM}}{d^4 q}\text{,}
\label{eq:f_QGP}
\end{equation}
where $\frac{d^4 R_{QGP}}{d^4 q}$ is the partonic dilepton rate and $\frac{d^4 R_{HM}}{d^4 q}$ is the hadronic dilepton rate, which are both defined in the following two sections. Last, $f_{QGP}$ is the QGP fraction is chosen such that $f_{QGP}=1$ for temperature $T>0.22$ GeV, $f_{QGP}=0$ for $T<0.184$ GeV and is linearly rising with temperature for $0.184<T<0.22$ GeV. Dilepton rates are integrated for all temperatures above $T_{FO}$. 

\subsection{Isotropic (inviscid) dilepton production rates}

The general expression for the rates, in the local rest frame, takes an elegant form:  
\begin{eqnarray}
\frac{d^4 R^{\ell^+\ell^-}}{d^4 q}=-\frac{L(M)}{M^2}\frac{\alpha^2_{EM}}{\pi^3 } \frac{\mathrm{Im}\Pi^R_{EM}(M,|{\bf q}|;T,\mu_B)}{e^{q^0/T}-1}\text{,} \label{eq:dilep_rate}
\end{eqnarray}
where $\mu_B=0$ in our hydrodynamical simulation, $M^2=q_{\mu}q^{\mu}$, $q^0=\sqrt{M^2+{\bf q}^2}$, $\alpha_{EM}=\frac{e^2}{4\pi}\approx \frac{1}{137}$, $L(M)=\left(1+\frac{2m^2_\ell}{M^2}\right)\sqrt{1-\frac{4m^2_\ell}{M^2}}$, $m_\ell$ is the lepton mass, $T$ is the temperature, and $\mathrm{Im}\Pi^R_{EM}$ is the imaginary part of the trace of the retarded (virtual) photon self-energy.

Recently, the perturbative thermal dilepton rates in the QGP have been computed at next-to-leading (NLO) \cite{Laine:2013vma,Ghisoiu:2014mha,Ghiglieri:2014kma} within a phenomenologically-interesting kinematic region. In a strongly-coupled setting, the Anti-de Sitter and conformal field theory correspondence has been used to compute emission rates of EM probes from non-Abelian plasmas exhibiting features similar to QCD plasmas \cite{CaronHuot:2006te}, while lattice calculations for thermal EM production \cite{Ding:2010ga,Ding:2013qw} are also available. However, all those rates are currently not amenable to a dissipative description of the medium, hence this study will focus on the QGP dilepton rate within the Born approximation. 

\subsubsection{Dilepton radiation from the QGP}

The Born dilepton rate can be written as: 
\begin{eqnarray}
\frac{d^4 R_{0}}{d^4 q} &=& \int \frac{d^3 k_1 d^3 k_2}{(2\pi)^6 k^0_1 k^0_2 } f_{\bf k_1} f_{\bf k_2} \frac{q^2}{2} \sigma \delta^4(q-k_1-k_2)\text{,} \nonumber\\
                 \sigma &=& \frac{16\pi \alpha_{\rm EM}^2 \left(\sum_{f'} e^2_{f'}\right) N_c}{3 q^2}\text{,}
\label{eq:R_born}
\end{eqnarray} 
where $f_{\bf k}$ is the quark/anti-quark distribution functions, $\sigma$ is the leading-order quark-antiquark annihilation (into a lepton pair) cross section, $N_c$ is the number of colors, the number of flavors is labeled by $f'$, where only the low-mass ones are considered, i.e., $f'=u,d,s$. Extending the isotropic dilepton rate in Eq. (\ref{eq:R_born}) to include shear-viscous effects, amounts to modifying the quark/antiquark Fermi-Dirac distribution functions ($f_{\bf k}$) to include anisotropic deformations. As shear viscosity increases in the QGP sector throughout $\eta/s(T)$, the dilepton rates become more sensitive to the form of the anisotropic correction to the dilepton rate ($\delta R$), and a systematic expansion of the anisotropic (or viscous) correction to the Born rates is presented in Section \ref{Sec:New_dR}. 

\subsubsection{Dilepton rates from the anisotropic Hadronic medium} \label{sec:HM_dileptons}

In the hadronic sector, we use the vector meson dominance Model (VDM), first proposed by Sakurai \cite{Gounaris:1968mw}, to relate the virtual photon self-energy $\mathrm{Im}\Pi^R_{EM}$ to the imaginary part of the retarded vector meson propagator ${\rm Im} D^{R}_V$, or, equivalently, the spectral function: 
\begin{equation}
{\rm Im} \Pi^R_{EM}=\sum_{V=\rho,\omega,\phi}\left(\frac{m^2_V}{g_V}\right)^2 {\rm Im} D^R_V. 
\end{equation}
In the above equation, vector mesons are denoted by $V=\rho,\omega,\phi$, with mass $m_V$, while their coupling to the photon is $g_V$. Since the Schwinger-Dyson equation relates the vector meson self-energy to the vector meson spectral function \cite{Roberts:1994dr}, it is sufficient to compute the vector meson self-energy to fully describe medium-induced modifications to the vector meson spectral function. Our approach to calculating the vector meson self-energy follows that of Eletsky {\it et al.} \cite{Eletsky:2001bb}. The vacuum piece of the self-energy is computed through chiral effective Lagrangians. On the other hand, the finite temperature contribution has been computed through the forward scattering amplitude approach, which includes experimentally observed resonances and Regge physics to account for scattering not going through resonances. Further details about the dilepton rates in the hadronic sector used within this work, including viscous corrections, are explored in detail in Ref. \cite{Vujanovic:2013jpa}.   

\subsection{A systematic expansion of the anisotropic (viscous) correction to dilepton production rate in the partonic medium} \label{Sec:New_dR}

Dilepton emission rates were recently extended to take into account deviations from local thermodynamic equilibrium in both the hadronic \cite{Vujanovic:2013jpa} and QGP sector, the latter being done in the Born limit \cite{Dusling:2008xj}. Such extensions are essential for a consistent calculation of dilepton production when a viscous fluid describes the evolution of the medium. In those calculations, the authors have generalized the single-quark distribution function to include anisotropic (viscous) correction using the 14-moment Israel-Stewart (IS) approximation. In the current calculation, we systematically expand the single-quark momentum distribution function to go beyond the IS approximation used in \cite{Dusling:2008xj,Vujanovic:2013jpa}, by solving the Boltzmann equation using the constant cross-section approximation. The same constant cross-section approximation was used when computing the transport coefficients in Eq. (\ref{eq:shear_relax}). The generalized version of the quark distribution function $f_{\bf k}$, present in the dilepton rate, takes the form:
\begin{eqnarray}
f_{\bf k}=\left[\exp\left(y_{\bf k}\right)+1\right]^{-1}\text{,} \label{eq:gen_df}
\end{eqnarray} 
where $y_{\bf k}=y(k^\nu, u^\nu; T, \mu)$. Assuming $y_{\bf k}=y_{0,{\bf k}}+\delta y_{\bf k}+\mathcal{O}\left((\delta y_{\bf k})^2\right)$, where $y_{0,{\bf k}}=(u^\nu k_\nu-\mu)/T$ and $\delta y_{\bf k} \ll y_{0,{\bf k}}$, we expand Eq. (\ref{eq:gen_df}) to linear order in $\delta y_{\bf k}$ obtaining 
\begin{eqnarray}
f_{\bf k} &=& f_{0,{\bf k}}+\delta f_{\bf k}\text{,}\nonumber\\
\delta f_{\bf k} &=& f_{0,{\bf k}}\left[1-f_{0,{\bf k}}\right]\delta y_{\bf k}\text{,}
\end{eqnarray}
where $f_{0,\bf k}=\left[\exp\left(y_{0,\bf k}\right)+1\right]^{-1}$. $\delta y_{\bf k}$ can be further expanded as  
\begin{eqnarray}
\delta y_{\bf k} = \mathcal{G}_{\bf k} \frac{\pi^{\mu\nu}k_\mu k_\nu}{\left[2T^2(\varepsilon+P)\right]},
\end{eqnarray}
where
\begin{eqnarray}
\mathcal{G}_{\bf k} = \left[\frac{1-\tanh(2x-2x_0)}{2}\right] \mathcal{G}^{\rm low}_{\bf k} + \left[\frac{1+\tanh(2x-2x_0)}{2}\right] \mathcal{G}^{\rm high}_{\bf k},
\end{eqnarray}
while $x=\frac{k\cdot u}{T}$ and $x_0=11.2$. The formal details of the solving the Boltzmann equation in the constant cross-section approximation are presented in Appendix A. Here, we simply quote the final result for $\mathcal{G}^{\rm low/high}_{\bf k}$: 
\begin{eqnarray}
\mathcal{G}^{\rm low}_{\bf k} &=& \frac{5.120}{0.1+x}\left[0.4046+0.1559x-7.405\times 10^{-3} x^2+1.693\times 10^{-4} x^3\right]\text{,} \,\, \nonumber \\
\mathcal{G}^{\rm high}_{\bf k} &=& \frac{161.8}{(0.1+x)^4}\left[-0.2587+0.4705x-0.2418x^2+6.547\times 10^{-2}x^3\right]. \,\,                                   
\label{eq:G_l-h_k}
\end{eqnarray}

Given that the functional form of $\delta y_{\bf k}$ is the same as in Refs. \cite{Dusling:2008xj,Vujanovic:2013jpa}, the same projection operator can be employed to compute the viscous correction to the dilepton rate in the QGP. That projection operator is:
\begin{eqnarray}
P_{\alpha\beta} & = & \frac{1}{2}\frac{g_{\alpha\beta}}{(u\cdot q)^2-q^2} + \frac{1}{2}\left[\frac{q^2+2(u\cdot q)^2}{\left[q^2-(u\cdot q)^2\right]^2}\right] u_{\alpha} u_{\beta} + \frac{3}{2}\frac{q_\alpha q_\beta}{\left[q^2-(u\cdot q)^2\right]^2} - \frac{3}{2}\left[\frac{u\cdot q}{\left[q^2-(u\cdot q)^2\right]^2}\right]\left(u_\alpha q_\beta + u_\beta q_\alpha \right) \text{,}
\end{eqnarray}
where $q^\mu$ is the four-momentum of the virtual photon. Using $P_{\alpha\beta}$, the viscous correction to the dilepton rate in the local rest frame of the medium is 
\begin{eqnarray}
\frac{d^4 \delta R}{d^4 q} & = & \frac{q^\alpha q^\beta \pi_{\alpha\beta}}{2T^2(\epsilon + P)} b_2\left(\frac{q^0}{T},\frac{|{\bf q}|}{T}\right), \nonumber\\
\frac{d^4 \delta R}{d^4 q} & = & \frac{q^\alpha q^\beta \pi_{\alpha\beta}}{2T^2(\epsilon + P)}\left\{C_q \frac{q^2}{2} \frac{\sigma}{(2\pi)^5} \frac{T^5}{|{\bf q}|^5} \int^{\frac{E_+}{T}}_{\frac{E_-}{T}} \frac{dE_{\bf k}}{T} f_{0,\bf k} \left[1-f_{0,\bf k}\right] f_{0}(q^0-E_{\bf k}) D'\right\},
\label{eq:dR_born}
\end{eqnarray} 
where $\frac{E_{\pm}}{T}=\frac{q^0\pm |{\bf q}|}{2T}$, $f_{0,\bf k}=\left\{\exp\left(E_{\bf k}/T\right)+1\right\}^{-1}$, $f_{0}(q^0-E_{\bf k})=\left\{\exp\left[(q^0-E_{\bf k})/T\right]+1\right\}^{-1}$, $D' = \left\{T^{-4}\left[(3q_0^2-|{\bf q}|^2) E^2_{\bf k}-3 q^0 E_{\bf k} q^2 + \frac{3}{4} q^4\right]\right\} \mathcal{G}_{\bf k}$, and $C_{q} \approx 0.99$. IS $\delta R$ is recovered by setting $\mathcal{G}_{\bf k}=1$. The complete Born rate can therefore be expressed as $\frac{d^4 R}{d^4 q}=\frac{d^4 R_0}{d^4 q}+\frac{d^4 \delta R}{d^4 q}$, where the first and second terms are found in Eqs. (\ref{eq:R_born}) and (\ref{eq:dR_born}), respectively.

To appreciate the improvement the generalized $\delta y_{\bf k}$ in Eq. (\ref{eq:dR_born}) brings relative to the IS viscous correction, one cannot compare $\delta R$ to $R_0$ directly, as the viscous correction depends on the size of $\pi^{\mu\nu}$ at every space-time point. However, one can compare the envelope of the viscous correction $b_2$ to the ideal QGP dilepton rate. So, intuition on the behavior of the viscous correction will instead be acquired through the ratio 
\begin{eqnarray}
A\left(\frac{q^0}{T},\frac{|{\bf q}|}{T}\right)=\frac{b_2\left(\frac{q^0}{T},\frac{|{\bf q}|}{T}\right)}{\frac{d^4 R_0}{d^4 q}}
\label{eq:ratio_A} 
\end{eqnarray}
evaluated in the local rest frame. The ratio $A$ has a very weak dependence on $\frac{|{\bf q}|}{T}$, hence evaluating it at $\frac{|{\bf q}|}{T}=0$ is sufficient. 
\begin{figure}[!h]
\begin{center}
\includegraphics[width=0.52\textwidth]{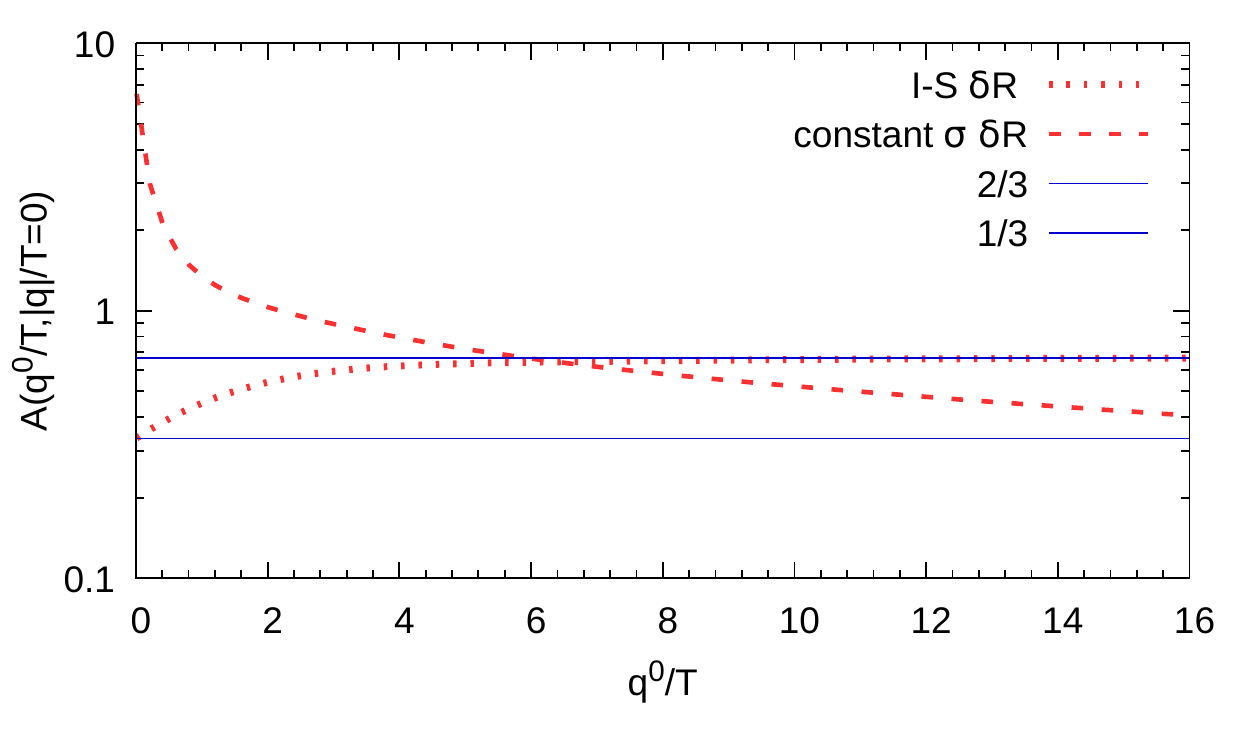}
\end{center}
\caption{(Color online) Relative size of the envelope of the viscous correction relative to the (ideal) isotropic rate in the local rest frame.}
\label{fig:size_of_dR}
\end{figure}
Figure \ref{fig:size_of_dR} clearly shows that $A$ for the IS viscous correction is bounded between $\frac{1}{3}$ and $\frac{2}{3}$. Since $\frac{q^\alpha q^\beta \pi_{\alpha\beta}}{2T^2(\epsilon + P)}$ is well behaved in the vanishing $q^{\mu}$ limit, the lower bound on $A$ is not a source of concern. Using only the upper bound, the IS correction to the QGP dilepton rate becomes ill-behaved when $\frac{q^\alpha q^\beta \pi_{\alpha\beta}}{2T^2(\epsilon + P)}>\frac{3}{2}$, thus making $\delta R > R_0$. In that respect, the viscous correction that we have computed is better behaved at large $q^0/T$ as $A\sim T/q^0$, and is furthermore is finite at $q^0/T=0$. This suppression at large $q^0/T$ is needed to ensure that $\delta R$ is well behaved when a large $\pi^{\mu\nu}$ is present, due to a $\eta/s(T)$. The effects of the constant cross-section anisotropic $\delta R$ correction and the IS $\delta R$ on the dilepton differential yield will be explored in Appendix B.  
       
In the Hadronic medium (HM), the IS viscous correction to the dilepton rate, presented in Ref. \cite{Vujanovic:2013jpa}, has been shown to be small, relative to the inviscid contribution, and thus well behaved. This statement remains true once a temperature-dependent specific shear viscosity is introduced, which affects the HM dilepton rate in the region $0.18<T<0.22$ GeV. Hence, an improved description of viscous correction in the HM is not warranted.  

\section{Results}
\label{results}

Before disclosing the effects of $\eta/s(T)$ on dilepton flow, it is important to specify the manner in which dilepton flow coefficients are computed. Earlier dilepton calculations using smooth initial conditions have computed the dilepton elliptic flow coefficient using the event plane method \cite{Chatterjee:2007xk,Vujanovic:2013jpa}. A recent dilepton study using MC Glauber initial conditions \cite{Vujanovic:2016anq} employs the scalar product method to compute flow coefficients. The present study continues to use the scalar product method, such that
\begin{eqnarray}
v^{\gamma^*}_n(X)&=&\frac{\frac{1}{N_{ev}}\sum^{N_{ev}}_{i=1}v^{\gamma^{*}}_{n,i}(X)  v^{h}_{n,i} \cos \left[n\left(\Psi^{\gamma^{*}}_{n,i}(X)-\Psi^h_{n,i} \right)\right]}{\sqrt{\frac{1}{N_{ev}}\sum^{N_{ev}}_{i=1} (v^{h}_{n,i})^2}}\nonumber\\
                &=&\frac{\left\langle v^{\gamma^{*}}_{n,i}(X)  v^{h}_{n,i} \cos \left[n\left(\Psi^{\gamma^{*}}_{n,i}(X)-\Psi^h_{n,i} \right)\right] \right\rangle_{{\rm ev}, i}}{\sqrt{\left\langle (v^h_{n,i})^2 \right\rangle_{{\rm ev},i}}},
\label{eq:vnSP}
\end{eqnarray}
where $N_{ev}=200$, $X$ is any dynamical variable such as $M$ or $p_T$, and $\langle \ldots \rangle_{{\rm ev}, i}$ is the average over events $i$. In a single event $i$, the hadronic $v^h_{n,i}$ and $\Psi^h_{n,i}$ are given by
\begin{equation}
v^h_{n,i} e^{i n \Psi^h_{n,i}} = \frac{\int d p_T  dy d\phi p_T \left[ p^0 \frac{d^3 N^h_i}{d^3 p} \right] e^{i n\phi}}{\int d p_T dy  d\phi p_T \left[ p^0 \frac{d^3 N^h_i	}{d^3 p} \right]},
\label{eq:vn_ev}
\end{equation}
where the charged hadron distribution is integrated over $-0.35<\eta<0.35$ and $0.035<p_T<3$~GeV to simulate acceptance used by the PHENIX experiment at RHIC. The dilepton $v^{\gamma^{*}}_{n,i}$ and $\Psi^{\gamma^{*}}_{n,i}$ are computed using the same approach, with the more general distribution $\frac{d^4 N_{i}^{\gamma^{*}}}{d^4 p}$. 

\subsection{Linear $\eta/s (T)$}\label{sec:lin_eta_s_T}

The goal of this section is to investigate the sensitivity of thermal dileptons to the size of $\eta/s(T)$'s slope. Since the effects a temperature-dependent $\eta/s$ induces on the evolution of the medium are rather complicated, keeping identical initial/freeze-out conditions, regardless of any entropy production that $\eta/s(T)$ introduces, is important for the purpose of a comparison. 

\begin{figure}[!h]
\begin{center}
\includegraphics[width=0.495\textwidth]{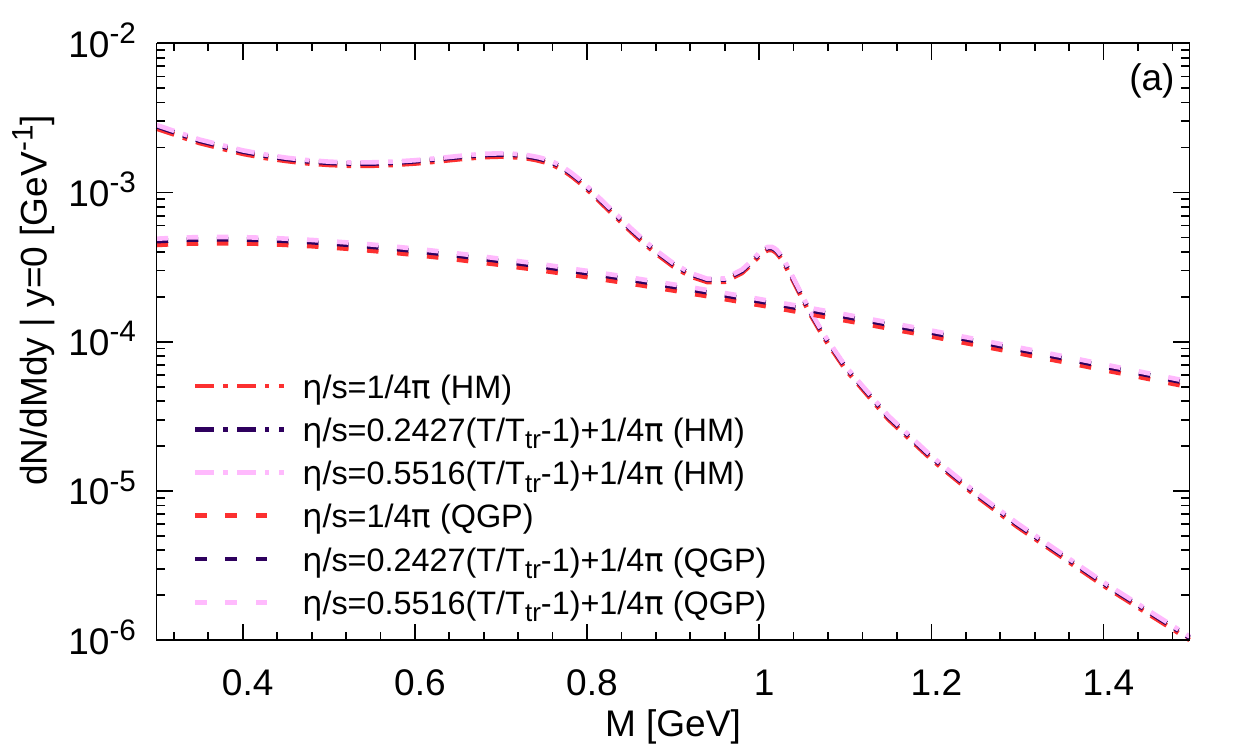}
\includegraphics[width=0.495\textwidth]{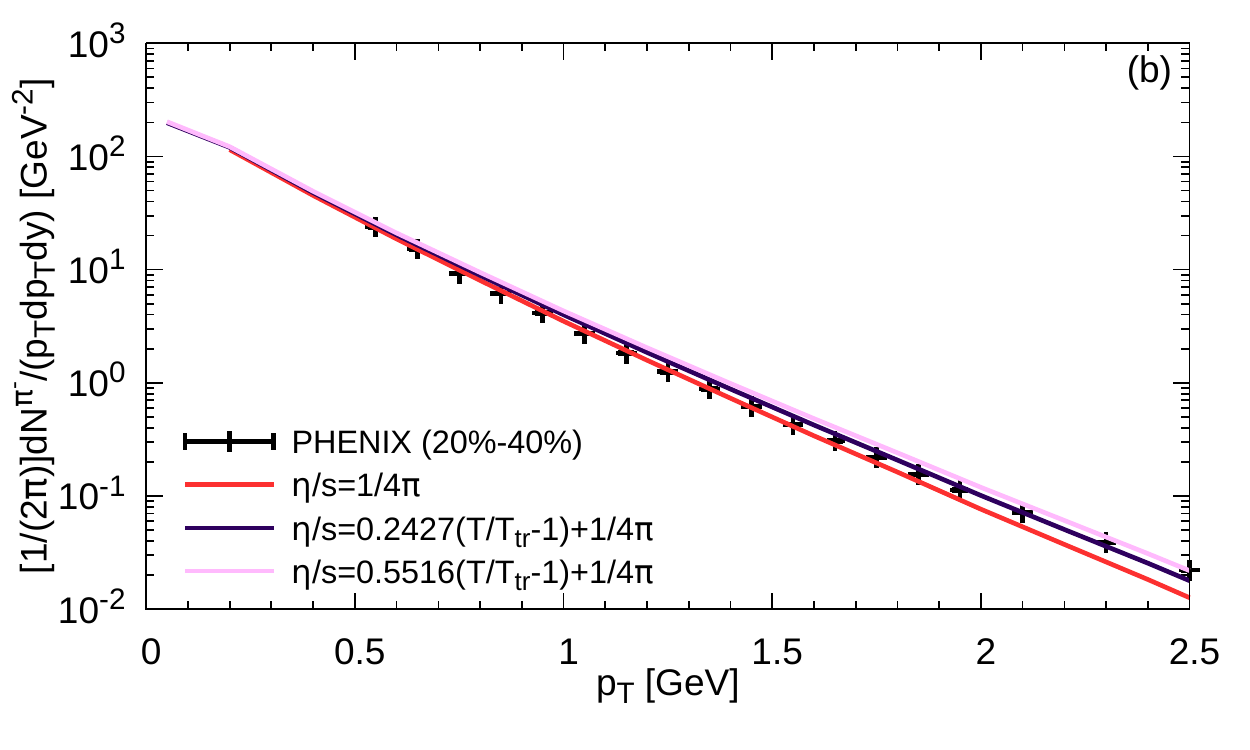}
\end{center}
\caption{(Color online) Yield of thermal dileptons (a) and pions (b) for various values of $\eta/s(T)$. All results are in the 20--40\% centrality class.}
\label{fig:entropy_prod_radial_exp_lin}
\end{figure}

\begin{figure}[!h]
\begin{center}
\includegraphics[width=0.495\textwidth]{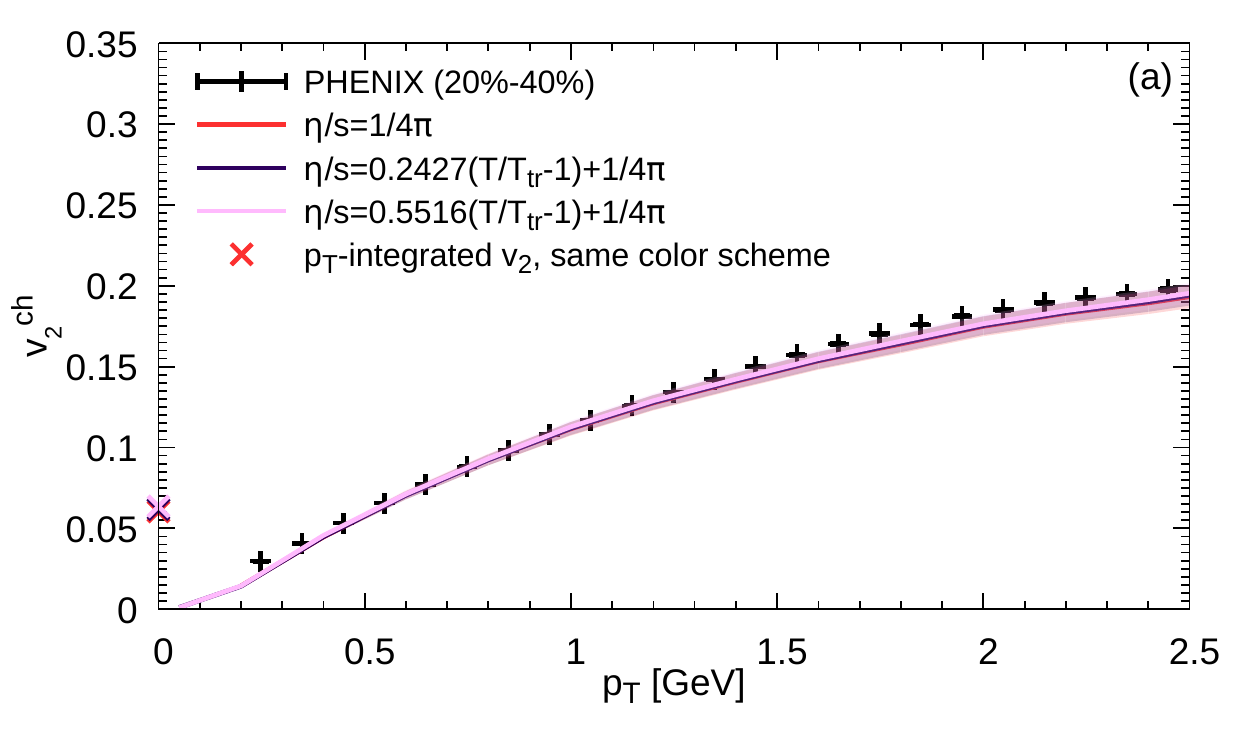}
\includegraphics[width=0.495\textwidth]{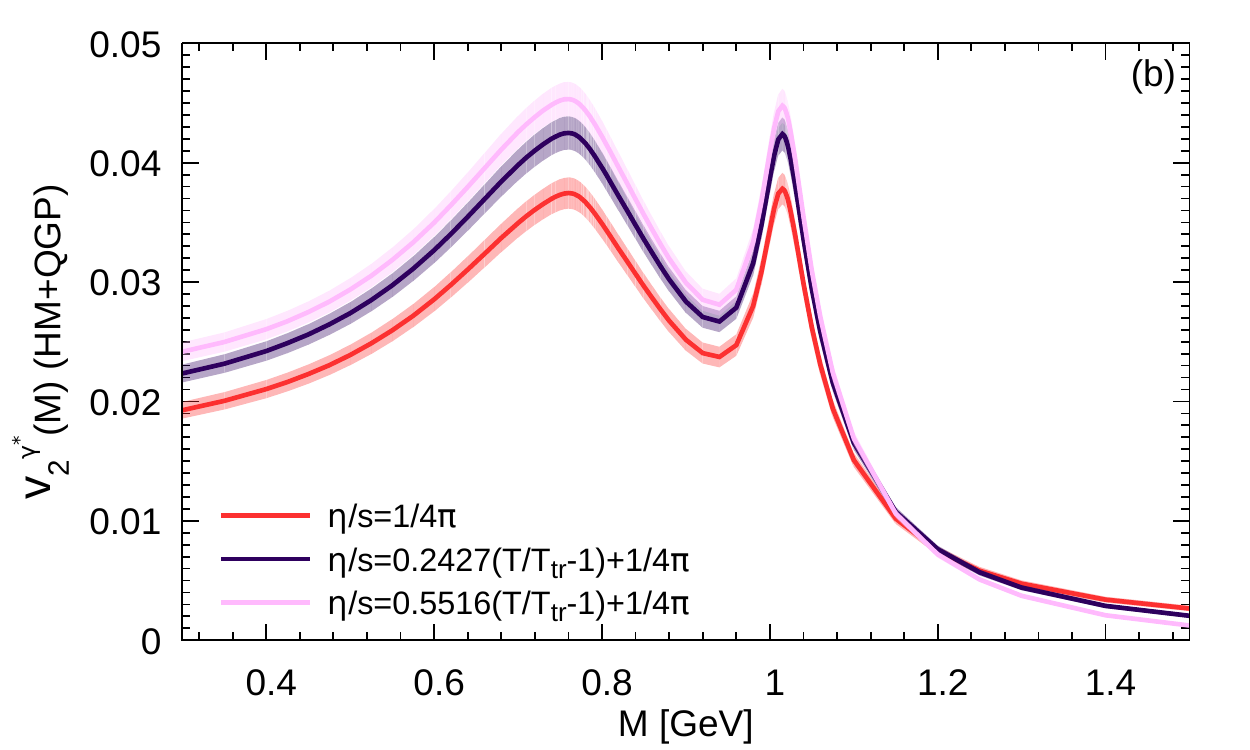}
\end{center}
\caption{(Color online) Elliptic flow of charged hadrons (a) and thermal dileptons (b) with different slopes of $\eta/s(T)$. Each colored band represents the statistical error associated with the 200 events run. All results are in the 20--40\% centrality class.}
\label{fig:v2_ch_th_dilep_lin}
\end{figure}

To quantify the amount of entropy, and radial flow generated via a linearly dependent $\eta/s(T)$, as well as the importance of $\delta f$ effects, the yield of thermal dileptons as a function of $M$ and the yield of pions as a function of $p_T$ is plotted in Fig. \ref{fig:entropy_prod_radial_exp_lin}. 
The invariant mass thermal dilepton yield is very slightly modified owing to $\eta/s(T)$ as seen in Fig. \ref{fig:entropy_prod_radial_exp_lin} (a). Indeed, the yield is increased by 5\% in the HM region while the QGP region receives an increase of 10\%. Since $M$ is a Lorentz-invariant quantity, while the invariant mass yield is unaffected by viscous corrections, the increase in the dilepton invariant mass yield is a consequence of the entropy production of a dissipative system. The somewhat larger increase in the pion yield at higher $p_T\gtrsim 1$ GeV [see Fig. \ref{fig:entropy_prod_radial_exp_lin} (b)] is dominated by a combination of a greater radial flow and larger $\delta f$ contribution when $\eta/s(T)$ is present relative to $\eta/s=1/(4\pi)$, while at low $p_T\lesssim 1$ GeV greater entropy production and radial flow give the main contribution to the increase in pion yield. The larger radial flow generated by $\eta/s(T)$ is however not affecting the elliptic flow of charged hadrons at top RHIC energy as can be seen in Fig. \ref{fig:v2_ch_th_dilep_lin} (a), and was first noticed in Ref. \cite{Niemi:2011ix}.  
On the other hand, a linearly dependent $\eta/s(T)$ changes the elliptic flow of thermal dileptons quite substantially [see Fig. \ref{fig:v2_ch_th_dilep_lin} (b)], with the effect being so large that it may potentially be measured in experiment. At this point, it is important to highlight the features that distinguish the effects of $\eta/s(T)$ from our earlier study in Ref. \cite{Vujanovic:2016anq}, where the manner in which relaxation time $\tau_\pi$ and the initial condition of $\pi^{\mu\nu}$ affect the $v_2(M)$ of thermal dileptons was investigated. As can be seen in Fig. \ref{fig:v2_ch_th_dilep_lin} (b), $\eta/s(T)$ causes an increase in the thermal $v_2(M)$ in the region where HM dileptons dominate, namely for $M\lesssim 1.15$ GeV. For $M\gtrsim 1.15$ GeV, where QGP dilepton production becomes the main source, a temperature dependent specific shear viscosity decreases $v_2(M)$. On the other hand, the effects on the $v_2(M)$ observed by increasing $\tau_\pi$, as explored in Ref. \cite{Vujanovic:2016anq}, go in the opposite direction, namely the $v_2(M)$ is decreased for $M\lesssim 1.15$ GeV and increased for $M\gtrsim 1.15$ GeV. Thus the effects of $\eta/s(T)$ are distinct from those associated with $\tau_\pi$. If, on the other hand, one compares the effects of initial conditions of $\pi^{\mu\nu}$ on dilepton $v_2 (M)$, as also studied in Ref. \cite{Vujanovic:2016anq}, then one notices that increasing initial $\pi^{\mu\nu}$ increases $v_2(M)$ of dileptons, which is not what is observed in the present study. Hence, the effects of $\eta/s(T)$ are different from those due to $\tau_\pi$ or initial conditions of $\pi^{\mu\nu}$. However, the next generation of fluid dynamical approaches should see dynamically-calculated initial shear pressure tensor in conjunction with temperature-dependent transport parameters. This will enable a new level of characterization of the initial states present in models of hadronic collisions. 

\begin{figure}[!h]
\begin{center}
\includegraphics[width=0.495\textwidth]{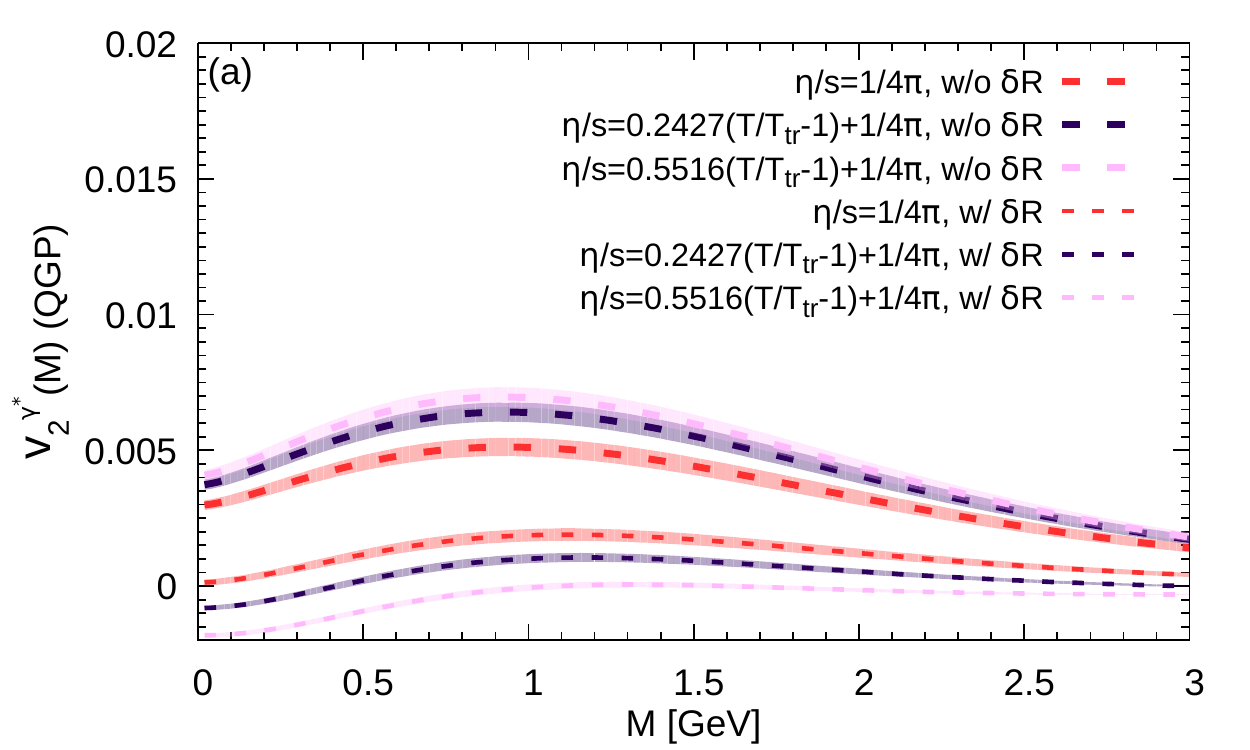}
\includegraphics[width=0.495\textwidth]{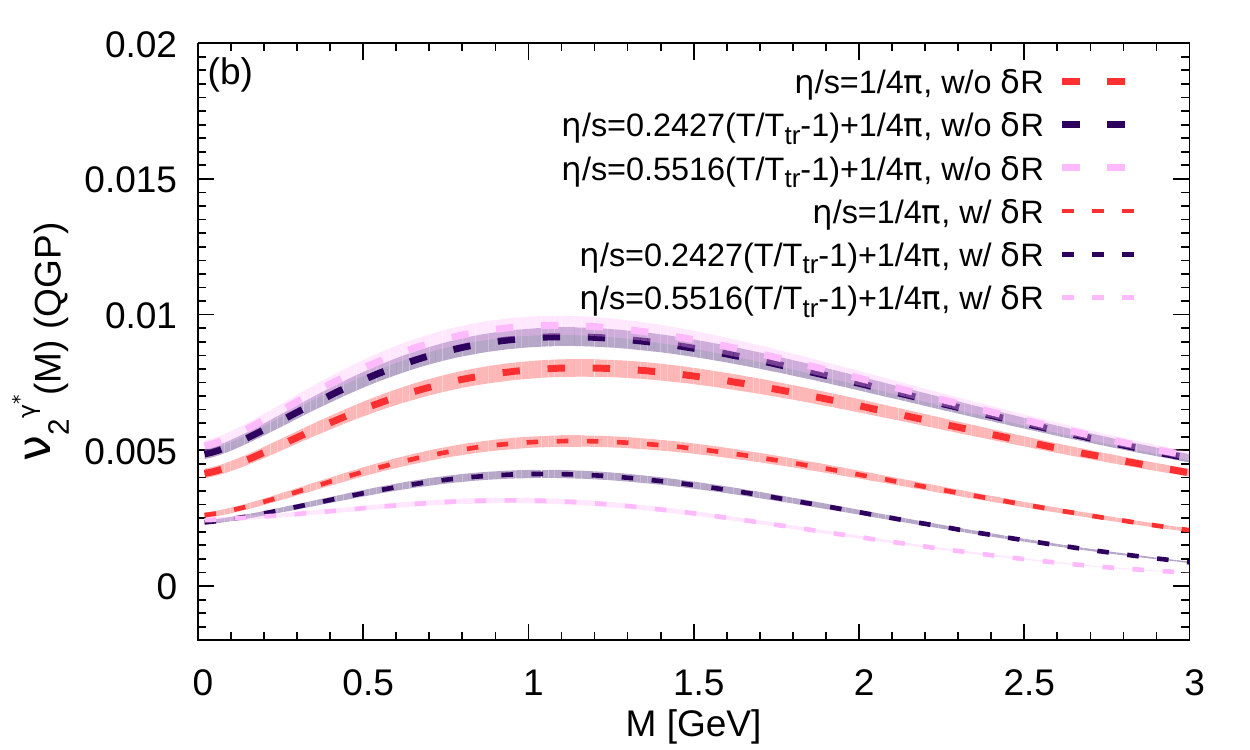}
\end{center}
\caption{(Color online) Elliptic flow $v_2(M)$ (a) and $\nu_2(M)$ (b) of QGP dileptons for different slopes of $\eta/s(T)$ using the constant cross section $\delta R$. The definition of $\nu_2(M)$ is in Eq. (\ref{eq:vn_pseudoSP}). Each colored band represents the statistical error associated with the 200 events generated. All results are in the 20--40\% centrality class.}
\label{fig:v2_nu2_M_qgp_dilep_lin}
\end{figure}

Having established the features that are associated with $\eta/s(T)$, we isolate in Fig. \ref{fig:v2_nu2_M_qgp_dilep_lin} the QGP contribution to dilepton anisotropic flow in order to explore how QGP dileptons are influenced by $\eta/s(T)$. For the moment only the constant cross-section $\delta R$ is being used since it is this viscous correction that is used in Fig. \ref{fig:v2_ch_th_dilep_lin} (b). To better appreciate all the effects of the constant cross section $\delta R$, a new variable $\nu^{\gamma^*}_2(M)$ is defined: 
\begin{equation}
\nu^{\gamma^*}_2 (M) =\frac{\left\langle v^{h}_{2,i} v^{\gamma^{*}}_{2,i} (M) \right\rangle_{{\rm ev},i}}{\sqrt{\left\langle (v^h_{2,i})^2 \right\rangle_{{\rm ev},i}}}
\label{eq:vn_pseudoSP}
\end{equation}
where $\langle \ldots \rangle_{{\rm ev},i}$ is an average over events as defined in Eq. (\ref{eq:vnSP}), and the sum over $i$ has implicitly been performed. $\nu^{\gamma^*}_2(M)$ is constructed such that it is not sensitive to event plane angle misalignment between $\Psi^{\gamma^{*}}_2 (M)$ and $\Psi^h_2$ and is therefore only sensitive to the manner in which magnitude of the $v_2$ of charged hadrons and dileptons is affected by $\eta/s(T)$. On the other hand, $v^{\gamma^*}_2(M)$ on the left hand side of Eq. (\ref{eq:vnSP}) is sensitive to both the overall magnitude of $v^{\gamma^*}_2(M)$ and $v^h_2$, as well as the change in the relative angle between $\Psi^{\gamma^{*}}_2(M)$ and $\Psi^h_2$. 
Note that the $p_T$-integrated charged hadron $v_2$ is essentially unaffected by whether the medium has a constant or a temperature-dependent specific shear viscosity [see Fig. \ref{fig:v2_ch_th_dilep_lin} (a)] and therefore, any effects of $\eta/s(T)$ are coming from the numerator of Eqs. (\ref{eq:vnSP}) and (\ref{eq:vn_pseudoSP}). With that in mind, including the viscous correction $\delta R$ to the dilepton rate, a temperature-dependent specific shear viscosity has two effects: one on the magnitude of $v^h_2$ and $v^{\gamma^*}_2(M)$ present in the numerator on the right hand side of Eq. (\ref{eq:vnSP}) and the other on orientation between the event planes denoted by $\Psi^{\gamma^{*}}_2(M)$ and $\Psi^h_2$. On average, $\eta/s(T)$ reduces the overall magnitude of the product $v^{\gamma^*}_2(M) v^h_2$ once $\delta R$ is included, as is clearly depicted in Fig. \ref{fig:v2_nu2_M_qgp_dilep_lin} (b). The change in the preferential emission direction of charged hadrons versus that of dileptons can be appreciated by comparing $v_2(M)$ and $\nu_2(M)$ presented in Fig. \ref{fig:v2_nu2_M_qgp_dilep_lin} (a) and  \ref{fig:v2_nu2_M_qgp_dilep_lin} (b), respectively. Indeed, a temperature-dependent $\eta/s$ can have such a strong effect on the misalignment of $\Psi^{\gamma^{*}}_2(M)$ and $\Psi^h_2$, that instances of ``anti-correlation'', i.e., regions of invariant mass where $\pi/2<2(\Psi^{\gamma^{*}}_2(M)-\Psi^h_2)<3\pi/2$, occur and generate negative $v_2(M)$. Note that without $\delta R$, the event plane angles $\Psi^{\gamma^{*}}_2(M)$ and $\Psi^h_2$ are not aligned, however there are no instances of ``anticorrelation''. In sum, both effects, namely the reduction, on average, in the overall magnitude of the product $v^{\gamma^*}_2(M) v^h_2$ and the anticorrelation between $\Psi^{\gamma^{*}}_2(M)$ and $\Psi^h_2$ depicted in Fig. \ref{fig:v2_nu2_M_qgp_dilep_lin}, are generated by including the constant cross-section $\delta R$ in the QGP dilepton rate. Figure \ref{fig:v2_nu2_M_qgp_dilep_lin} is also showing that the larger the absolute value of $\pi^{\mu\nu}$ is [see Fig. \ref{fig:T_munu_aniso_and_pi_munu_lin} (b)], the larger the viscous correction is, which manifests itself in large effects on $v_2(M)$ and $\nu_2(M)$. 

The effects of inserting a different the envelope function, i.e., a different coefficient $b_2$ in Eq. (\ref{eq:dR_born}), on $v^{\gamma^*}_2(M)$ and $\nu^{\gamma^*}_2(M)$ are explored in Fig. 6. Two cases are presented: one where the IS $\delta R$ is used, which is obtained by setting $\mathcal{G}_{k}=1$ in Eq. (14), and the other where the constant cross-section $\delta R$ is used, with $b_2$ defined in Eq. (14).
\begin{figure}[!h]
\begin{center}
\includegraphics[width=0.495\textwidth]{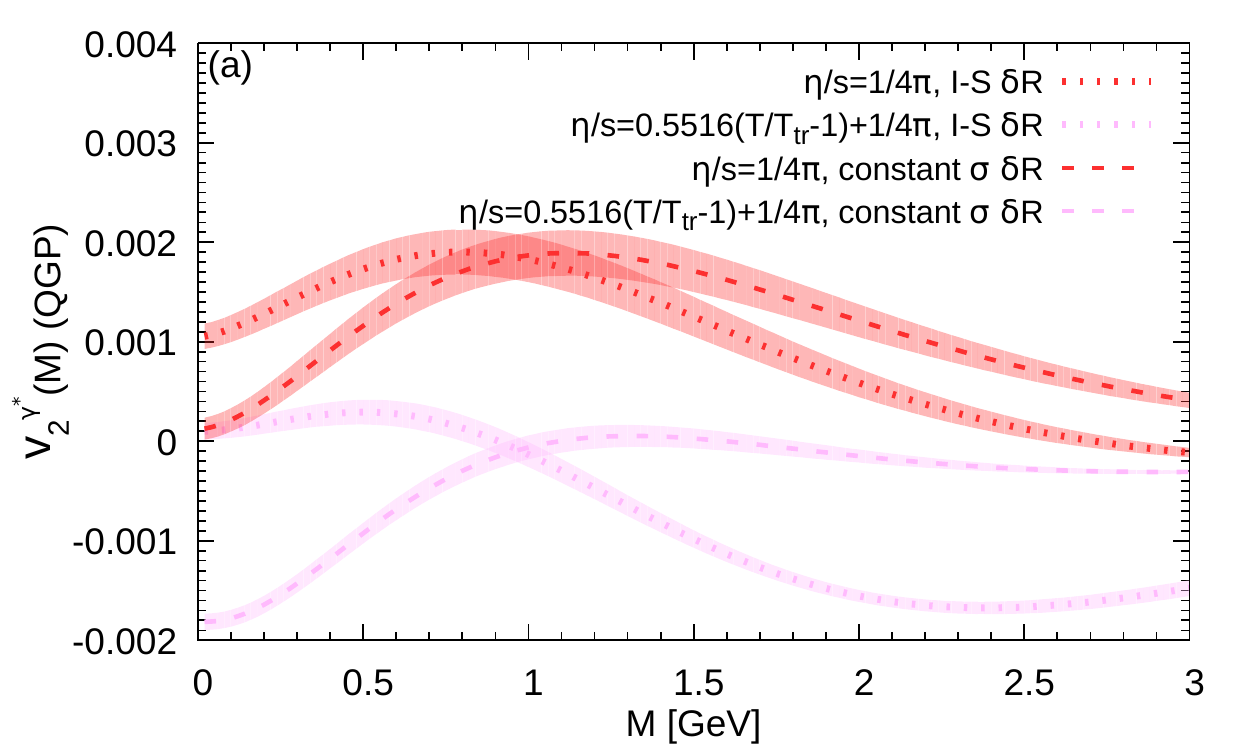}
\includegraphics[width=0.495\textwidth]{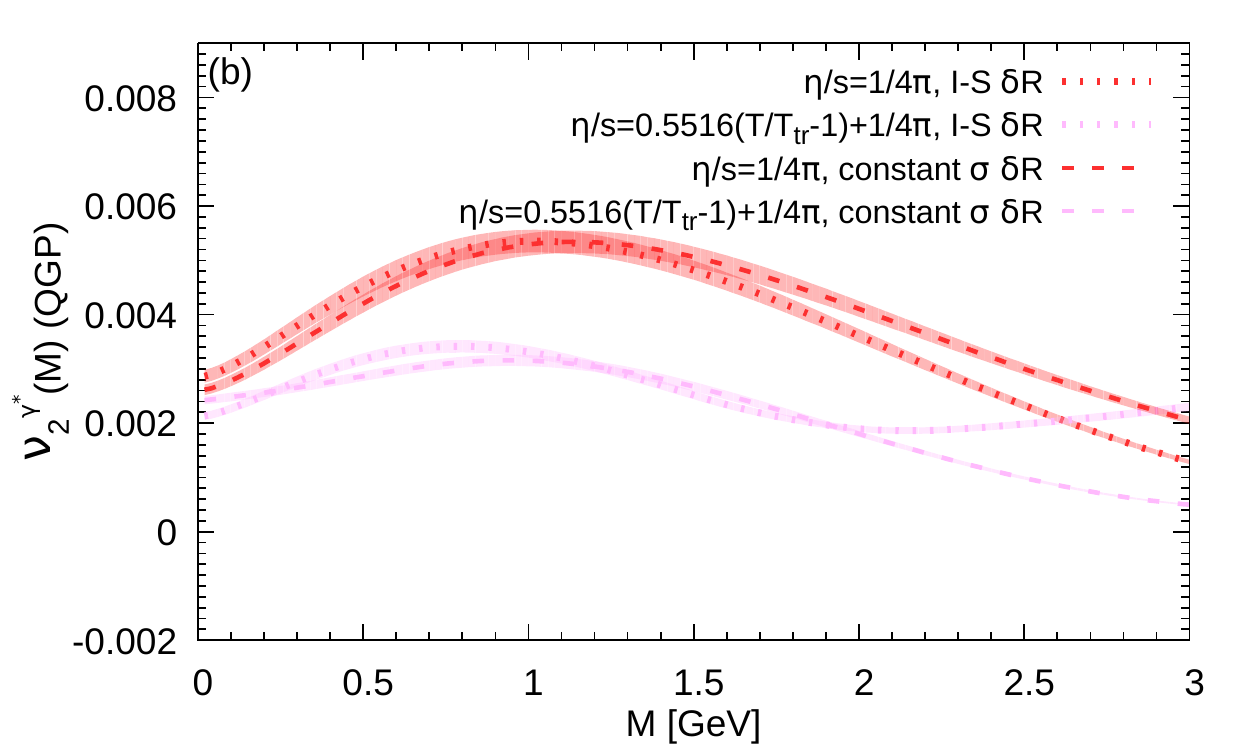}
\end{center}
\caption{(Color online) Elliptic flow $v_2(M)$ (a) and $\nu_2(M)$ (b) of QGP dileptons for IS $\delta R$ and constant cross-section $\delta R$ for both $\eta/s=1/(4\pi)$ and $\eta/s(T)$.  Each colored band represents the statistical error associated with the 200 events generated. All results are in the 20--40\% centrality class.}
\label{fig:v2_nu2_M_qgp_dilep_old_new_dR_lin}
\end{figure}
Since the envelope, denoted by $b_2$, doesn't affect the $v_2$ of charged hadrons, the effects of the envelope on the magnitude on dilepton flow anisotropy can be appreciated by first focusing on $\nu^{\gamma^*}_2(M)$ at $\eta/s=1/(4\pi)$. Comparing to the result without $\delta R$ (see Fig. \ref{fig:v2_nu2_M_qgp_dilep_lin} (b)), Fig. \ref{fig:v2_nu2_M_qgp_dilep_old_new_dR_lin} (b) shows that the constant cross section $\delta R$ suppresses the $\nu^{\gamma^*}_2(M)$ in the $M<1$ GeV region more than the IS $\delta R$ does. For $M>1$ GeV, the constant cross section $\delta R$ suppresses the $\nu^{\gamma^*}_2(M)$ less than the IS $\delta R$. So, for the case $\eta/s=1/(4\pi)$, the entire invariant mass behavior of $\nu^{\gamma^*}_2$ for both $\delta R$s is consistent with what one would expect by examining Fig. \ref{fig:size_of_dR}.  

The energy dependence of $\delta R$ also affects the final $v_2(M)$ of QGP dileptons as shown in Fig. \ref{fig:v2_nu2_M_qgp_dilep_old_new_dR_lin} (a), thus emphasizing the effects of $\delta R$ on the relative angle between $\Psi^{\gamma^{*}}_2(M)$ and $\Psi^h_2$. Recall that most of the contribution to the invariant mass distribution of dilepton yield and $v_2$ is dominated by the low $p_T$ region, with higher $p_T$ regions being exponentially suppressed. The larger the correction to the dilepton yield is [see Fig. \ref{fig:deltaN_over_N0_vs_M} of Appendix \ref{sec:size_of_dR}], the larger the relative angle between $\Psi^{\gamma^{*}}_2(M)$ and $\Psi^h_2$ is. So, the constant cross-section $\delta R$ has the strongest effect on the $2\left[\Psi^{\gamma^{*}}_2(M)-\Psi^h_2\right]$ at low $M$, while for IS $\delta R$ this happens at larger $M$, which is consistent with Fig. \ref{fig:size_of_dR}. 

Having explored the effects of the four-momentum dependence of the constant cross-section $\delta R$, the effects of $\eta/s(T)$ on $v_2(M)$ of QGP dileptons are now investigated by inspecting the manner in which the evolution of the hydrodynamic momentum anisotropy $\langle (T^{xx}-T^{yy})/(T^{xx}+T^{yy})\rangle$ is modified under the influence of viscosity. The hydrodynamic momentum anisotropy $\langle (T^{xx}-T^{yy})/(T^{xx}+T^{yy})\rangle$ is computed in a way that represents, as closely as possible, how this quantity is probed by dilepton radiation. Indeed, dileptons are sensitive to the sum/difference of $T^{xx}_i (\tau, x,y,\eta_s)$ and $T^{yy}_i(\tau, x,y,\eta_s)$ in every fluid cell of every hydrodynamical event $i$. Since dilepton rates are being space-time integrated for each hydrodynamical simulation before the individual events are combined, the hydrodynamical momentum anisotropy is computed in that order as well. Furthermore, as temperature goes down, an interpolation between the QGP and HM dilepton rates occurs. So, the hydrodynamic momentum anisotropy is calculated taking into account that interpolation. Thus, identification of the  hydrodynamical momentum anisotropy in the QGP sector is possible through:
\begin{equation}
\left\langle \frac{T^{xx}-T^{yy}}{T^{xx}+T^{yy}}\right\rangle=\frac{1}{N_{ev}}\sum^{N_{ev}}_{i=1} \left\{\frac{\int \tau d\eta_s dy dx f_{QGP}(T) \left[T^{xx}_i\left(x^\mu\right) - T^{yy}_i\left(x^\mu\right)\right]}{\int \tau d\eta_s dy dx f_{QGP}(T) \left[T^{xx}_i \left(x^\mu \right) + T^{yy}_i \left(x^\mu\right)\right]}\right\}.
\end{equation} 
where $x^\mu=(\tau, x,y,\eta_s)$, $f_{QGP}$ is defined in Eq. (\ref{eq:f_QGP}) and represents the fraction of the cell in the QGP sector, $T$ is the temperature, and $N_{ev}=200$ events. When studying the HM sector, one simply uses $(1-f_{QGP})$ when computing $\langle (T^{xx}-T^{yy})/(T^{xx}+T^{yy})\rangle$. The anisotropy on the freeze-out surface will be computed via
\begin{eqnarray}
\left\langle \frac{T^{xx}-T^{yy}}{T^{xx}+T^{yy}}\right\rangle&=&\frac{1}{N_{ev}}\sum^{N_{ev}}_{i=1} \left\{\frac{\int d^3 \Sigma_\mu \left(x^\mu \right) u^\mu \left(x^\mu \right) B(\tau) \left[T^{xx}_i \left(x^\mu \right) - T^{yy}_i\left(x^\mu \right)\right]}{\int d^3\Sigma_\mu \left(x^\mu \right) u^\mu \left(x^\mu \right) B(\tau)  \left[T^{xx}_i \left(x^\mu \right) + T^{yy}_i\left(x^\mu \right)\right]}\right\}, \nonumber\\
B(\tau)&=&\left\{ \begin{array}{rl}
                                  1 & \tau \in [\tau_j-\frac{\Delta \tau}{2}, \tau_j+\frac{\Delta \tau}{2})\\
                                  0 & {\rm otherwise}
                               \end{array}
                        \right.
\end{eqnarray}
where $\tau_j=\tau_0+j\Delta\tau$ with $j\in\mathbb{N}$, $d^3 \Sigma_\mu$ is the infinitesimal volume element orthogonal to the freeze-out hypersurface, $u^\mu$ is the flow profile on the freeze-out hypersurface and $\Delta \tau=0.03$ fm/$c$ is the hydrodynamical time step used to propagate the fluid equations forward in time (see Section \ref{sec:hydro}). Figure \ref{fig:T_munu_aniso_and_pi_munu_lin} (a) shows the hydrodynamical momentum anisotropy in the QGP. In that figure, ideal $T^{\mu\nu}$ refers to the momentum anisotropy $\langle (T^{xx}-T^{yy})/(T^{xx}+T^{yy})\rangle$ computed using only $T^{\mu\nu}_0$ of a viscous evolution, while the full $T^{\mu\nu}$ curves also include $\pi^{\mu\nu}$. 
\begin{figure}[!h]
\begin{center}
\includegraphics[width=0.495\textwidth]{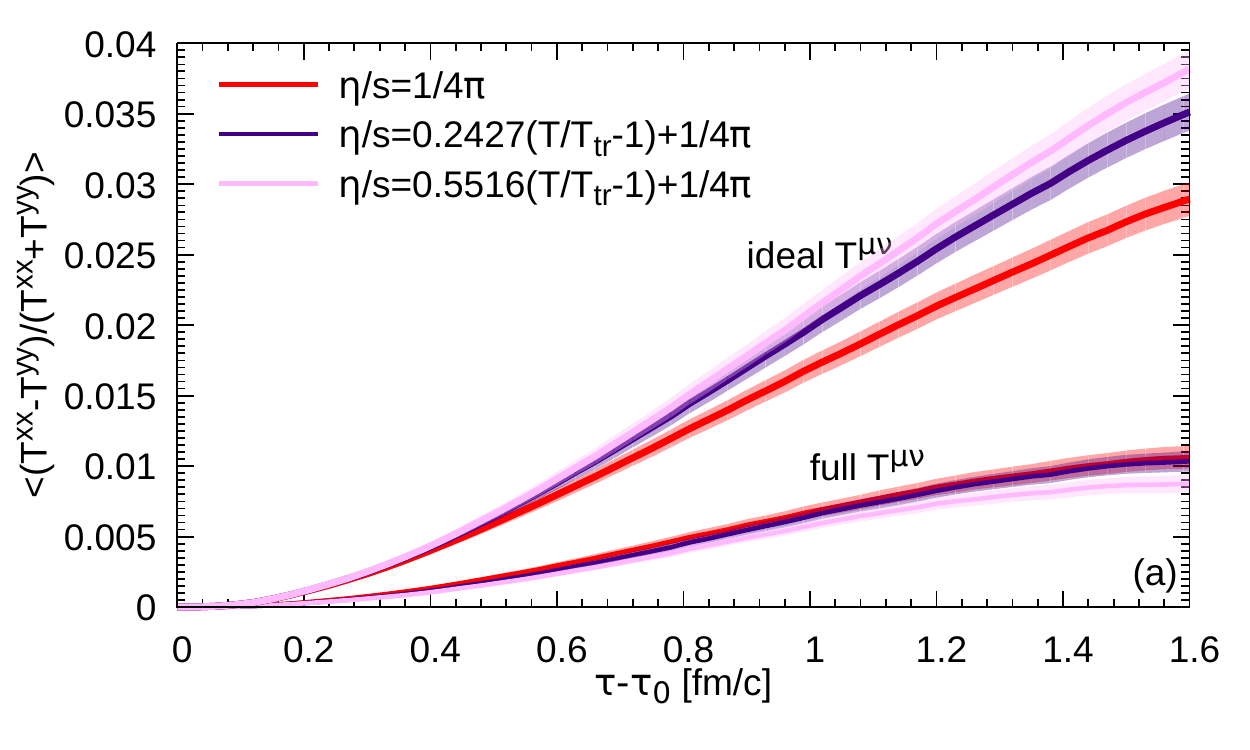}
\includegraphics[width=0.495\textwidth]{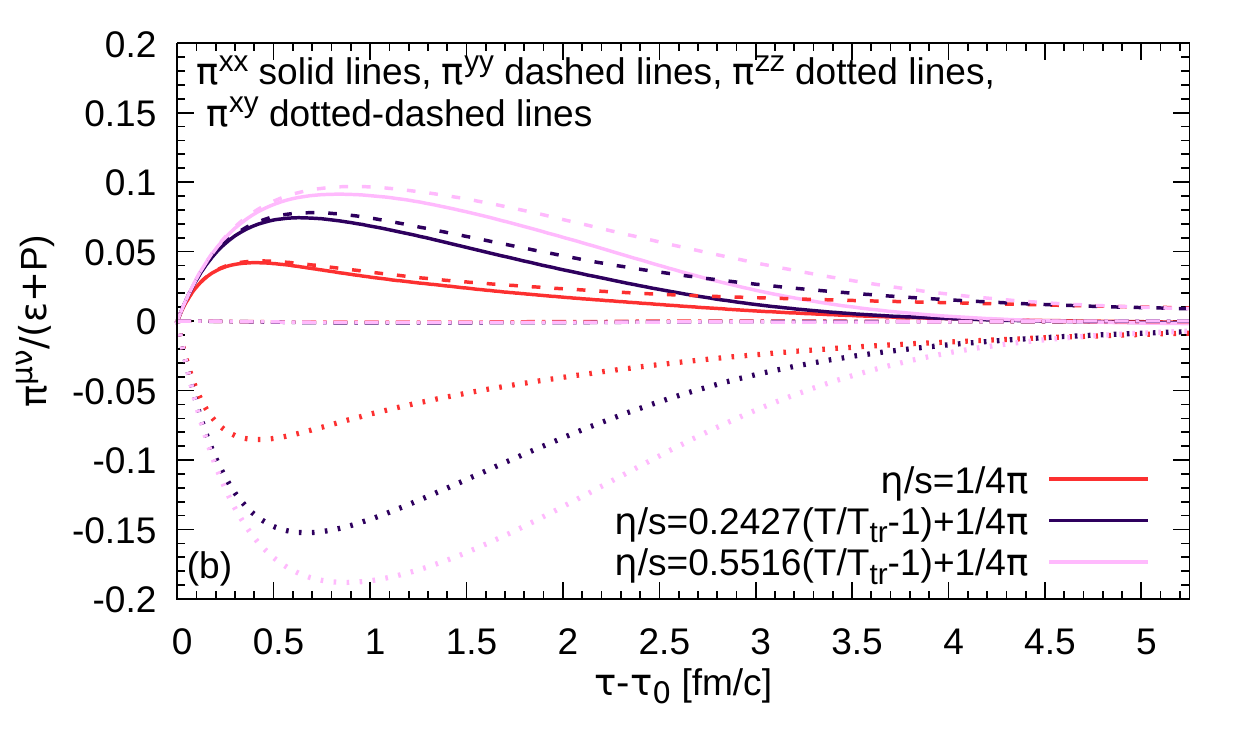}
\end{center}
\caption{(Color online) (a) Momentum anisotropy of the fluid where QGP dilepton rates are used. The meaning of ideal/full $T^{\mu\nu}$ is explained in the text. The development of $\pi^{\mu\nu}/(\varepsilon+P)$ in the local rest frame at $(x,y,\eta_s)=(0,0,0)$ is shown in (b) where an average over 200 events was performed.}
\label{fig:T_munu_aniso_and_pi_munu_lin}
\end{figure}
Recall that $R_0$ solely couples to fluid velocity $u^\mu$ and temperature $T$ and hence is directly sensitive to modification of these two quantities owing to the presence of $\pi^{\mu\nu}$ in the hydrodynamical evolution, while $\delta R$ couples to $\pi^{\mu\nu}$ in addition to $u^\mu$ and $T$. The elliptic flow and $\nu_2(M)$ of QGP dileptons in Fig. \ref{fig:v2_nu2_M_qgp_dilep_lin} without viscous correction $\delta R$, is increased with $\eta/s(T)$, owing to the fact that $\pi^{\mu\nu}$ at early times increases the transverse velocity gradients of the fluid which then generates a larger radial flow and hydrodynamical momentum anisotropy. This increase in the momentum anisotropy can be seen in the top three curves of Fig. \ref{fig:T_munu_aniso_and_pi_munu_lin} (a), where $\pi^{\mu\nu}$ was removed when computing $\langle (T^{xx}-T^{yy})/(T^{xx}+T^{yy})\rangle$ and hence are labeled as ideal $T^{\mu\nu}$. On the other hand, the coupling to $\pi^{\mu\nu}$ via $\delta R$ is responsible for decreasing the elliptic flow as shown in Ref. \cite{Vujanovic:2013jpa} (and references therein), while $\pi^{\mu\nu}$ also reduces the hydrodynamic momentum anisotropy seen in the bottom three curves of Fig. \ref{fig:T_munu_aniso_and_pi_munu_lin} (a). Thus one notices that the order of the $v_2(M)$ curves obtained without/with the constant cross-section $\delta R$ [see Fig. \ref{fig:v2_nu2_M_qgp_dilep_lin} (a)] follows the order of the curves of the momentum anisotropy obtained by using ideal/full $T^{\mu\nu}$ [see Fig. \ref{fig:T_munu_aniso_and_pi_munu_lin} (a)]. There is also a correlation between higher $M$ dileptons being are more sensitive to the early time dynamics while lower $M$ dileptons are more sensitive to the later time evolution.   

It should also be noted that the effect of $\eta/s(T)$ on the evolution of $\pi^{\mu\nu}/(\varepsilon+P)$, shown in Fig. \ref{fig:T_munu_aniso_and_pi_munu_lin} (b), is in contrast with that of $\tau_\pi$ shown in Ref. \cite{Vujanovic:2016anq}. Indeed, starting from zero initial $\pi^{\mu\nu}$, increasing the relaxation time results in decreasing $\pi^{\mu\nu}(\tau)$ for early $\tau-\tau_0$ probed by QGP dileptons, which in turn allows for a faster anisotropic flow development, thus increasing $v_2(M)$ of QGP dileptons. The effect $\eta/s(T)$ on $\pi^{\mu\nu}/(\varepsilon+P)$ shown Fig. \ref{fig:T_munu_aniso_and_pi_munu_lin} (b) is the opposite in the early stages of the evolution: a large $\pi^{\mu\nu}$ at early times slows down the development of anisotropic flow, thus reducing $v_2(M)$ of QGP dileptons. 

\begin{figure}[!h]
\begin{center}
\includegraphics[width=0.495\textwidth]{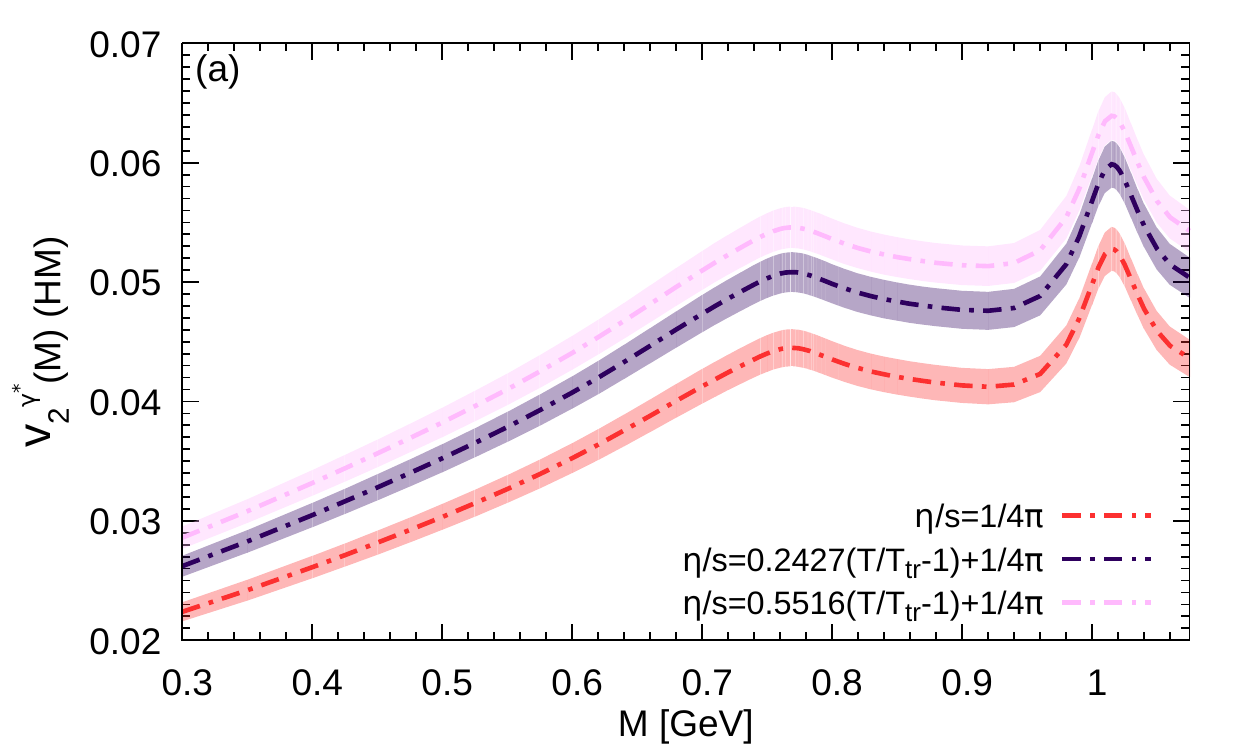}
\includegraphics[width=0.497\textwidth]{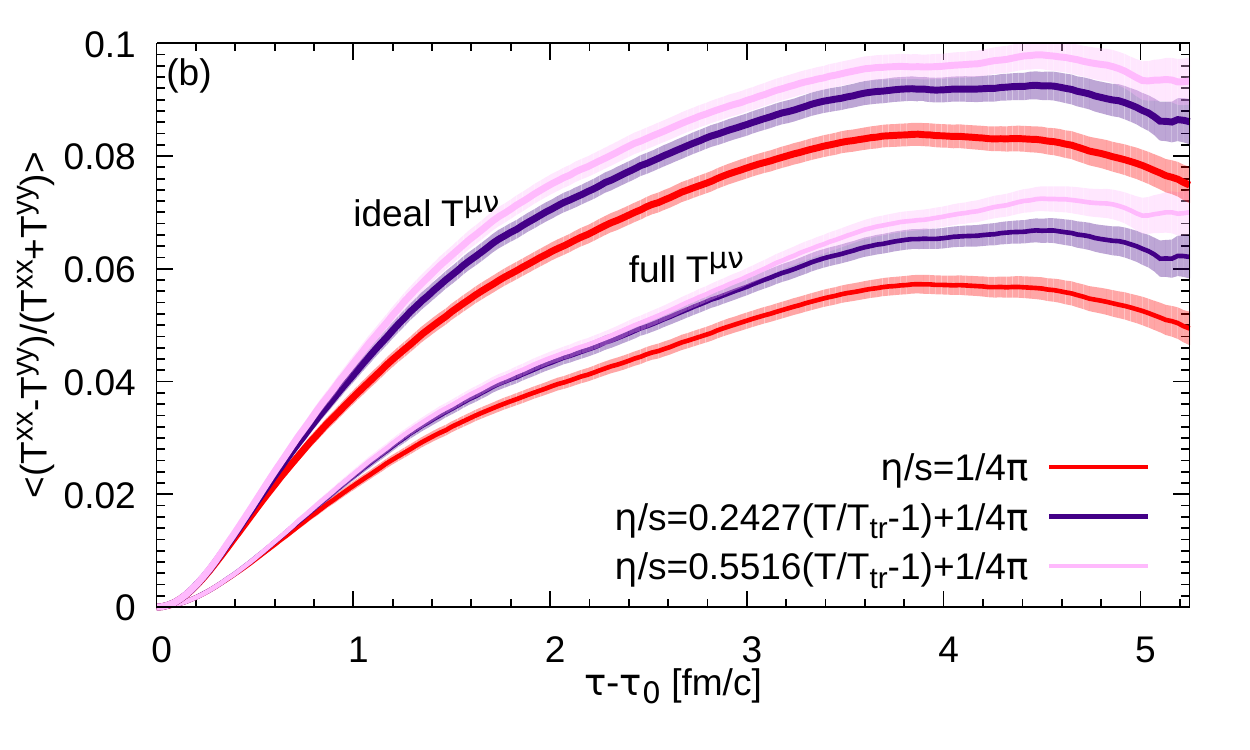}
\end{center}
\caption{(Color online) $v_2(M)$ of HM dileptons under the influence of a $\eta/s(T)$ is shown in (a), while the hydrodynamic momentum anisotropy evolution in the HM is displayed in (b). All results shown are within the 20--40\% centrality class.}
\label{fig:v2_M_hm_T_munu_hm}
\end{figure}

Having explored the effects of a $\eta/s(T)$ on $v_2(M)$ of QGP dileptons, Fig. \ref{fig:v2_M_hm_T_munu_hm} (a) focuses on the $v_2(M)$ of HM dileptons. There, one notices that $\eta/s(T)$ at high temperatures causes an increase in the $v_2(M)$ of HM dileptons as well as an increase in the development of flow anisotropy in the HM sector as quantified by the hydrodynamics momentum anisotropy depicted in Fig. \ref{fig:v2_M_hm_T_munu_hm} (b). Note that the hydrodynamic momentum anisotropy obtained using both ideal and full $T^{\mu\nu}$ increases when a temperature-dependent $\eta/s$ is present relative to $\eta/s=1/(4\pi)$, thus $v_2(M)$ of HM dileptons should increase as well, and indeed it does so. Note that the $v_2(M)$ of HM dileptons are little affected by the viscous correction to the dilepton rate \cite{Vujanovic:2013jpa}. So, effects with and without viscous corrections are not shown in Fig. \ref{fig:v2_M_hm_T_munu_hm} (a), as the curves would lie nearly on top of one another, thus only the full calculation with viscous corrections is depicted. Examining more closely the $v_2(M)$ of HM dileptons, one notices that it tracks the development of hydrodynamic momentum anisotropy obtained by using ideal $T^{\mu\nu}$, as expected. 

\begin{figure}[!h]
\begin{center}
\includegraphics[width=0.495\textwidth]{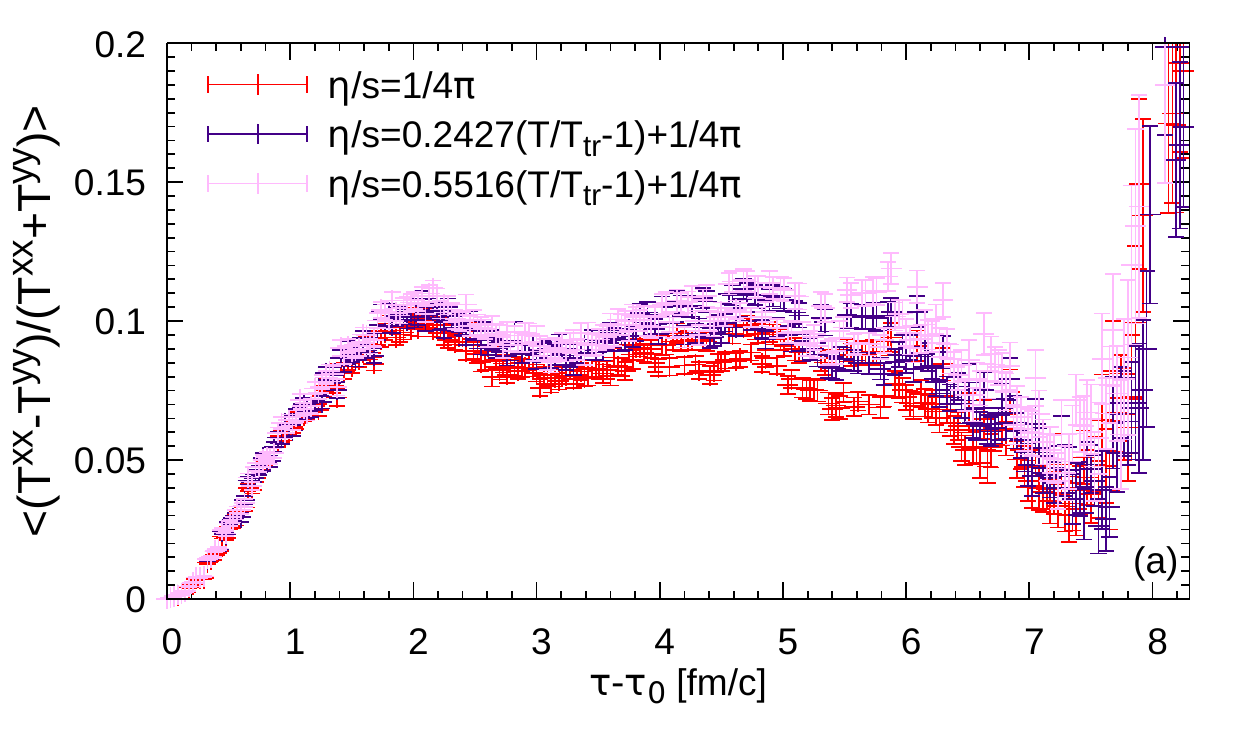}
\includegraphics[width=0.495\textwidth]{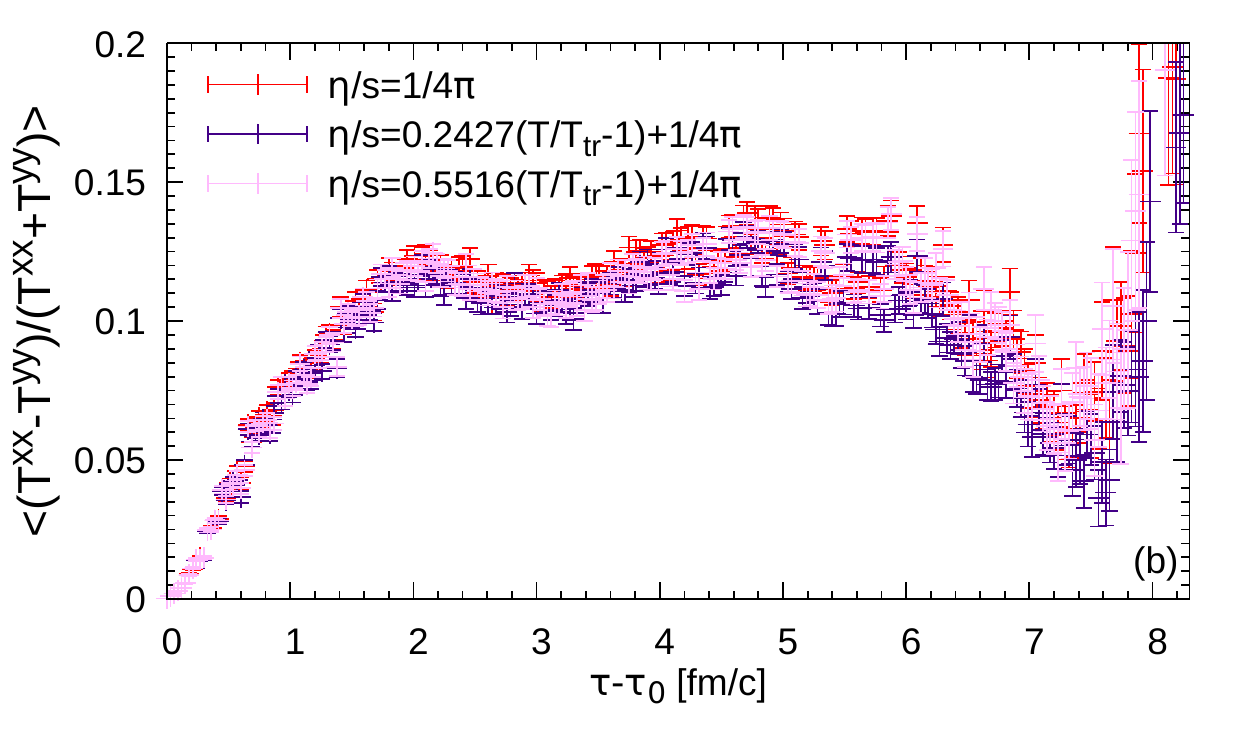}
\end{center}
\caption{(Color online) (a) The hydrodynamical momentum anisotropy evolution on the freeze-out surface using the full $T^{\mu\nu}$. (b) Presents the same quantity as (a) except only the inviscid part of $T^{\mu\nu}$ is being used to compute the hydrodynamical momentum anisotropy. All results shown are within the 20--40\% centrality class.}
\label{fig:T_munu_FO}
\end{figure}

The hydrodynamic momentum anisotropy on the freeze-out surface shown in Fig. \ref{fig:T_munu_FO} behaves differently than at higher temperatures. Though the curve with $\eta/s=1/(4\pi)$ in Fig. \ref{fig:T_munu_FO} (a) seems to have a smaller hydrodynamical momentum anisotropy than the other two cases having $\eta/s(T)$, that difference isn't very significant given the uncertainties. As far as Fig. \ref{fig:T_munu_FO} (b) is concerned, there one notices that the hydrodynamical momentum anisotropy on the freeze-out surface is the same, within the uncertainties, for all three media considered. Therefore the hydrodynamic momentum anisotropy that builds up at higher temperatures, and is thus affecting the $v_2(M)$ of dileptons, doesn't seem to propagate to the freeze-out surface and affect significantly the hydrodynamical momentum anisotropy there, hence leaving the $v_2(p_T)$ of charged hadrons largely unaffected. 
\begin{figure}[!h]
\begin{center}
\begin{tabular}{ccc}
\includegraphics[width=0.33\textwidth]{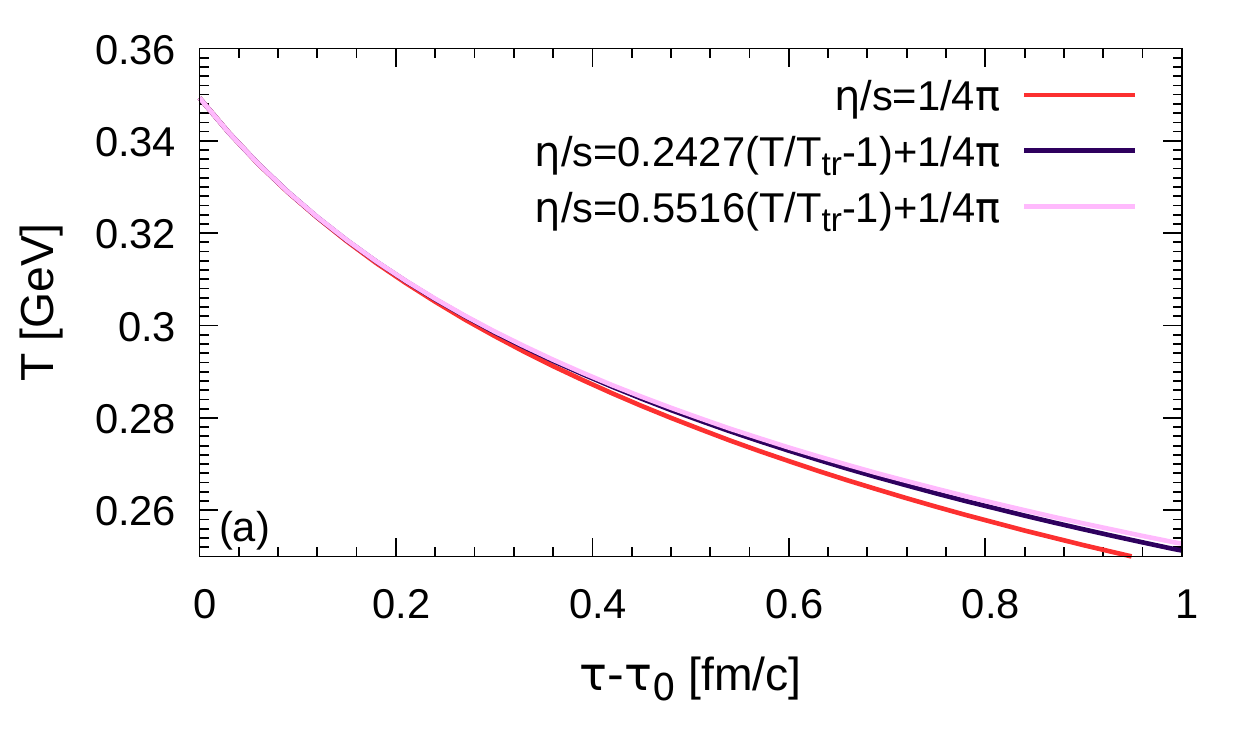} & \includegraphics[width=0.33\textwidth]{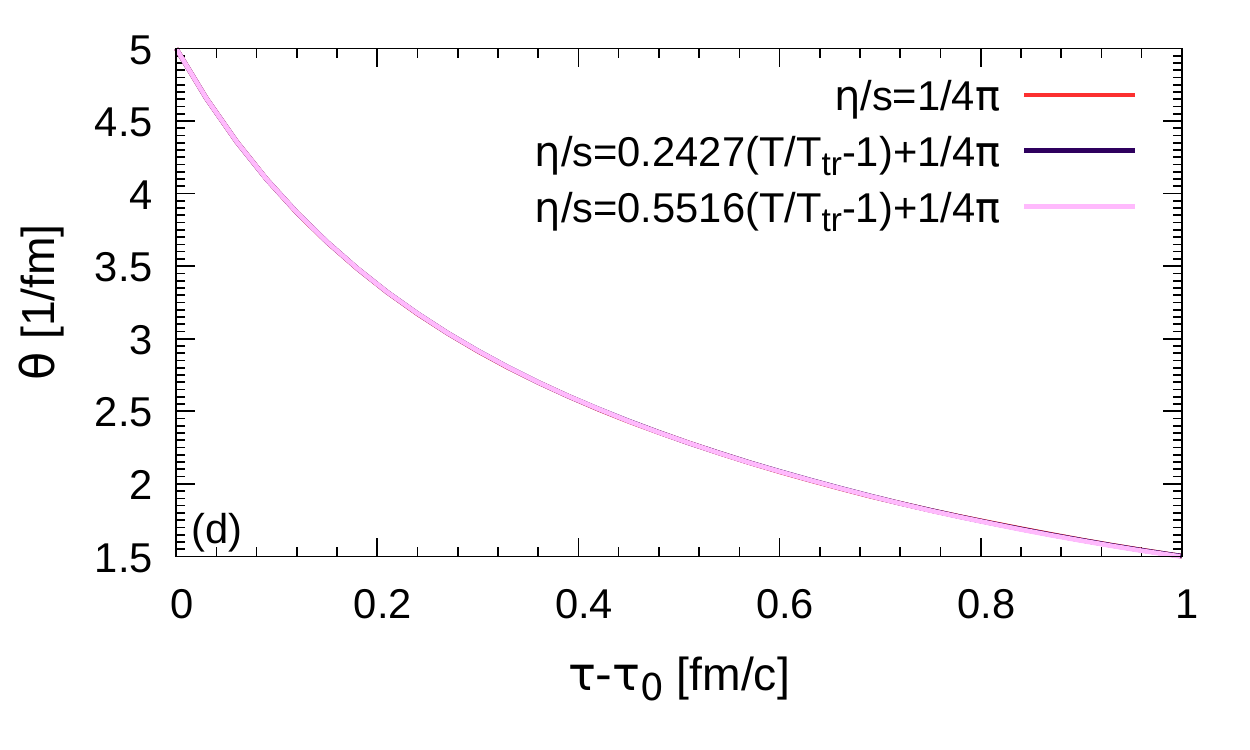} & \includegraphics[width=0.33\textwidth]{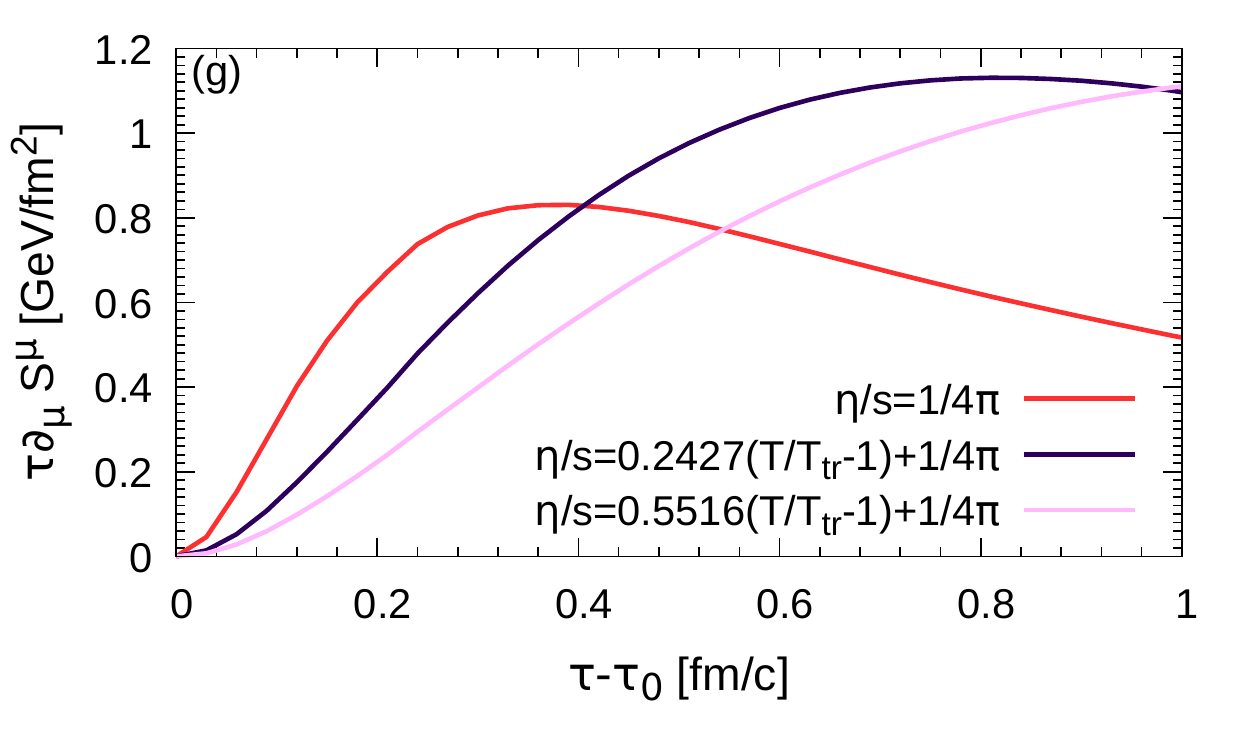} \\ 
\includegraphics[width=0.33\textwidth]{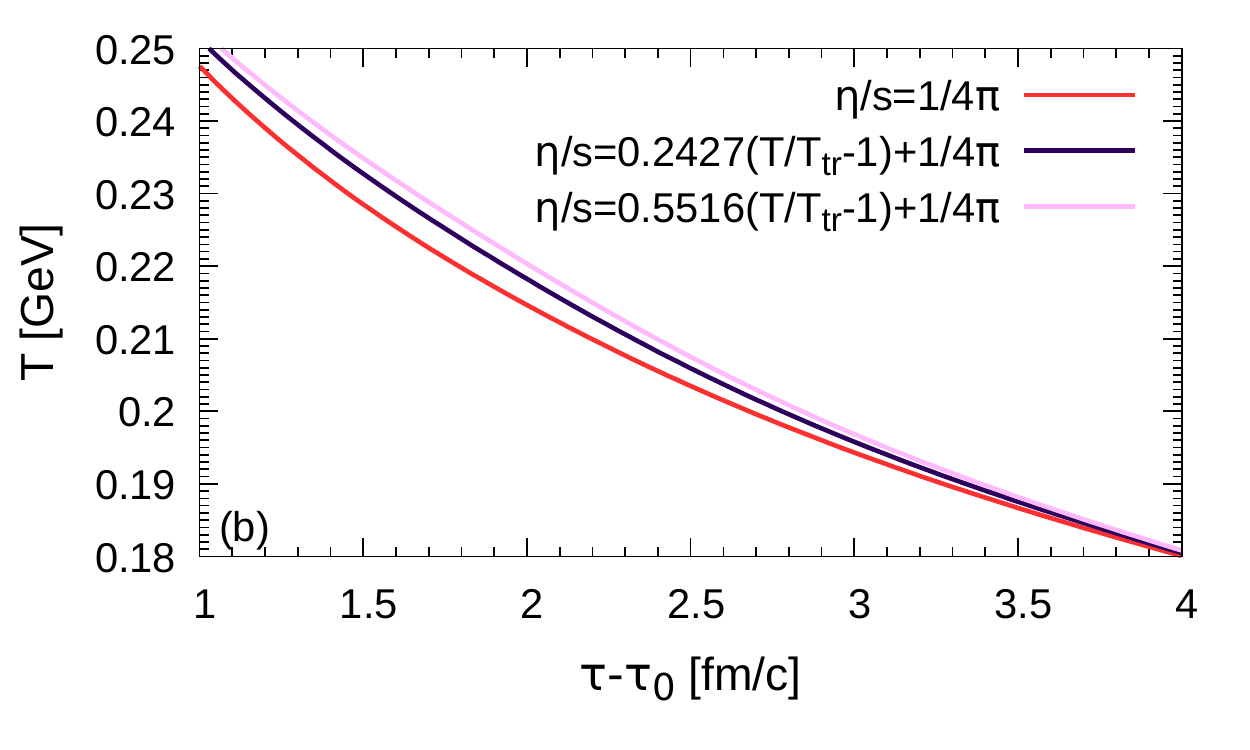}  & \includegraphics[width=0.33\textwidth]{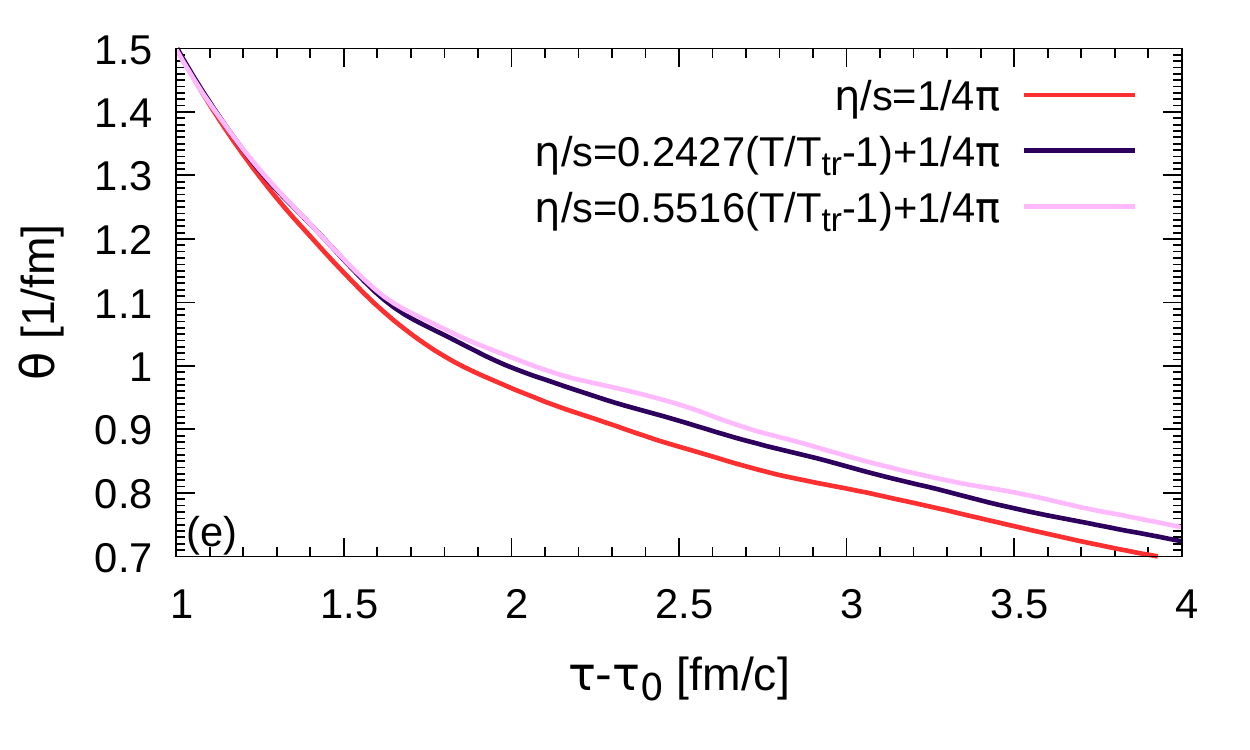}  & \includegraphics[width=0.33\textwidth]{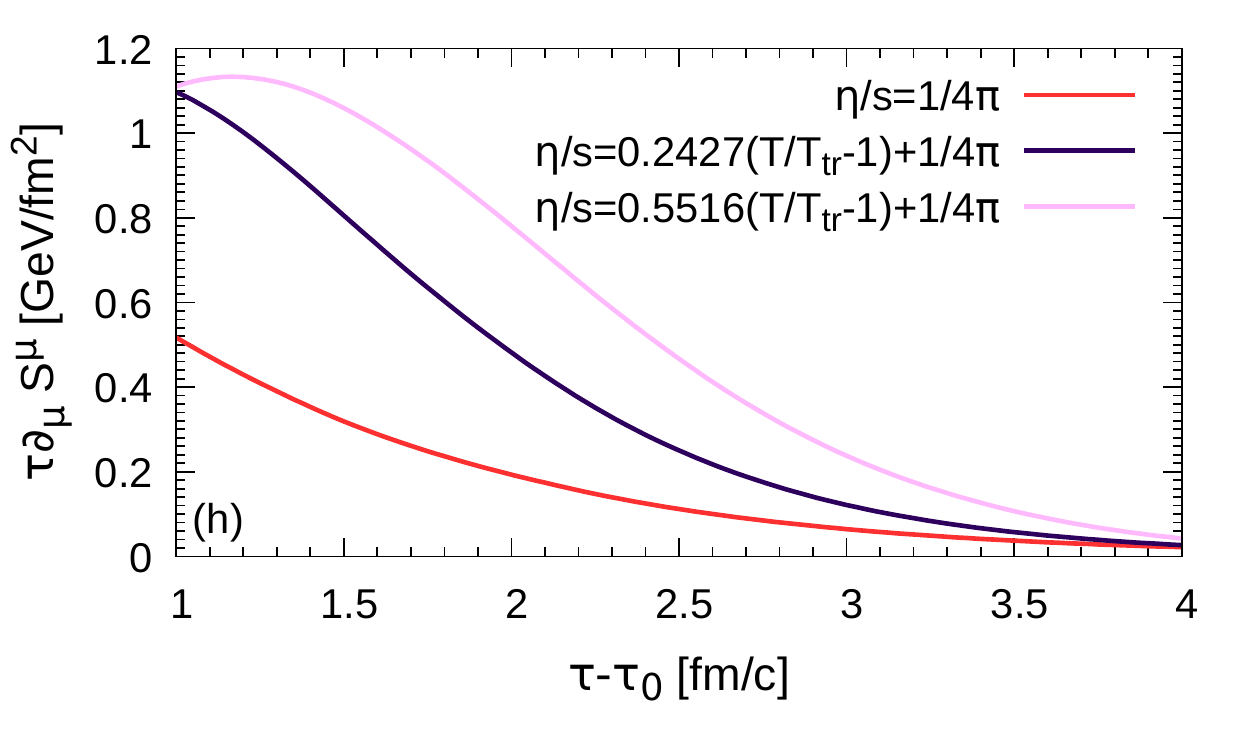}\\ 
\includegraphics[width=0.33\textwidth]{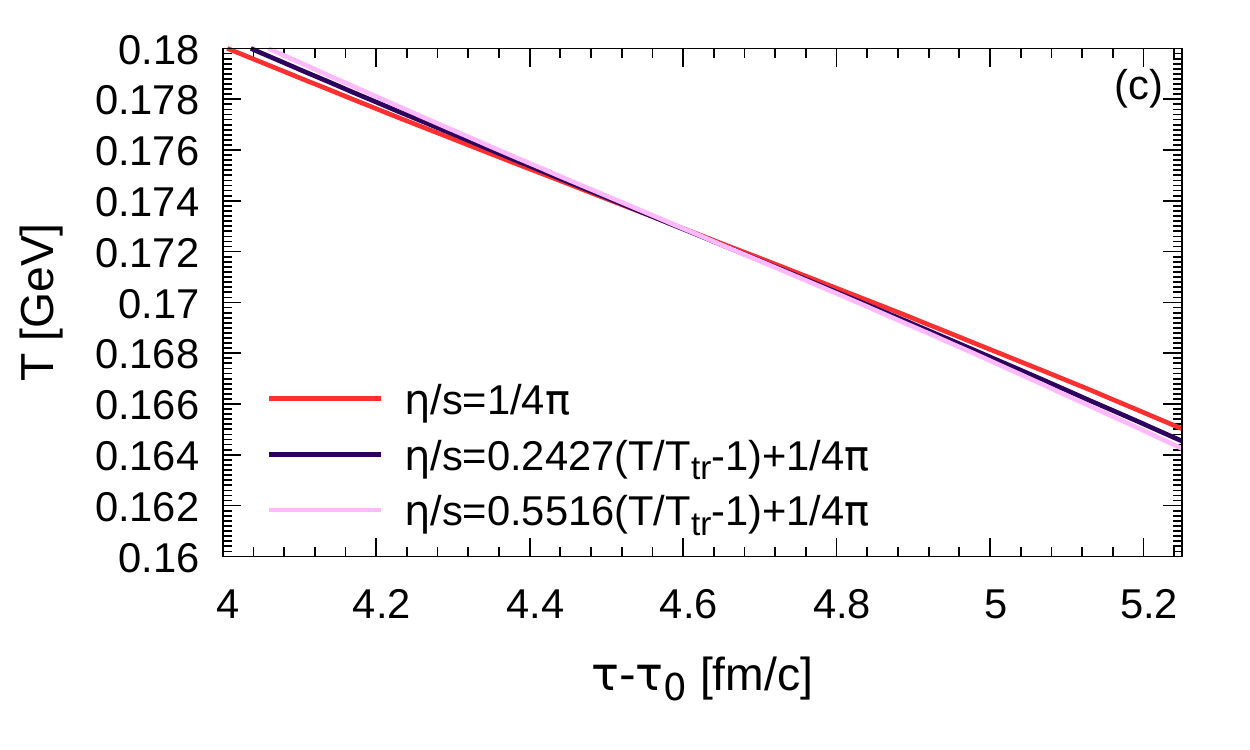}     & \includegraphics[width=0.33\textwidth]{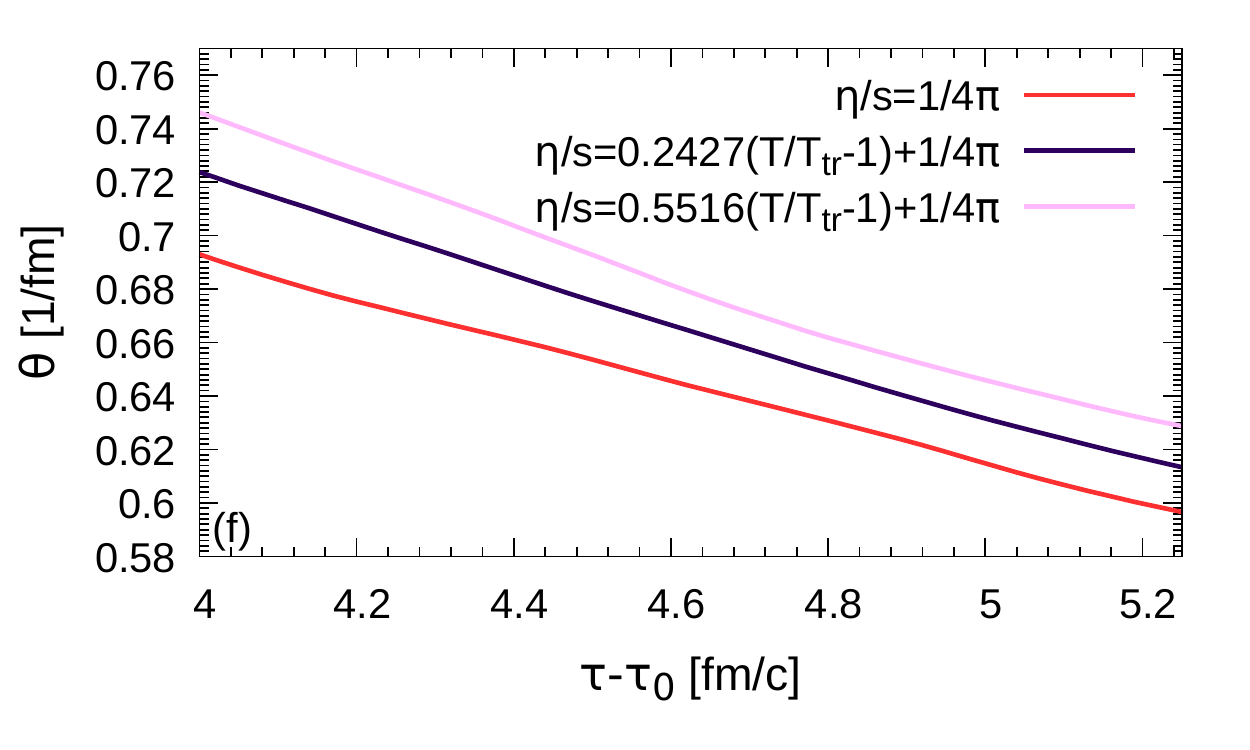} & \includegraphics[width=0.33\textwidth]{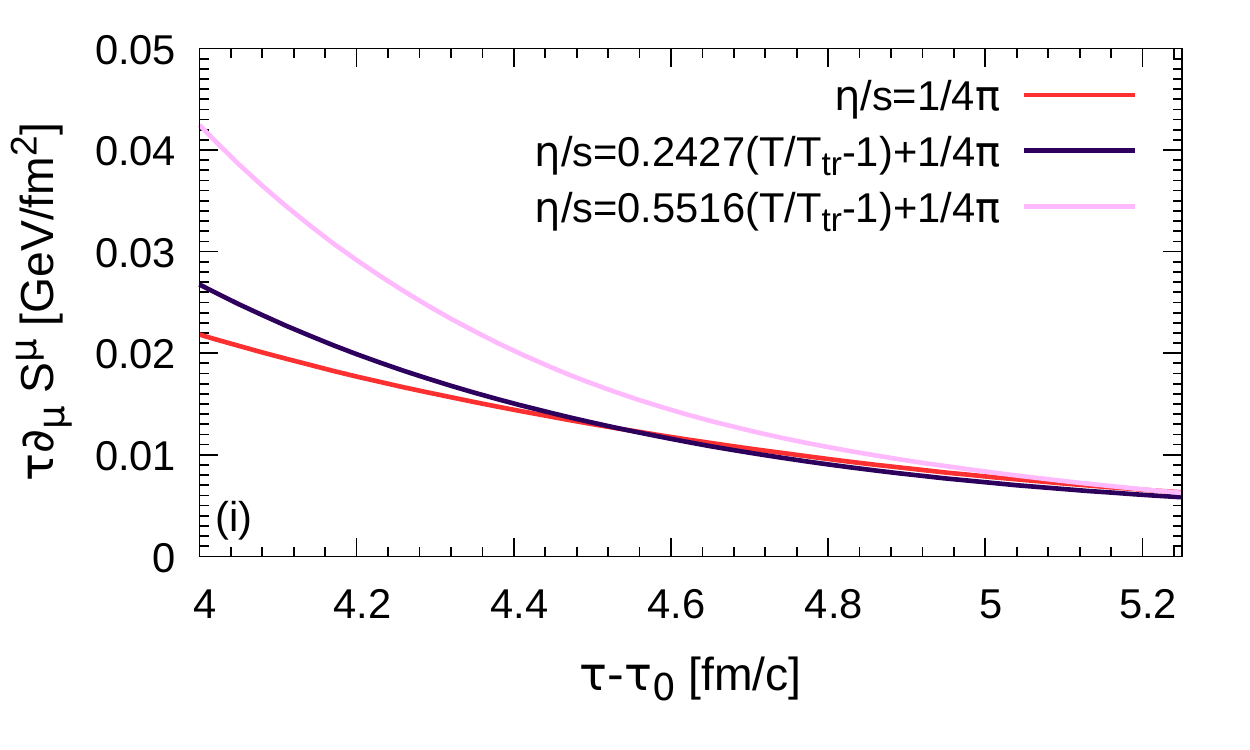}\\
\end{tabular}
\end{center}
\caption{(Color online) Event-averaged temperature for the cell $(x,y,\eta_s)=(0,0,0)$ during the first fm/$c$ of evolution (a) and at later stages (b) and (c). Event-averaged expansion rate $\theta$ during the first fm/$c$ of evolution (d) and at later stages (e) and (f). Entropy production rate $\partial_\mu S^\mu$ rescaled by $\tau$ during the first fm/$c$ of evolution (g) and at later stages (h) and (i).}
\label{fig:Temp_vs_tau_lin}
\end{figure}

Given that the initial and the freeze-out conditions are the same for all three media (see Sec. \ref{sec:IC} for details), the cooling rate [see Figs. \ref{fig:Temp_vs_tau_lin} (a)--\ref{fig:Temp_vs_tau_lin} (c)] of the medium is a competition between the entropy production rate $\partial_\mu S^\mu=\pi^{\mu\nu}\pi_{\mu\nu}/(2\eta T)$ and expansion rate $\theta=\partial_\mu u^\mu$ of the system. The entropy production $\partial_\mu S^\mu$ has been rescaled by $\tau$ such that the amount of entropy produced in the cell located at $x=y=\eta_s=0$ is given by
\begin{eqnarray}
S(\tau)=\int^{\tau}_{\tau_0} d\tau' \int^{\Delta x/2}_{-\Delta x/2} dx' \int^{\Delta y/2}_{-\Delta y/2} dy' \int^{\Delta \eta_s/2}_{-\Delta \eta_s/2} d\eta_s' \left[\tau' \partial_\mu S^\mu\right], 
\end{eqnarray}
and hence one can use the area under the curves in Figs. \ref{fig:Temp_vs_tau_lin} (d)--\ref{fig:Temp_vs_tau_lin} (f) to estimate $S(\tau)$ for the central cell. Focusing on the dynamics for happening during the first fm/$c$ of evolution, we see that the expansion rate is large and the same regardless of whether the medium has $\eta/s=1/(4\pi)$ or $\eta/s(T)$. In fact, during the first $\sim 0.3$ fm/$c$ of evolution, the differences in entropy production rate do not affect the temperature profile, which seems to be driven by the expansion rate $\theta$. Once the expansion rate is less strong, and the entropy production rate of the media with $\eta/s(T)$ becomes stronger than that of $\eta/s=1/(4\pi)$ occurring after $0.3$ fm/$c$, then the extra entropy production present for media $\eta/s(T)$ [see Fig. \ref{fig:Temp_vs_tau_lin} (g)] causes a slower temperature reduction for the media with $\eta/s(T)$ relative to the one with $\eta/s=1/(4\pi)$, at early times $0.3 \lesssim \tau-\tau_0\leq 1$ fm/$c$ [see Fig. \ref{fig:Temp_vs_tau_lin} (a)]. A more quantitative exploration of the dynamics present during the first fm/$c$ is reserved for a later study.

At later times presented in Figs. \ref{fig:Temp_vs_tau_lin} (b), \ref{fig:Temp_vs_tau_lin} (e), and \ref{fig:Temp_vs_tau_lin} (h), the expansion rate remains the same for all three media considered until $\tau-\tau_0\sim 1.3$ fm/$c$ while the entropy production rate is larger for the medium with larger $\eta/s$, and the order of the curves in Fig. \ref{fig:Temp_vs_tau_lin} (b) reflects this. However, such a situation cannot be maintained indefinitely, since the hotter media with $\eta/s(T)$, will have larger pressure gradients than the one with $\eta/s=1/(4\pi)$. So, as soon as $\eta/s(T)$ allows for these pressure gradients to be more efficiently converted into a larger expansion rate, which according to Fig. \ref{fig:Temp_vs_tau_lin} (e) happens when $\tau-\tau_0\gtrsim 1.3$ fm/$c$, the fluids with $\eta/s(T)$ will start cooling at a faster rate than the one with $\eta/s=1/(4\pi)$ and this is reflected by the temperature profile [see Fig. \ref{fig:Temp_vs_tau_lin} (b)]. Also, Fig. \ref{fig:Temp_vs_tau_lin} (h) shows that the entropy production for all three media stops being relevant by $\tau-\tau_0\sim4$ fm/$c$. The cooling at $\tau-\tau_0>4$ fm/$c$ in Fig. \ref{fig:Temp_vs_tau_lin} (c) is dominated by the faster expansion rate of the media with $\eta/s(T)$ relative to the one with $\eta/s=1/(4\pi)$; with $\theta$ at $\tau-\tau_0 \sim 4$ fm/$c$ being about half the value it had at $\tau-\tau_0 \sim 1$ fm/$c$ and dropping another $\sim 15$\% in the interval $4\leq\tau-\tau_0\leq 5.25$ fm/$c$. Entropy production becomes negligible the interval $4\leq\tau-\tau_0\leq 5.25$ fm/$c$ as shown in Fig. \ref{fig:Temp_vs_tau_lin} (i). Ultimately, the medium with $\eta/s=1/(4\pi)$ will freeze out later than the other two media. This is not shown in Fig. \ref{fig:Temp_vs_tau_lin} (c), since hydrodynamical events start freezing out right after $\tau-\tau_0=5.25$ fm/$c$, and at that point the event-averaged temperature becomes ill-behaved.
\begin{figure}[!h]
\begin{center}
\begin{tabular}{ccc}
\includegraphics[width=0.33\textwidth]{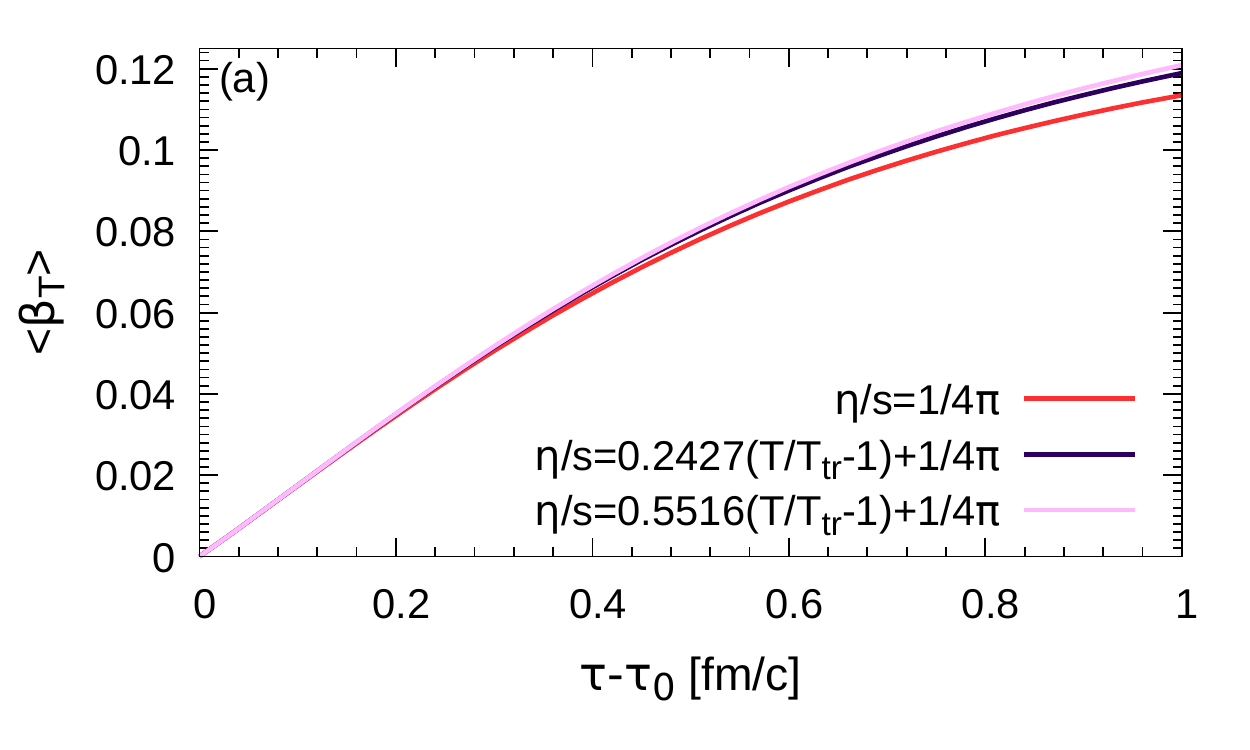} & \includegraphics[width=0.33\textwidth]{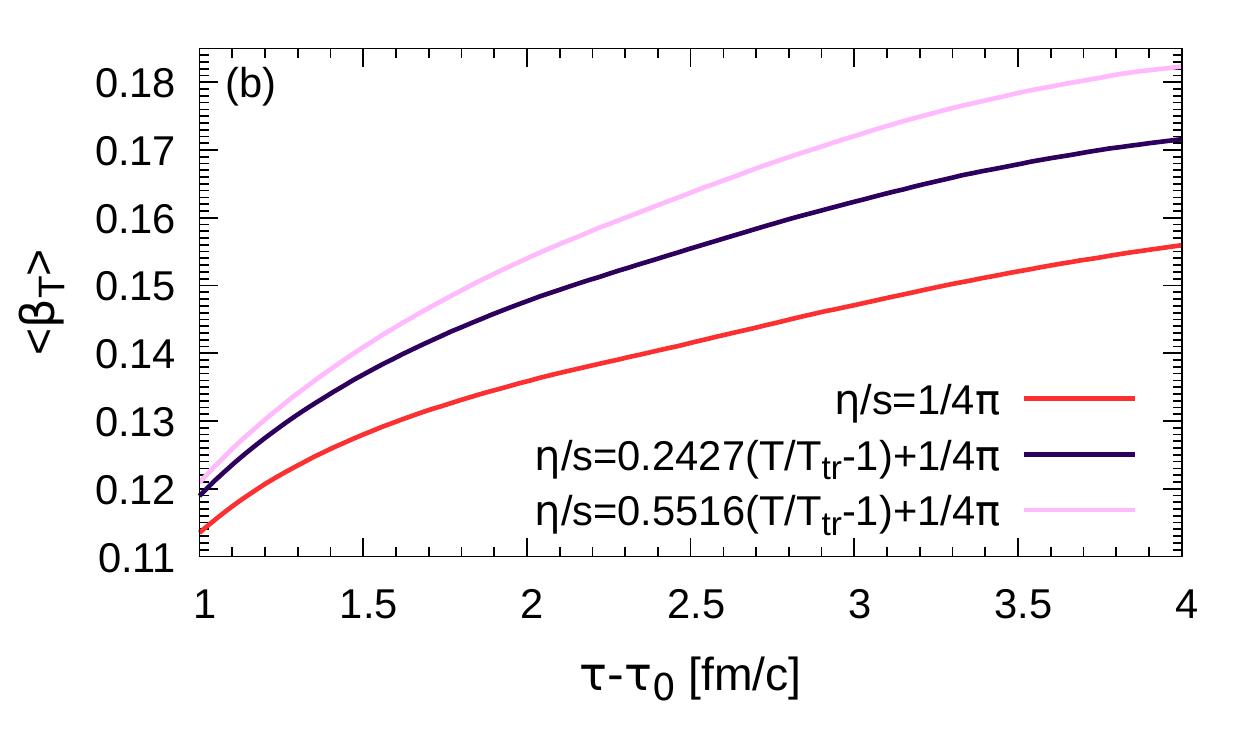} & \includegraphics[width=0.33\textwidth]{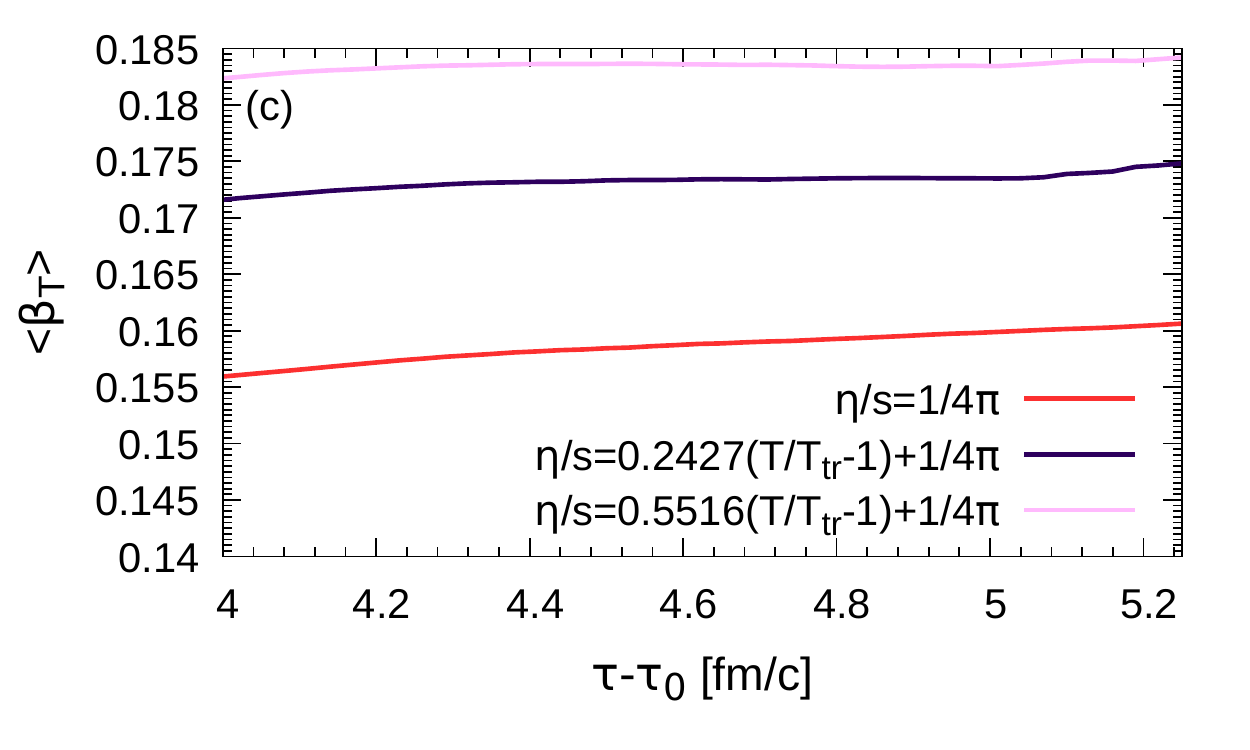}
\end{tabular}
\end{center}
\caption{(Color online) Event-averaged $\beta_T$ during the first fm/$c$ of evolution (a) and at later stages (b), (c).}
\label{fig:vT_vs_tau_lin}
\end{figure}

The transverse flow profile $\langle \beta_T (\tau) \rangle$ shown in Fig. \ref{fig:vT_vs_tau_lin}, qualitatively behaves as expected from the expansion rate. $\langle \beta_T (\tau) \rangle$ was computed via:
\begin{eqnarray}
\langle \beta_T (\tau) \rangle &=& \frac{1}{N_{ev}}\sum_{i=1}^{N_{ev}}\left\{\frac{\int \tau d\eta_s dy dx f_{FO}(T) \left[\left(\beta^x_i \left(x^\mu\right) \right)^2 + \left(\beta^y_i \left(x^\mu\right)\right)^2\right]^{1/2}}{\int \tau d\eta_s dy dx f_{FO}(T)}\right\}\nonumber\\
f_{FO}(T)&=&\left\{ \begin{array}{rl}
                                  1 & T>T_{FO}\\
                                  0 & {\rm otherwise}
                               \end{array}
                        \right.,
\label{eq:}
\end{eqnarray}
where $x^\mu=(\tau,x,y,\eta_s)$, $\beta^j=u^j/u^0$, while $u^j$ and $u^0$ are the spatial and temporal components of the flow $u^\mu$, respectively. 

\begin{figure}[!h]
\begin{center}
\includegraphics[width=0.495\textwidth]{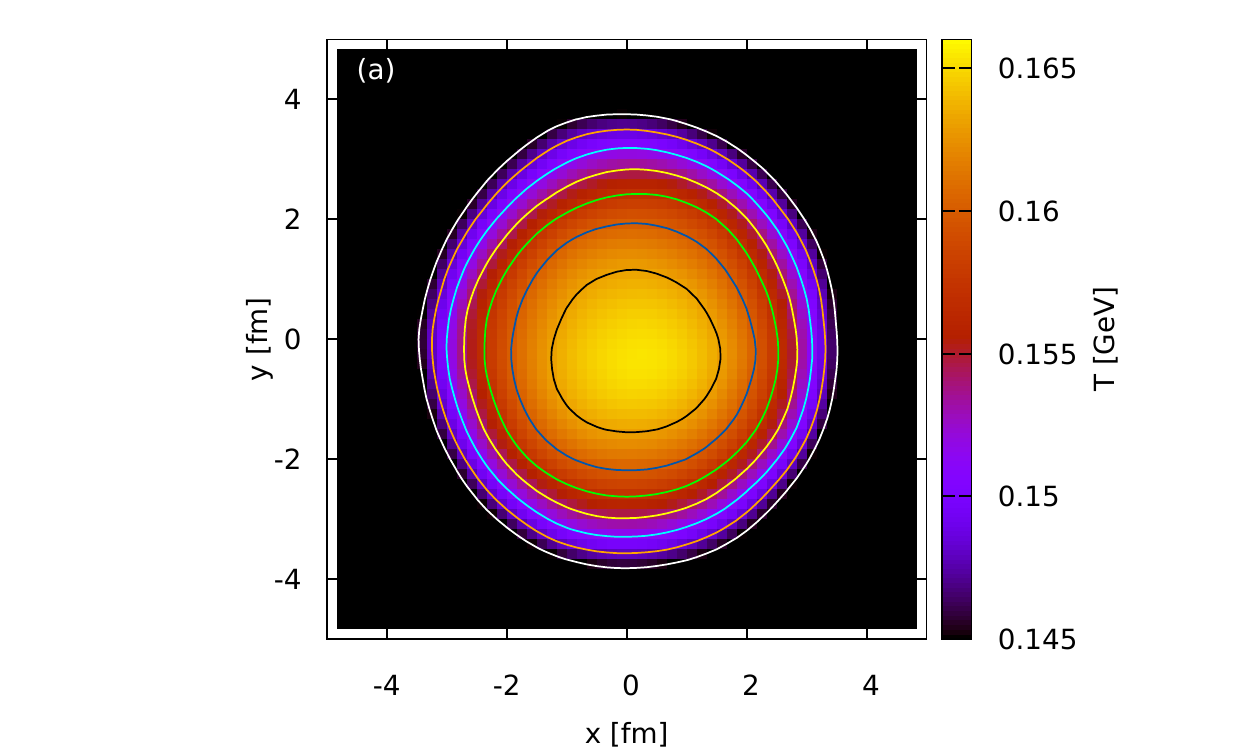}
\includegraphics[width=0.495\textwidth]{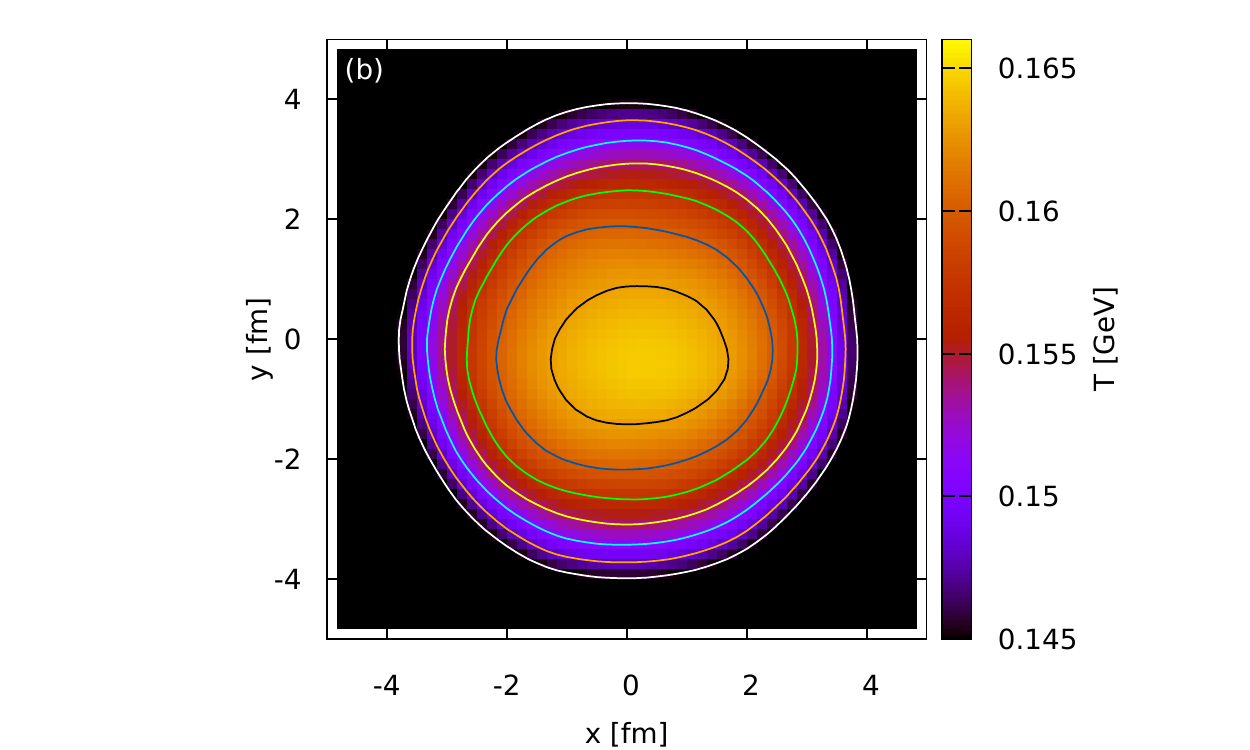}
\end{center}
\caption{(Color online) Event-averaged temperature in the transverse plane at $z=0$ at $(\tau-\tau_0)=5.25$ fm/$c$ for $\frac{\eta}{s}=\frac{1}{4\pi}$ (a) and $\frac{\eta}{s}= 0.4513\left(\frac{T}{T_{tr}}-1\right)+\frac{1}{4\pi}$ (b). The constant temperature contours ranging from 163 MeV for the inner most contour all the way down to $T_{F0}=145$ MeV, are separated by 3 MeV intervals.}
\label{fig:Temp_vs_xy_lin}
\end{figure}

Having discussed cooling as a competition between expansion rate and entropy production rate, while also showing that these dynamics affect transverse flow buildup, the focus is now given to the development of anisotropic flow. Figure \ref{fig:T_munu_aniso_and_pi_munu_lin} (a) shows that relative to the medium with $\eta/s=1/(4\pi)$, a medium with $\eta/s(T)$ suppresses more the conversion of the original geometrical anisotropy into a momentum anisotropy of the QGP. So, for the first $\sim 1$ fm/$c$ of evolution, the QGP with a temperature-dependent $\eta/s$ develops anisotropic flow slower and is hotter, than the QGP with a constant $\eta/s$. However, inspecting Fig. \ref{fig:v2_M_hm_T_munu_hm} (b) shows that the anisotropic flow buildup in the hadronic sector, where the viscosity is lower than that in the QGP, is significantly faster than in the QGP, thus more efficiently converting pressure gradients into hydrodynamic momentum anisotropy. Because the dilepton HM rates are not particularly sensitive to viscous correction of their production rates, they track more closely the buildup of the momentum anisotropy originating from the ideal part of $T^{\mu\nu}$ as can be seen by comparing the order of the curves in Figs. \ref{fig:v2_M_hm_T_munu_hm} (a) and \ref{fig:v2_M_hm_T_munu_hm} (b). This difference in the development of the anisotropic flow between the media with $\eta/s(T)$ and one with $\eta/s=1/(4\pi)$ is really established during the first $\sim 1.8$ fm/$c$ of evolution and happens above the freeze-out surface. Because dileptons are emitted throughout the entire evolution of the medium, they are sensitive to the difference in anisotropic flow buildup shown in Fig. \ref{fig:v2_M_hm_T_munu_hm} (b), as can be seen in Fig. \ref{fig:v2_M_hm_T_munu_hm} (a). This difference in the early anisotropic flow build-up is also imprinted on the temperature profile of the system in the $x$-$y$ plane, at temperatures above the freeze-out surface. At $\tau-\tau_0=5.25$ fm/$c$ shown in Fig. \ref{fig:Temp_vs_xy_lin}, when all three systems have already started to reduce their momentum anisotropy obtained from the ideal part of $T^{\mu\nu}$ [see Fig. \ref{fig:v2_M_hm_T_munu_hm} (b)], the high temperatures contour lines (see $T=160,163$ MeV) show that the medium with a $\eta/s(T)$ produces a more elongated shape than $\eta/s=1/(4\pi)$. However, at the freeze-out temperature, that shape for both media is roughly the same.  

Given that the charged hadron $v_2$ in Fig. \ref{fig:v2_ch_th_dilep_lin} (a) is unaffected by $\eta/s(T)$ and the fact that the hydrodynamical momentum anisotropy on the freeze-out surface in Fig. \ref{fig:T_munu_FO} is less affected by $\eta/s(T)$ compared to higher temperatures, it seems that the larger anisotropic ``push'' generated by a temperature-dependent $\eta/s$, present at high temperatures, is mostly quenched by the time the system freezes out, and thus doesn't significantly affect the $v_2$ of charged hadrons. 

\begin{figure}[!h]
\begin{center}
\includegraphics[width=0.495\textwidth]{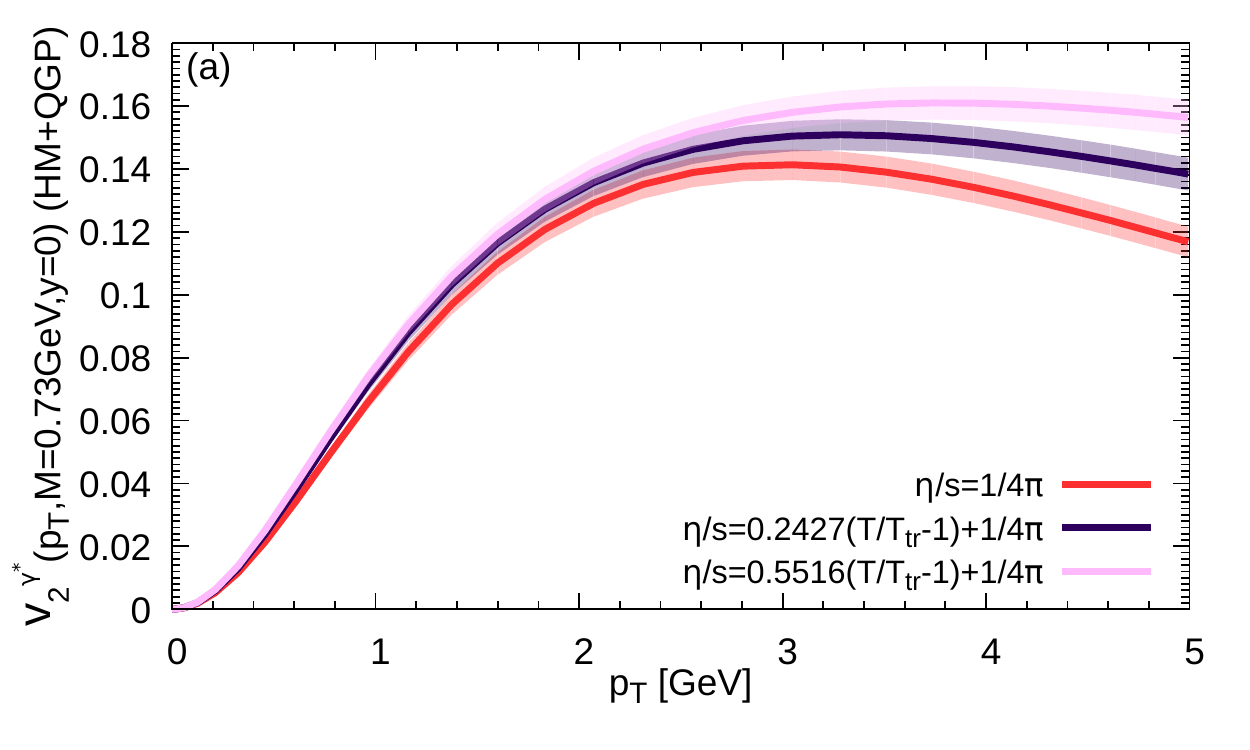}
\includegraphics[width=0.495\textwidth]{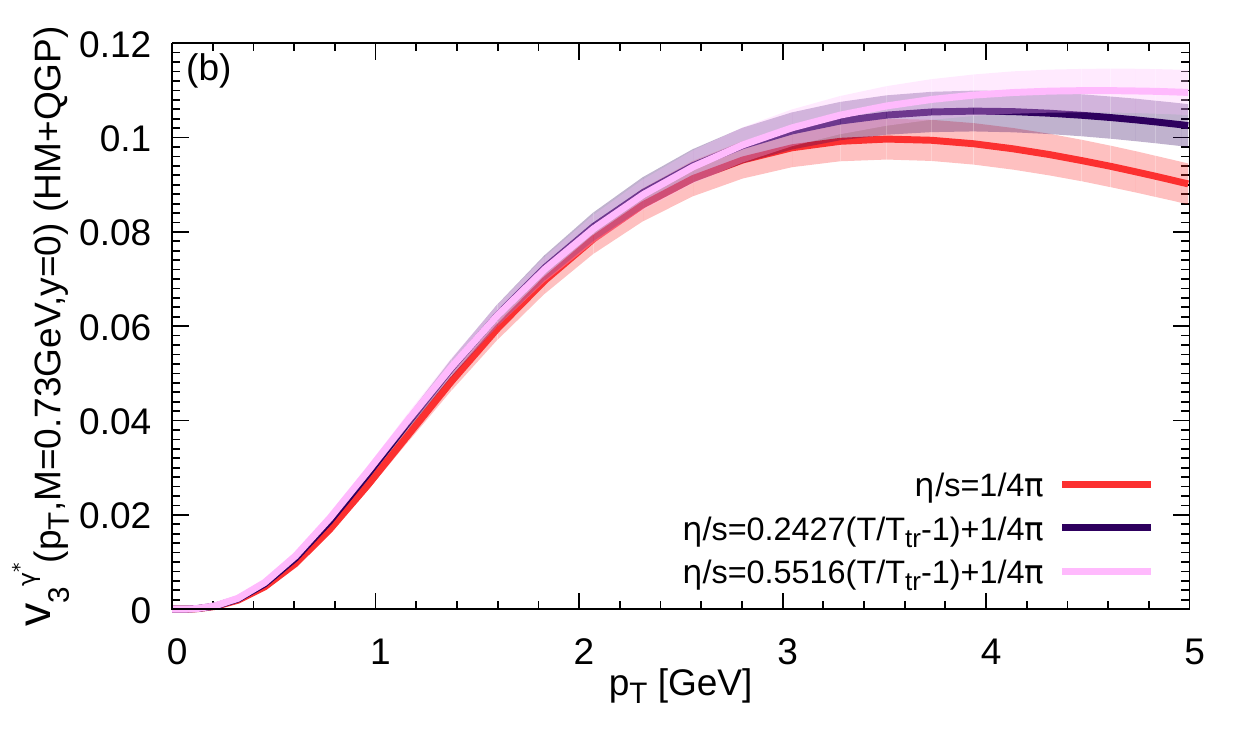}
\end{center}
\caption{(Color online) Elliptic (a) and triangular (b) flow of thermal dileptons for a linearly dependent $\eta/s(T)$.}
\label{fig:v2_v3_pt_lin}
\end{figure}

Last, we explore elliptic and triangular flow of thermal dileptons as a function of $p_T$ in Fig. \ref{fig:v2_v3_pt_lin}. To maximize the potential opportunity of constraining the size of $\eta/s(T)$ in experimental dilepton data, the invariant mass $M$ was chosen in a region where the thermal radiation dominates over all other sources \cite{Vujanovic:2013jpa}. 
In particular, notice the size of the difference in the dependence of flow harmonics---especially $v_2(p_T\gtrsim 2$ GeV) and $v_3(p_T\gtrsim$ 3 GeV)---when a temperature-dependent $\eta/s$ is being used. Such a prominent variation is ideal if the slope of $\eta/s(T)$ at high temperatures is to be experimentally constrained. The caveat, of course, is that one also needs to constrain $\eta/s(T)$ for $T<T_{tr}$ using, e.g., hadrons, which has shown sensitivity to $\eta/s(T)$ for $T<T_{tr}$ at top RHIC energy \cite{Niemi:2011ix}. However, our present goal is to show that the $v_2$ of dilepton at top RHIC energy can break the degeneracy seen in the behavior of charged hadron $v_2(p_T)$ at midrapidity \cite{Niemi:2011ix} towards the presence of an $\eta/s(T)$ at high temperatures. Thus dileptons and hadrons observables should be used simultaneously to put tighter constraints on the properties of the QCD medium at high temperatures.  
   
\subsection{Quadratic $\eta/s(T)$}

We now turn our attention towards the second derivative of $\eta/s(T)$. The initial and freeze-out conditions are unchanged.   
\begin{figure}[!h]
\begin{center}
\includegraphics[width=0.495\textwidth]{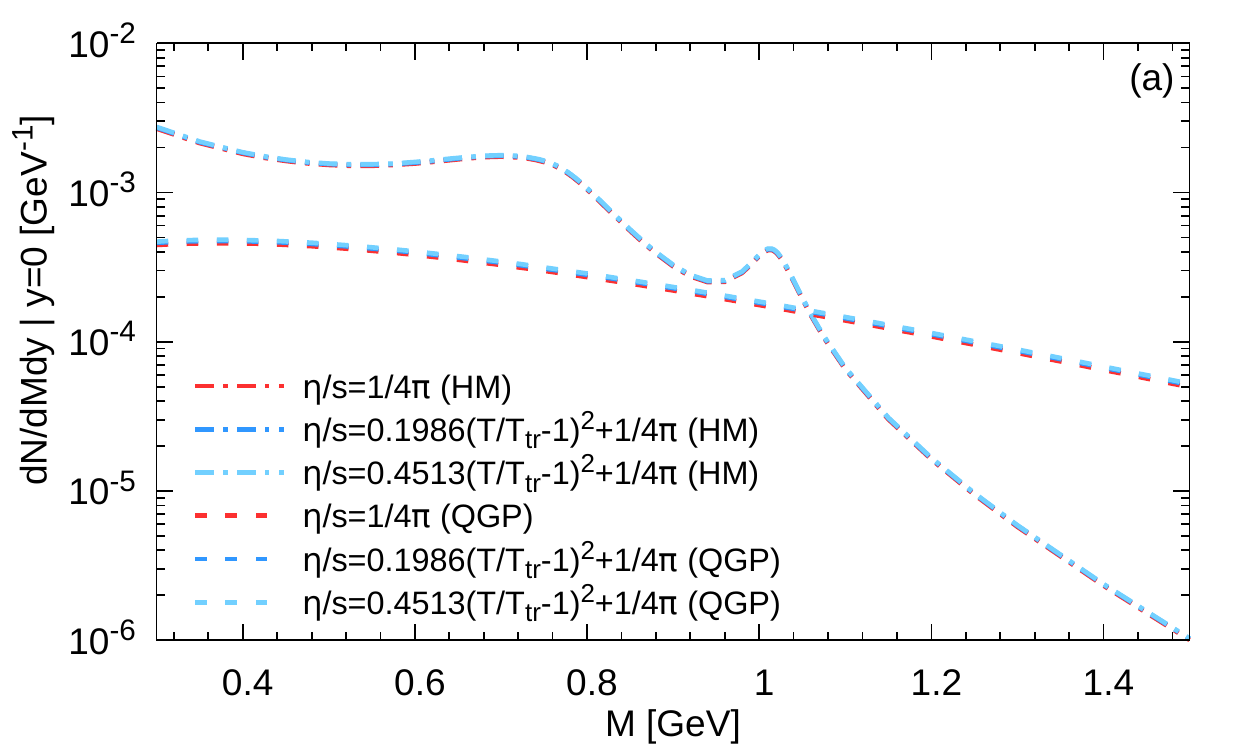}
\includegraphics[width=0.495\textwidth]{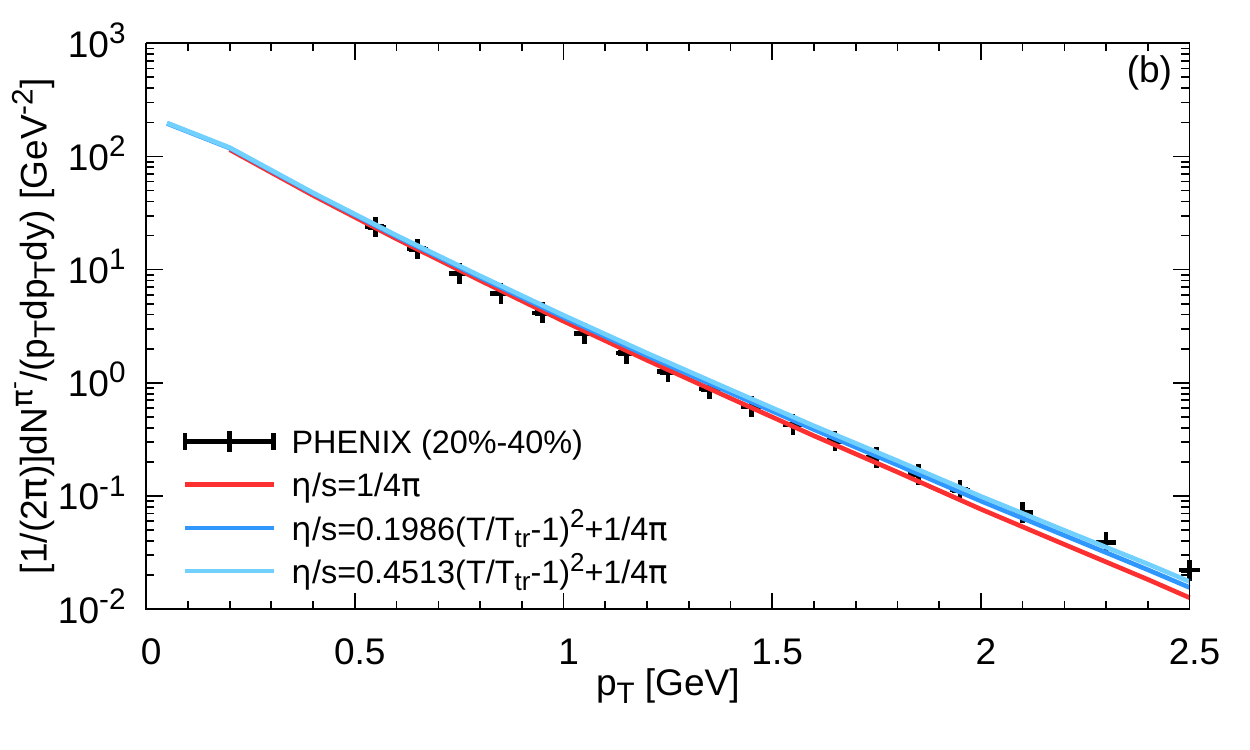}
\end{center}
\caption{(Color online) Yield of thermal dileptons (a) and pions (b) for various values of $\eta/s(T)$.}
\label{fig:entropy_prod_radial_exp_quad}
\end{figure}
As explained in Sec. \ref{sec:lin_eta_s_T}, the consequences of additional entropy production for media with a quadratic $\eta/s(T)$ relative to the one with $\eta/s=1/(4\pi)$ can be seen by examining the invariant mass dilepton yield in Fig. \ref{fig:entropy_prod_radial_exp_quad} (a). Indeed, the dilepton yield is increased by about 2\% in the HM and 6\% in the QGP regions, respectively. Those two percentages should be compared with the 5\% and 10\% increase quoted in the previous section. As a reference, Fig. \ref{fig:entropy_prod_radial_exp_quad} (b) shows the effects of the quadratic $\eta/s(T)$ on the $p_T$ spectrum of charged pions. 
\begin{figure}[!h]
\begin{center}
\includegraphics[width=0.495\textwidth]{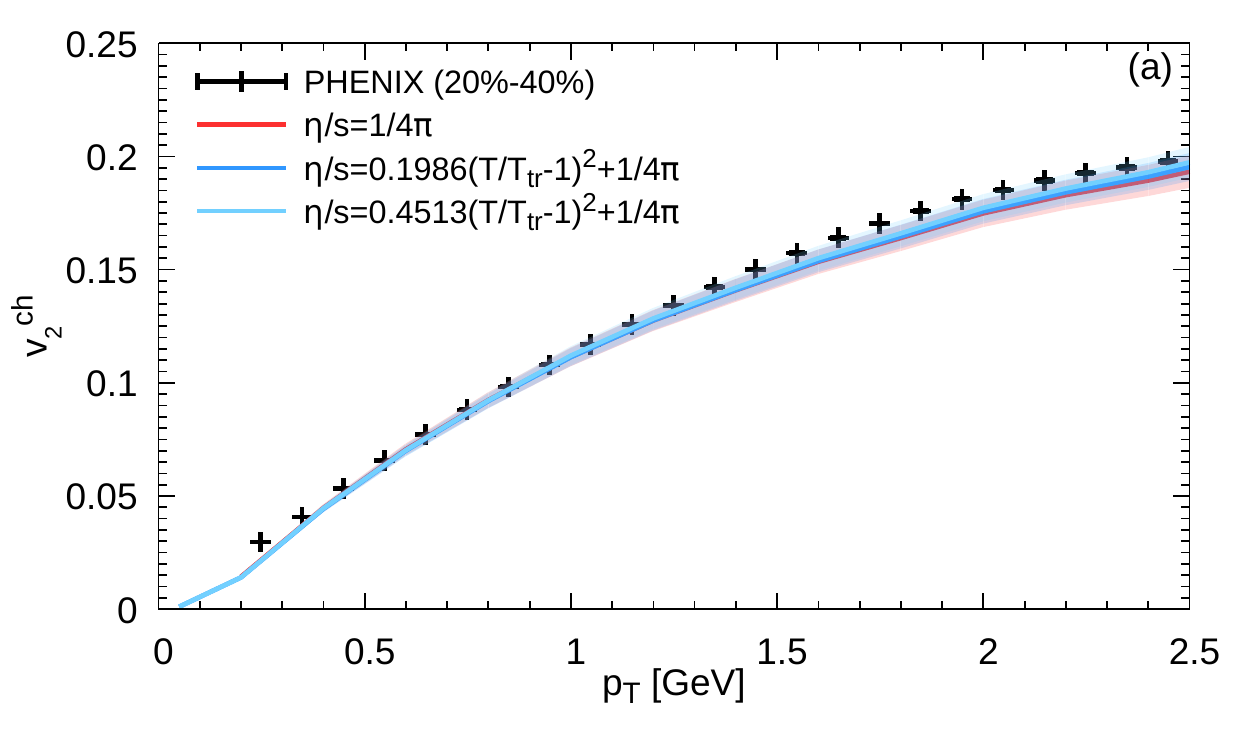}
\includegraphics[width=0.495\textwidth]{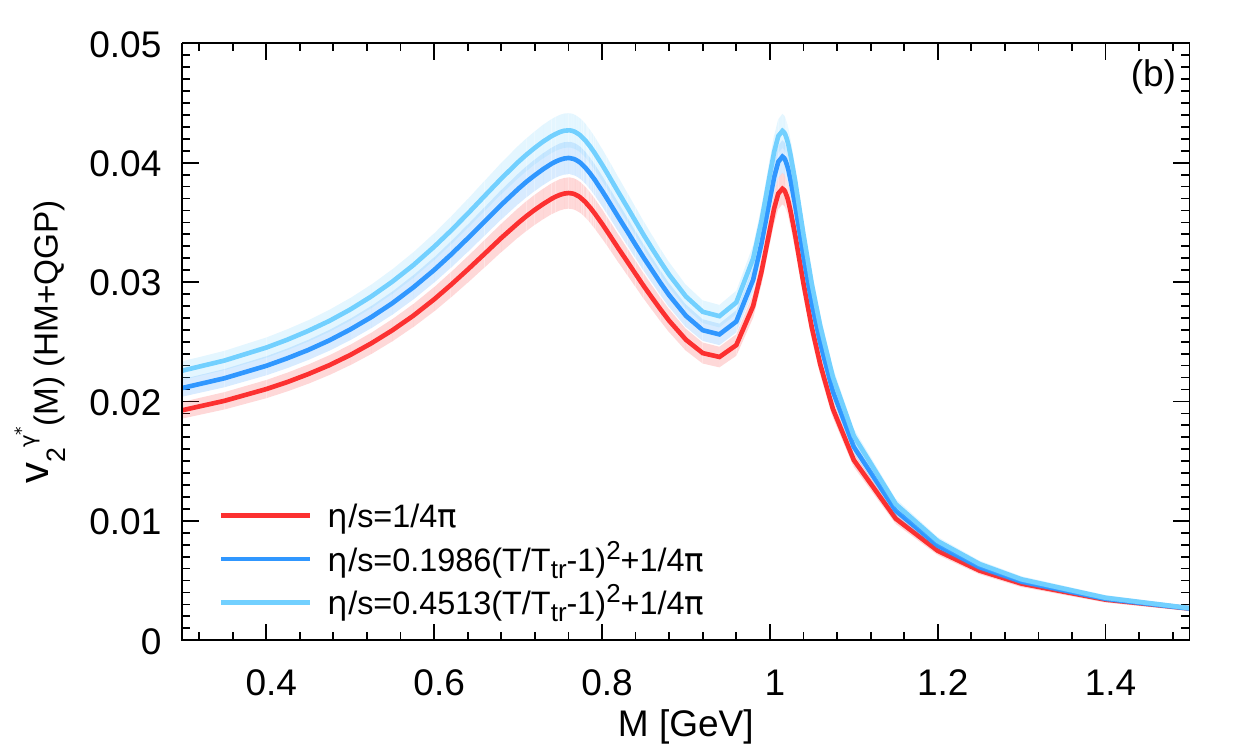}
\includegraphics[width=0.495\textwidth]{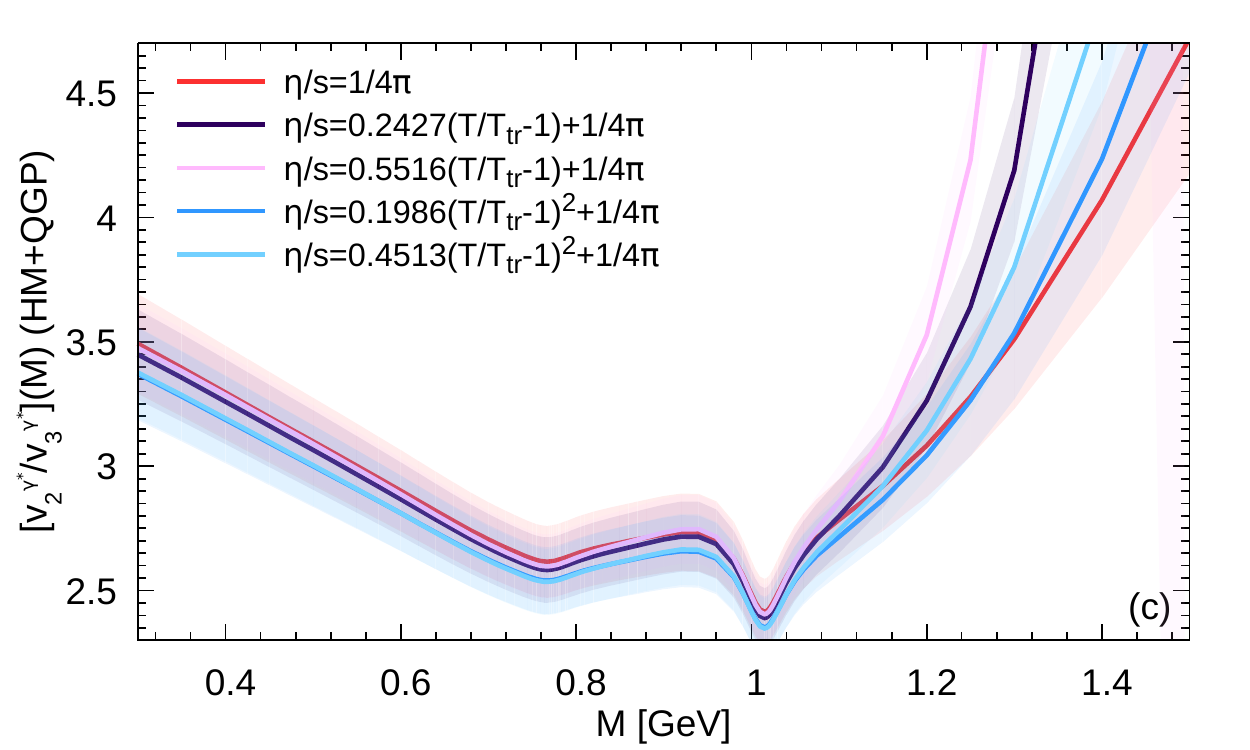}
\end{center}
\caption{(Color online) Elliptic flow of charged hadrons (a) and thermal dileptons (b) with different values for $a$ in Eq. (\ref{eq:m_a_lin_quad}). (c) $v_2/v_3$ ratio for linear and quadratic $\eta/s(T)$.}
\label{fig:v2_ch_th_dilep_quad}
\end{figure}
As in the previous section, the charged hadron elliptic flow is not affected by $\eta/s(T)$, while elliptic flow of thermal dileptons is [see Figs. \ref{fig:v2_ch_th_dilep_quad} (a) and \ref{fig:v2_ch_th_dilep_quad} (b), respectively]. 

Though the invariant mass distribution of thermal dilepton $v_2$ for a quadratic $\eta/s(T)$ is similar to a linear $\eta/s(T)$, the average value of the $v_2$ to $v_3$ ratio depicted in Fig. \ref{fig:v2_ch_th_dilep_quad} (c) seems to distinguish between a linear and a quadratic $\eta/s(T)$, especially at higher $M$. Of course, one should be mindful of the uncertainties around the average value displayed in Fig. \ref{fig:v2_ch_th_dilep_quad} (c). Nevertheless, both the invariant mass distribution of $v_2$ and the $v_2/v_3$ ratio are promising quantities to measure experimentally. 

The transverse momentum distribution of dilepton flow harmonics at different invariant masses is also an interesting quantity to consider, given that it can discern some features that are more akin to a linear versus quadratic $\eta/s(T)$. 
\begin{figure}[!h]
\begin{center}
\includegraphics[width=0.495\textwidth]{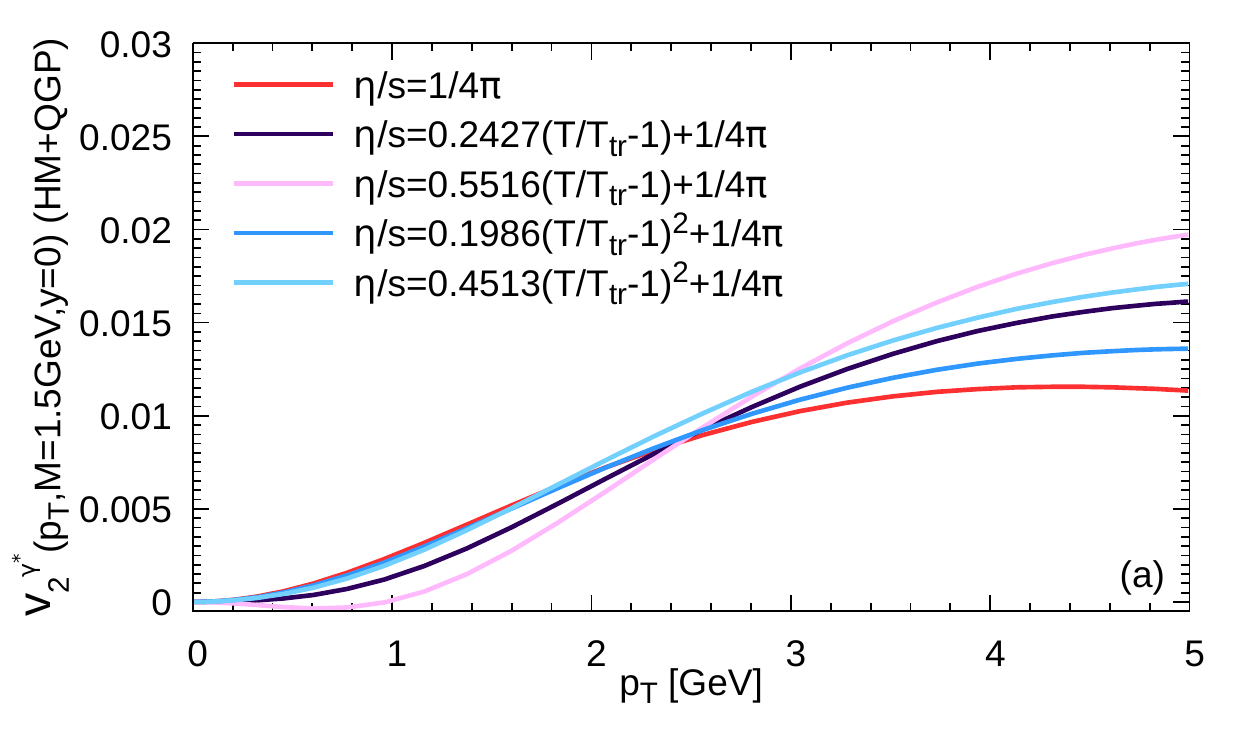}
\includegraphics[width=0.495\textwidth]{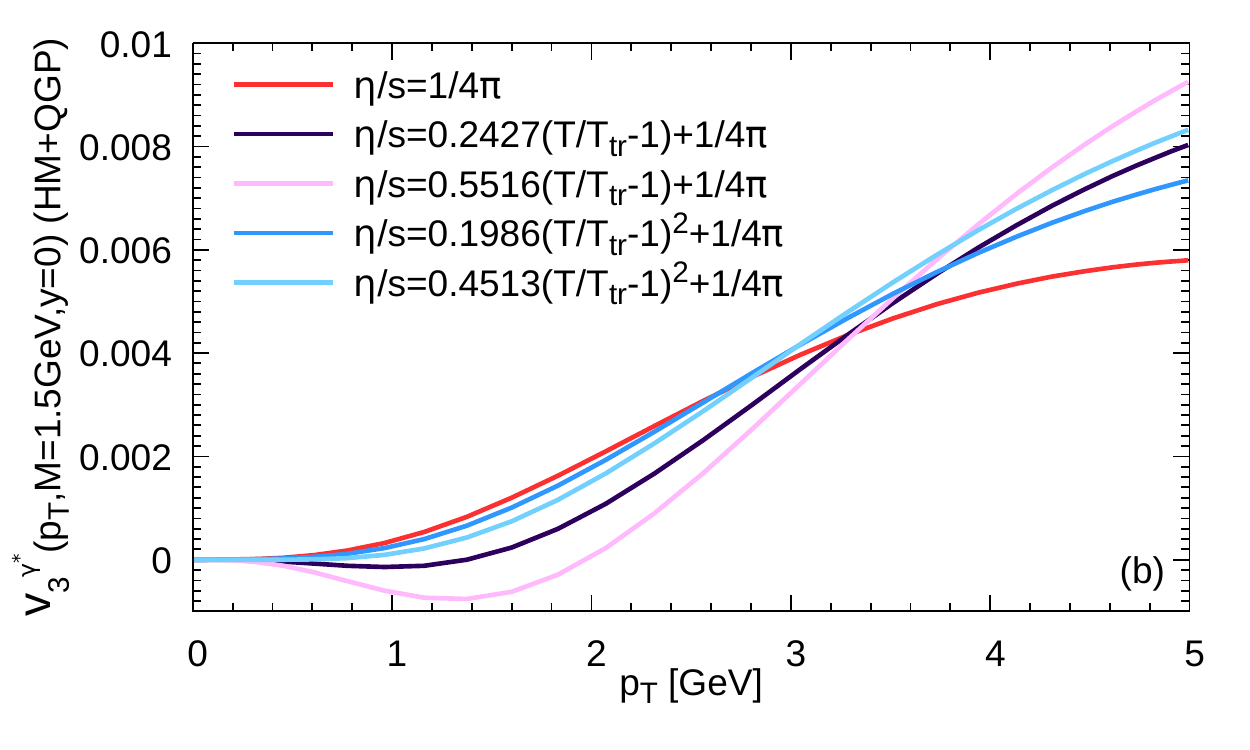}
\includegraphics[width=0.495\textwidth]{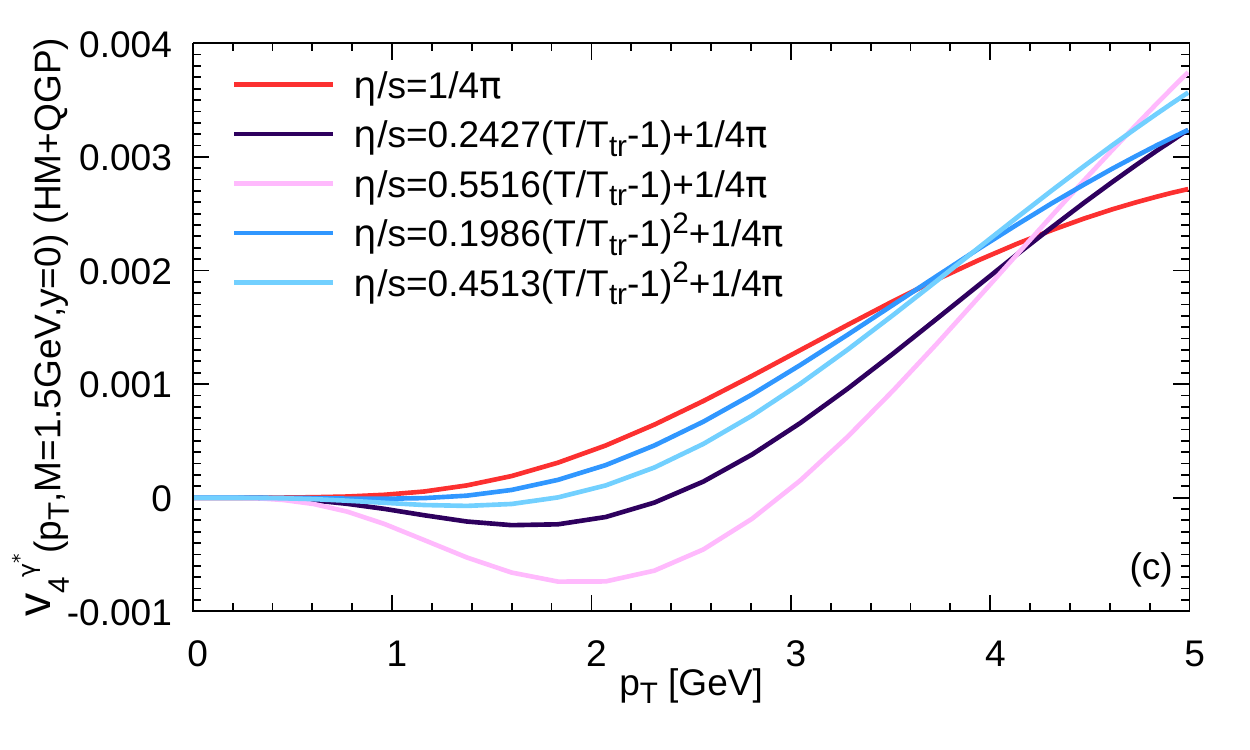}
\end{center}
\caption{(Color online) A comparison of $v_2$ (a), $v_3$ (b), and $v_4$ (c) of thermal dileptons using linear and quadratic $\eta/s(T)$ at an intermediate invariant mass. The uncertainties associated with the 200 events generated were removed to improve visual clarity.}
\label{fig:v2_pt_dilep_linquad_high_M}
\end{figure}
Starting at intermediate invariant masses, though the overall magnitude of the signal is small, Fig. \ref{fig:v2_pt_dilep_linquad_high_M} shows that $v_n(p_T)$ is different for the two functional forms for $\eta/s(T)$. 
\begin{figure}[!h]
\begin{center}
\includegraphics[width=0.495\textwidth]{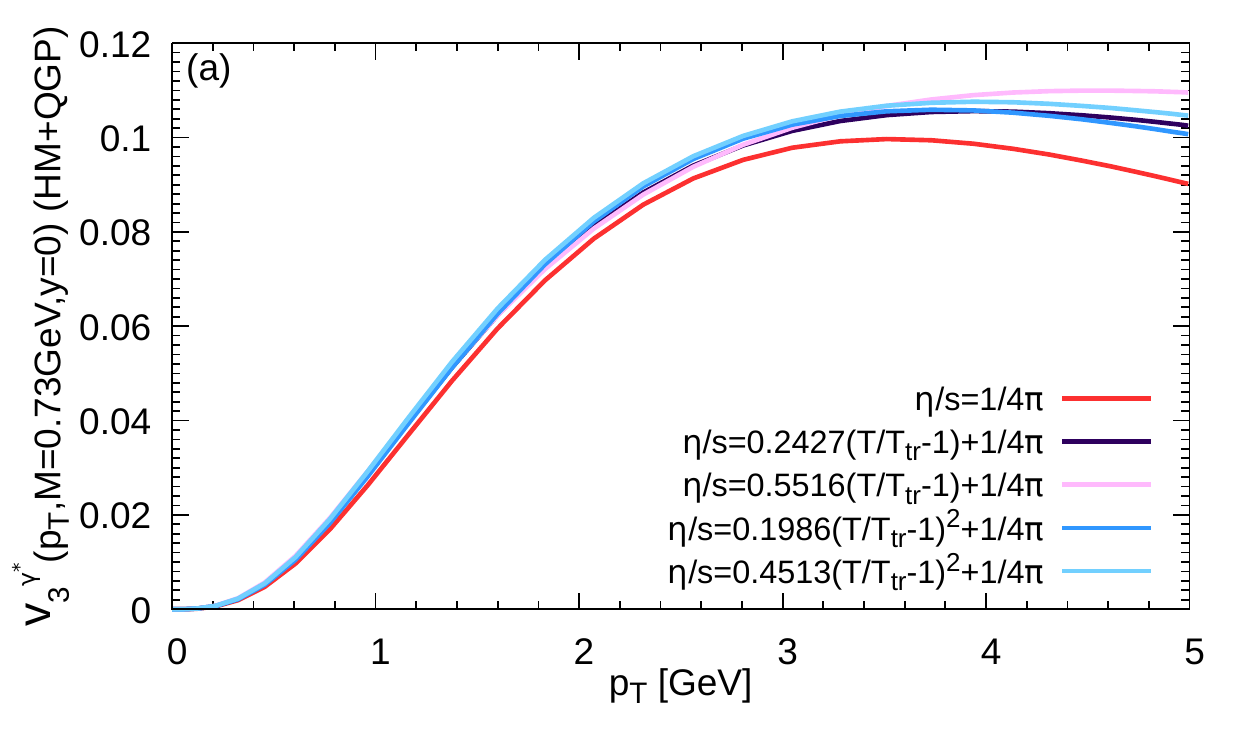}
\includegraphics[width=0.495\textwidth]{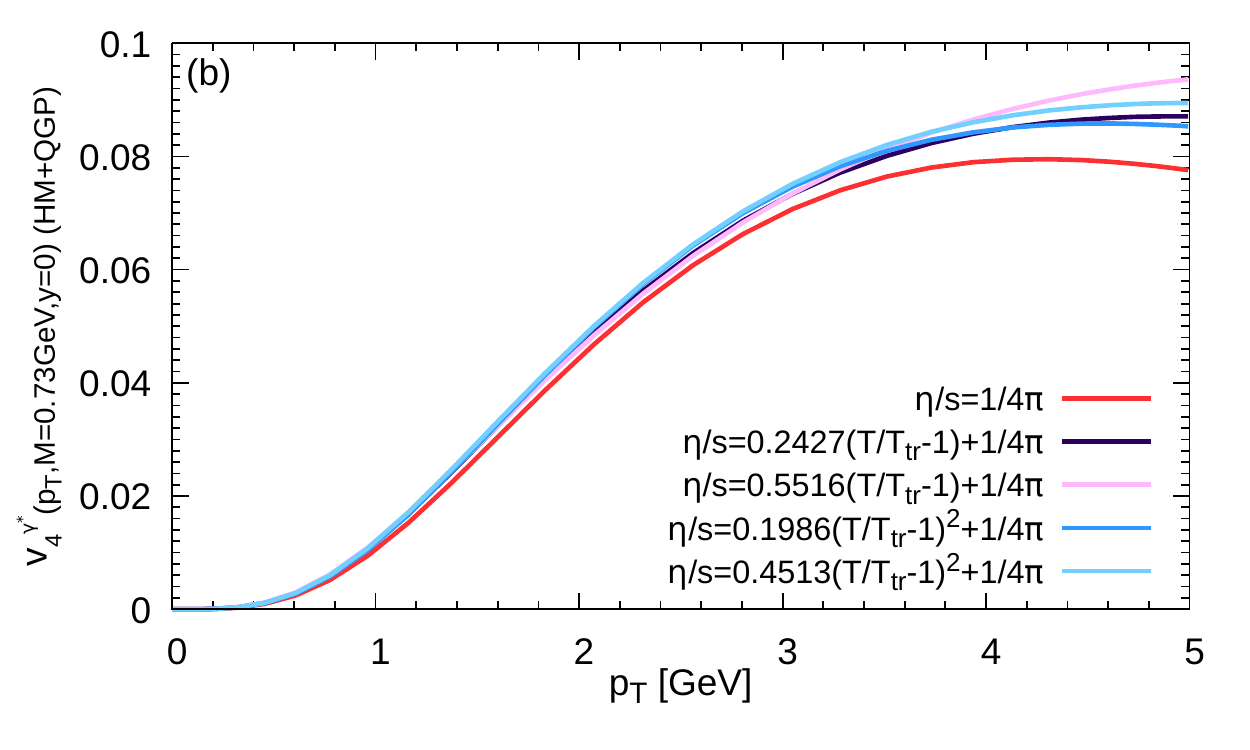}
\end{center}
\caption{(Color online) A comparison of $v_3$ (a) and $v_4$ (b) of thermal dileptons using linear and quadratic $\eta/s(T)$ at low invariant mass. The uncertainties associated with the 200 events generated were removed to improve visual clarity.}
\label{fig:v2_pt_dilep_linquad_low_M}
\end{figure}
At low invariant masses, a similar statement holds true for higher flow harmonics, especially at $p_T\gtrsim3$ GeV [see Fig. \ref{fig:v2_pt_dilep_linquad_low_M}]. In both cases, the differences seen in the $v_n(M,p_T)$ cannot be accounted for through a renormalization of the slope alone, for example. Experimentally distinguishing between the two forms of $\eta/s(T)$ using $v_n(M,p_T)$ will be challenging at RHIC given the sensitivity required, be it in the overall magnitude of the signal or in the relative difference between signals. In that regard, though the overall size of $\eta/s(T)$ may be constrained at RHIC, studying dilepton flow at LHC energies constitutes an auspicious avenue for constraining the shape of the functional dependence of $\eta/s$ at high temperatures. Such a study is currently underway. 

\section{Conclusions}
The goal of the present work is to investigate the sensitivity of thermal dileptons to a temperature-dependent $\eta/s$ at temperatures higher than 180 MeV, at top RHIC energy. We have studied the sensitivity of dilepton anisotropic flow coefficients to the slope of a linearly dependent $\eta/s(T)$ and the size of specific shear viscosity's second derivative with respect to temperature. Charged hadrons are found to be poorly sensitive to any temperature dependence, be it linear or quadratic, of $\eta/s$ at $T>180$ MeV, as was previously found in Ref. \cite{Niemi:2011ix}. We have shown that dileptons have sensitivity to a temperature-dependent $\eta/s$ at high temperatures. 

The STAR Collaboration at RHIC has recently acquired new dilepton data using its Muon Telescope Detector (MTD) and Heavy Flavor Tracker (HFT) \cite{Geurts:2016}. Having the MTD and HFT running at the same time allows one to remove the dilepton radiation coming from open heavy flavor hadrons in the low to intermediate invariant mass (i.e., $M\lesssim2.5$ GeV), thus allowing one to directly measure thermal dilepton radiation for $1.1\lesssim M\lesssim 2.5$ GeV and compare to to the results presented herein. Note that for $M \lesssim 1.1$ GeV, the open heavy flavor and the dilepton cocktail contribution needs to be removed to expose thermal radiation. As mentioned in Ref. \cite{Vujanovic:2013jpa}, the dilepton cocktail consists of late time Dalitz and vector meson decays, which are both present in the current RHIC data sets. Removing these two sources is possible, as the NA60 experiment at SPS has shown in the dimuon channel \cite{Arnaldi:2008er,Arnaldi:2008fw,Damjanovic:2008ta}, however the data at RHIC has an increased challenge of removing the open heavy flavor contribution given that the cross section for heavy flavor quark production is much larger at RHIC energy than at SPS. Therefore, given the challenges of removing both the open heavy flavor and cocktail in the low invariant mass sector, focusing on the intermediate mass region, where only open heavy flavor needs to be removed to expose thermal radiation, seems like a more promising avenue. 

The analysis of these dilepton data using the MTD and the HFT detectors at STAR is currently ongoing \cite{Geurts:2016}, with improved dilepton measurements of $v_2$ expected soon. As shown here, the ability to measure $v_2(M)$ of thermal dileptons opens the possibility of using thermal dileptons to resolve details in the overall magnitude of $\eta/s(T)$ of the QGP. Thus, extracting the temperature dependence of $\eta/s$ via dileptons seems to be a very promising prospect at RHIC. LHC, on the other hand, does not currently have the capabilities in place to accurately measure low to intermediate mass dileptons, with such measurements only being possible once the LHC detector upgrades are in place \cite{Jacobs-Roland:2016}.

Though both the slope and the size of the second derivative did influence the magnitude and shape of the dilepton flow harmonics, with appreciable effects on $v_2(M)$, distinguishing between a linear versus quadratic temperature dependence would be difficult at RHIC for $M\lesssim 2.5$ GeV using $v_2(M)$ alone, while the invariant mass distribution of the $v_2/v_3$ ratio is a more encouraging prospect to consider. As far as $v_n(M,p_T)$ is concerned, at fixed low invariant mass, though the shape of $v_3(p_T)$ and $v_4(p_T)$ is different within the linear and quadratic temperature dependence of $\eta/s$, that difference only becomes apparent at high transverse momenta. At intermediate invariant masses where the shape of $v_n(p_T)$ varies more significantly when comparing a linear to a quadratic $\eta/s(T)$, the overall magnitude of the signal is decidedly smaller. Given the differential nature of the $v_n(M,p_T)$ measurement, extracting the signal with enough statistics to be able to distinguish between a linear or quadratic $\eta/s (T)$ is experimentally challenging at low and intermediate invariant masses. Thus, the most promising dilepton candidate to learn about the temperature dependence of $\eta/s$ is $v_2(M)$, while the $v_2/v_3$ ratio offers a promising new route.   

\section*{Acknowledgments}
We are grateful to  J.-F. Paquet, C. Shen, B. Schenke, and U. Heinz for helpful discussions. This work was supported in part by the Natural Sciences and Engineering Research Council of Canada, in part by the Director, Office of Energy Research, Office of High Energy and Nuclear Physics, Division of Nuclear Physics, of the U.S. Department of Energy under Contracts No. DE-AC02-98CH10886, No. DE-AC02-05CH11231, and No. \rm{DE-SC0004286}, and in part by the National Science Foundation (in the framework of the JETSCAPE Collaboration) through Award No. 1550233. G.V. acknowledges support by the Fonds de Recherche du Qu\'ebec---Nature et Technologies (FRQNT), the Canadian Institute for Nuclear Physics, and by the Seymour Schulich Scholarship. G.S.D. acknowledges support through a Banting Fellowship from the Government of Canada. C.G. gratefully acknowledges support from the Canada Council for the Arts through its Killam Research Fellowship program. Computations were performed on the Guillimin supercomputer at McGill University under the auspices of Calcul Qu\'ebec and Compute Canada. The operation of Guillimin is funded by the Canada Foundation for Innovation (CFI), the Natural Sciences and Engineering Research Council (NSERC) of Canada, NanoQu\'ebec, and the Fonds de Recherche du Qu\'ebec---Nature et  Technologies (FRQNT).
\appendix

\section{Computing the viscous correction to the QGP rate via the Boltzmann Equation}

The discussion presented here follows Ref. \cite{Denicol:2014loa}. In order to derive the $\delta f_{\bf k}$ used in Eq. (\ref{eq:G_l-h_k}), the starting point is 
\begin{eqnarray}
\delta y_{\bf k}=\mathscr{G_{\bf k}}\phi_{\bf k},
\end{eqnarray}
where $\phi_{\bf k}$ is to be computed after performing an irreducible tensor decomposition and $\mathscr{G_{\bf k}}$ is an arbitrary function of $k^\mu u_\mu$. Indeed, one can decompose $\phi_{\bf k}$ as
\begin{eqnarray}
\phi_{\bf k}=\lambda^{(0)}_{\bf k}+\sum^{\infty}_{\ell=1}\lambda^{\langle\mu_1\ldots \mu_\ell\rangle}_{\bf k} k_{\langle\mu_1} \ldots k_{\mu_\ell\rangle} 
\end{eqnarray}
where $\lambda^{\langle\mu_1\ldots\mu_\ell\rangle}_{\bf k}=\Delta^{\mu_1\ldots\mu_\ell}_{\nu_1\ldots\nu_{\ell'}}\lambda^{\nu_1\ldots\nu_{\ell'}}_{\bf k}$ with $\Delta^{\mu_1\ldots\mu_\ell}_{\nu_1\ldots\nu_{\ell'}}$ being defined in Refs. \cite{DeGroot:1980dk,Denicol:2014loa}. For $\ell=1$ and $\ell=2$, the irreducible tensor $\Delta^{\mu_1\ldots\mu_\ell}_{\nu_1\ldots\nu_{\ell'}}$ simplifies to $\Delta^\mu_\nu$, and $\Delta^{\mu\nu}_{\alpha\beta}$, respectively. These two tensors were defined in Sec. \ref{sec:hydro}. The tensors $\lambda^{\langle\mu_1\ldots \mu_\ell\rangle}_{\bf k}$, being expanded in terms of the mutually orthogonal irreducible tensors $k_{\langle\mu_1} \ldots k_{\mu_\ell\rangle}$, can be further factorized into a linear combination of an orthonormal set of functions $P^{(\ell)}_{n,{\bf k}}$, that explicitly depend on $E_{\bf k}=u^\nu k_\nu$, and a set of rank-${\ell}$ tensor coefficient $c^{\langle\mu_1\ldots \mu_\ell\rangle}_n$ as 
\begin{eqnarray}
\lambda^{\langle\mu_1\ldots \mu_\ell\rangle}_{\bf k} = \sum^{N_\ell}_{n=0} c^{\langle\mu_1\ldots \mu_\ell\rangle}_n P^{(\ell)}_{n,{\bf k}}. 
\end{eqnarray}  
So, the expansion basis of the tensorial structure of $\phi_{\bf k}$ is $k_{\langle\mu_1} \ldots k_{\mu_\ell\rangle}$, which, analogous to spherical harmonics, contains the angular dependence of $\phi_{\bf k}$. The expansion coefficients are $c^{\langle\mu_1\ldots \mu_\ell\rangle}_n$. Using the spherical harmonics analogy, $\lambda^{(0)}_{\bf k}$, $\lambda^{\langle\mu_1\rangle}_{\bf k}$, and $\lambda^{\langle\mu_1\mu_2\rangle}_{\bf k}$ can be interpreted as monopole, dipole, and quadrupole contributions to $\phi_{\bf k}$, respectively, and so on for the higher order tensors. The irreducible tensors $k_{\langle\mu_1} \ldots k_{\mu_\ell\rangle}$ satisfy the orthogonality condition 
\begin{eqnarray}
\int dK n_{\bf k} k_{\langle\mu_1} \ldots k_{\mu_\ell\rangle} k^{\langle\mu_1} \ldots k^{\mu_{\ell'}\rangle} = \frac{\ell ! (2\ell +1)\delta_{\ell\ell'}}{(2\ell+1)!!} \int dK (\Delta^{\alpha\beta} k_\alpha k_\beta)^\ell n_{\bf k},
\label{eq:}
\end{eqnarray}
where 
\begin{eqnarray}
\int dK \equiv \int \frac{d^4 k}{(2\pi)^4}\delta\left(k^\nu k_\nu -m^2\right)\theta\left(k^0\right).
\label{eq:}
\end{eqnarray}
On the other hand, $\phi_{\bf k}$'s radial dependence is expanded using the orthonormal basis functions $P^{(\ell)}_{n,{\bf k}}$, which can be written as 
\begin{eqnarray}
P^{(\ell)}_{n,{\bf k}}=\sum^{n}_{r=0} a^{(\ell)}_{n,r} E^r_{\bf k}.
\label{eq:}
\end{eqnarray}
The orthonormal basis functions $P^{(\ell)}_{n,{\bf k}}$ satisfy
\begin{eqnarray}
\int dK P^{(\ell)}_{n,{\bf k}} P^{(\ell)}_{m,{\bf k}} \omega^{(\ell)} = \delta_{mn},
\label{eq:}
\end{eqnarray}
where 
\begin{eqnarray}
\omega^{(\ell)}= \frac{(-1)^\ell}{(2\ell+1)!!}\frac{(\Delta^{\alpha\beta} k_\alpha k_\beta)^\ell (1-f_{0,{\bf k}}) f_{0,{\bf k}} \mathscr{G}_{\bf k}}{\int dK (-\Delta^{\alpha\beta} k_\alpha k_\beta)^\ell f_{0,{\bf k}}(1-f_{0,{\bf k}}) \mathscr{G}_{\bf k}}
\label{eq:}
\end{eqnarray}
and has the property $\int dK \omega^{(\ell)}=1$. Being interested in computing $\delta y_{\bf k}$ for shear viscous stresses, the only term needed is $\ell=2$. Thus $\delta y_{\bf k}$ can be expressed as 
\begin{equation}
\delta y_{\bf k} = \mathscr{G}_{\bf k} \sum_{n=0}^{N_2} P^{(2)}_{n,{\bf k}} c^{\langle\mu \nu\rangle}_n k_{\langle\mu} k_{\nu\rangle},
\end{equation}
where, using the orthogonality condition of the irreducible tensors, 
\begin{eqnarray}
c^{\langle\mu \nu\rangle}_n = \frac{1}{2!} \frac{\int dK \mathscr{G}_{\bf k} P^{(2)}_{n,{\bf k}} k^{\langle\mu} k^{\nu\rangle} \delta f_{\bf k}}{\int dK (-\Delta^{\alpha\beta} k_\alpha k_\beta)^2 f_{0,{\bf k}} (1-f_{0,{\bf k}}) \mathscr{G}_{\bf k}}.
\label{eq:}
\end{eqnarray}
It is convenient to re-express $\delta y_{\bf k}$ in terms of irreducible moments of $\delta f_{\bf k}$,
\begin{eqnarray}
\rho^{\mu\nu}_{n} = \int dK E^n_{\bf k} k^{\langle\mu} k^{\nu\rangle} \delta f_{\bf k},
\label{eq:}
\end{eqnarray} 
such that
\begin{eqnarray}
\delta f_{\bf k} = f_{0,\bf k} (1-f_{0,{\bf k}}) \mathscr{G}_{\bf k} \sum_{n=0}^{N_2} \mathcal{H}^{(2)}_{n,{\bf k}} \rho^{\mu\nu}_{n} k_{\langle\mu} k_{\nu\rangle},
\label{eq:moment_expansion}
\end{eqnarray}
where 
\begin{eqnarray}
\mathcal{H}^{(2)}_{n,{\bf k}} = \frac{1}{2!} \frac{\sum^{N_2}_{m=n} a^{(2)}_{mn}P^{(2)}_{m,{\bf k}} \mathscr{G}_{\bf k}}{\int dK (-\Delta^{\alpha\beta} k_\alpha k_\beta)^2 f_{0,{\bf k}}(1-f_{0,{\bf k}}) \mathscr{G}_{\bf k}}.
\label{eq:}
\end{eqnarray}

At this point, we have expressed $\delta f_{\bf k}$ in terms of its moments $\rho^{\mu\nu}_n$. However, only the lowest of these moments, $\rho^{\mu\nu}_0=\pi^{\mu\nu}$, are described within  hydrodynamics. In order to apply this formula to describe the momentum distribution of particles within a fluid, it is still necessary to approximate the remaining moments in terms of the fluid dynamical degrees of freedom. In the hydrodynamical limit, one can assume that all moments $\rho^{\mu\nu}_{n}$ have sufficiently approached their asymptotic values and have relaxed to their Navier-Stokes limit. That is,
\begin{eqnarray}
\rho^{\mu\nu}_n &\approx& 2\eta_n \sigma^{\mu\nu}.
\label{eq:}
\end{eqnarray}
With this approximation it becomes possible to express all moments $\rho^{\mu\nu}_n$ in terms of $\pi^{\mu\nu}$, in the following way:
\begin{eqnarray}
\rho^{\mu\nu}_n\approx\frac{\eta_n}{\eta}\pi^{\mu\nu},
\label{eq:}
\end{eqnarray}
where we have used the Navier-Stokes limit for $\pi^{\mu\nu}$, namely $\pi^{\mu\nu}=2\eta\sigma^{\mu\nu}$. Here, $\eta_n$ is a set of transport coefficients which contain the microscopic information of the system. In fact, $\eta_0$ is nothing but the usual shear viscosity coefficient $\eta$ already discussed. The remaining transport coefficients are less known, but can be calculated within the framework of the Boltzmann equation (or kinetic theory). An estimate of these transport coefficients was derived in Ref. \cite{Denicol:2012cn} within the Boltzmann equation, assuming the colliding quarks are massless and that their $2\rightarrow2$ scattering cross section is constant.\footnote{Note that this approximation is not valid in the hadronic sector, where all colliding particles are massive.} Using the same constant cross-section approximation, the final expression for $\delta f_{\bf k}$ becomes
\begin{eqnarray}
\delta f_{\bf k} = f_{0,{\bf k}}(1-f_{0,{\bf k}}) \mathscr{G}_{\bf k} \left[\sum^{N_2}_{n=0}\mathcal{H}^{(2)}_{n,\bf k} \frac{\eta_n}{\eta}\right]\frac{\pi^{\mu\nu}}{2(\varepsilon+P)} \frac{k_\mu}{T} \frac{k_\nu}{T},
\label{eq:}
\end{eqnarray} 
where $\mathcal{G}_{\bf k}=\mathscr{G}_{\bf k} \left[\sum^{N_2}_{n=0}\mathcal{H}^{(2)}_{n,\bf k} \frac{\eta_n}{\eta}\right]$, and the temperature dependence was introduced by replacing all instances of $k^\mu$ with $\frac{k^\mu}{T}$ in the above derivation. Keeping terms up to $N_2=3$, to improve convergence of the series for $\delta f_{\bf k}$, two functions were chosen. In the low $x=\frac{k\cdot u}{T}$ limit (where $x<11.2)$, $\mathscr{G}_{\bf k}=\frac{1}{0.1+x}$, whereas $\mathscr{G}_{\bf k}=\frac{1}{(0.1+x)^4}$ is present in the high $x$ region, i.e., for $x>11.2$. Collecting powers of $x$ after expanding out the series $\sum^{N_2}_{n=0}\mathcal{H}^{(2)}_{n,\bf k} \frac{\eta_n}{\eta}$, one can derive Eq. (\ref{eq:G_l-h_k}). Furthermore, we have verified that the $\delta f_{\bf k}$ in Eq. (\ref{eq:G_l-h_k}) has converged by going to higher $N_2=4$ order, without significantly changing the coefficients the power series of $x$. Note that the coefficients in Eq. (\ref{eq:G_l-h_k}) were computed assuming that all chemical potentials are set to zero. If that is not the case, which happens when the net baryon number diffusion is considered, for example, then the coefficients depend on the chemical potential. Last note that setting $\mathscr{G}_{\bf k}=1$, and letting $N_2=0$, recovers the original IS viscous correction. 

We conclude this appendix with the following two remarks regarding $\delta f_{\bf k}$:  
\begin{itemize}
	\item Using perturbative QCD (pQCD) to derive $\delta f_{\bf k}$ would not be suitable. For one, the shear viscosity $\eta_0$ obtained through pQCD would be very large \cite{Arnold:2003zc} (possibly also leading to very large $\eta_n$). Implementing such a large $\eta_0$ in a hydrodynamical simulation would not only prevent any fit to experimental data, but would in fact violate the small Knudsen and inverse Reynolds numbers assumption of dissipative hydrodynamics.
	\item If one is to apply the above procedure to the hadronic sector, not only is the mass of hadrons participating in a particular interaction needed, but also the scattering cross section (or matrix element) associated with that particular interaction. So, every single hadron would have its own $\delta f_{\bf k}$. Given that a multitude of hadronic interaction cross sections are simply not known experimentally (and are poorly constrained theoretically), there is very little incentive to systematically expand $\delta f_{\bf k}$ in the hadronic sector. 
\end{itemize}

\section{Effects of $\delta R$ corrections on the differential dilepton yield of the QGP sector}\label{sec:size_of_dR}
In the light of the viscous correction to the quark distribution function in the QGP derived in the previous appendix, it is instructive to investigate the manner in which the differential dilepton yield is modified. To that end, similar to the ratio $A$ defined in Eq. (\ref{eq:ratio_A}), consider the quantity
\begin{eqnarray}
\frac{\delta N}{N_0}&=&\frac{\left\vert\langle \int^b_a p_T dp_T \int d\phi dy_s \tau d\eta_s \frac{d^4 \delta R}{d^4 p}\rangle_{\rm ev}\right\vert}{\langle\int^b_a p_T dp_T \int d\phi dy_s \tau d\eta_s \frac{d^4 R_0}{d^4 p}\rangle_{\rm ev}}\nonumber\\
                    &=&\frac{\left|\frac{d \delta N}{dMdydxd\tau} (M,a,b,y=0)\right|}{\frac{d N_0}{dMdydxd\tau} (M,a,b,y=0)},    
\label{eq:deltaN_N}
\end{eqnarray} 
where $y$ is the momentum rapidity whereas $y_s$ is the space-time $y$-coordinate. Equation (\ref{eq:deltaN_N}) quantifies the size of the contribution of the QGP yield coming from the anisotropic correction to the rate relative to the ideal rate \cite{Shen:2016zpp}. 
\begin{figure}[!h]
\begin{tabular}{cc}
\includegraphics[width=0.5\textwidth]{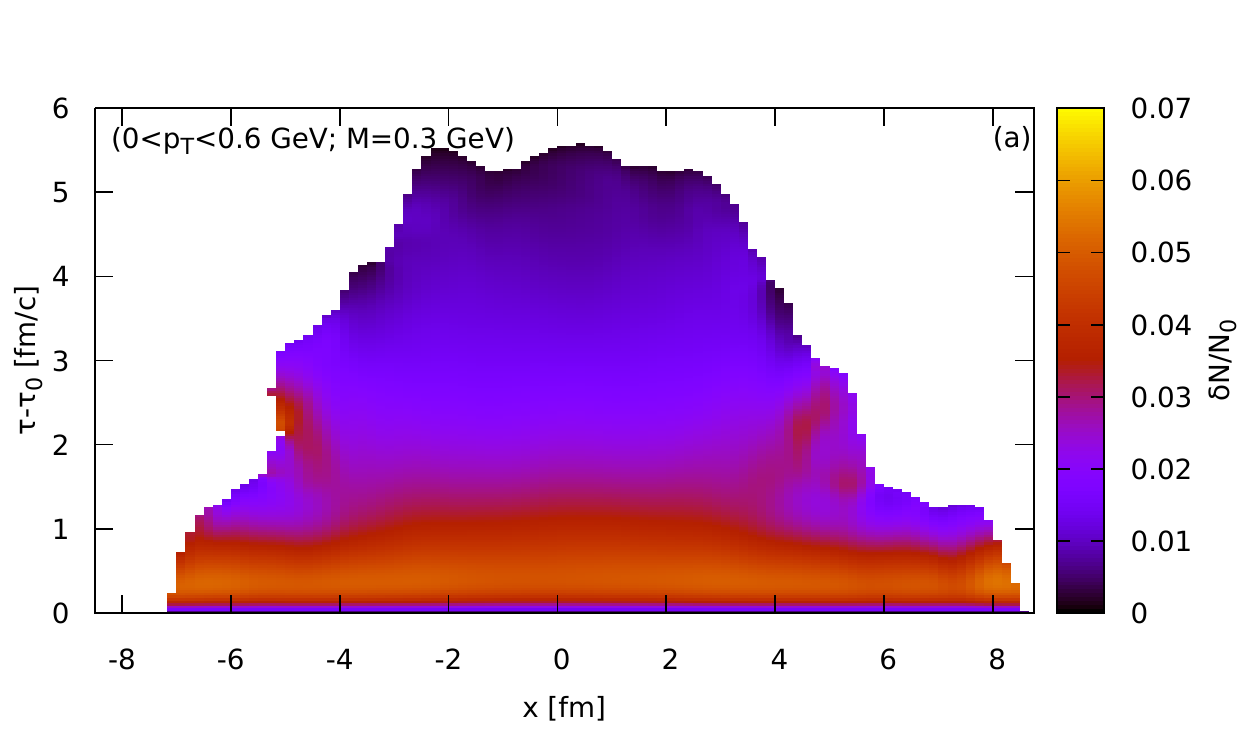}    & \includegraphics[width=0.5\textwidth]{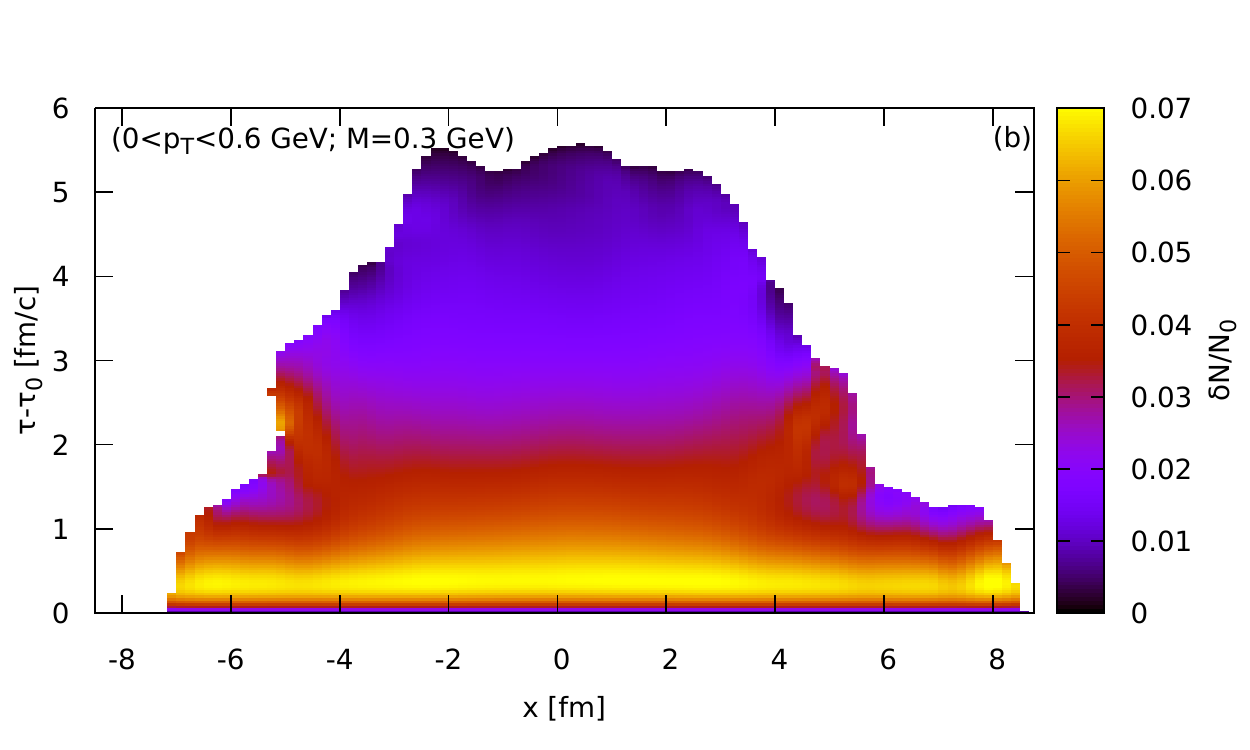}\\
\includegraphics[width=0.5\textwidth]{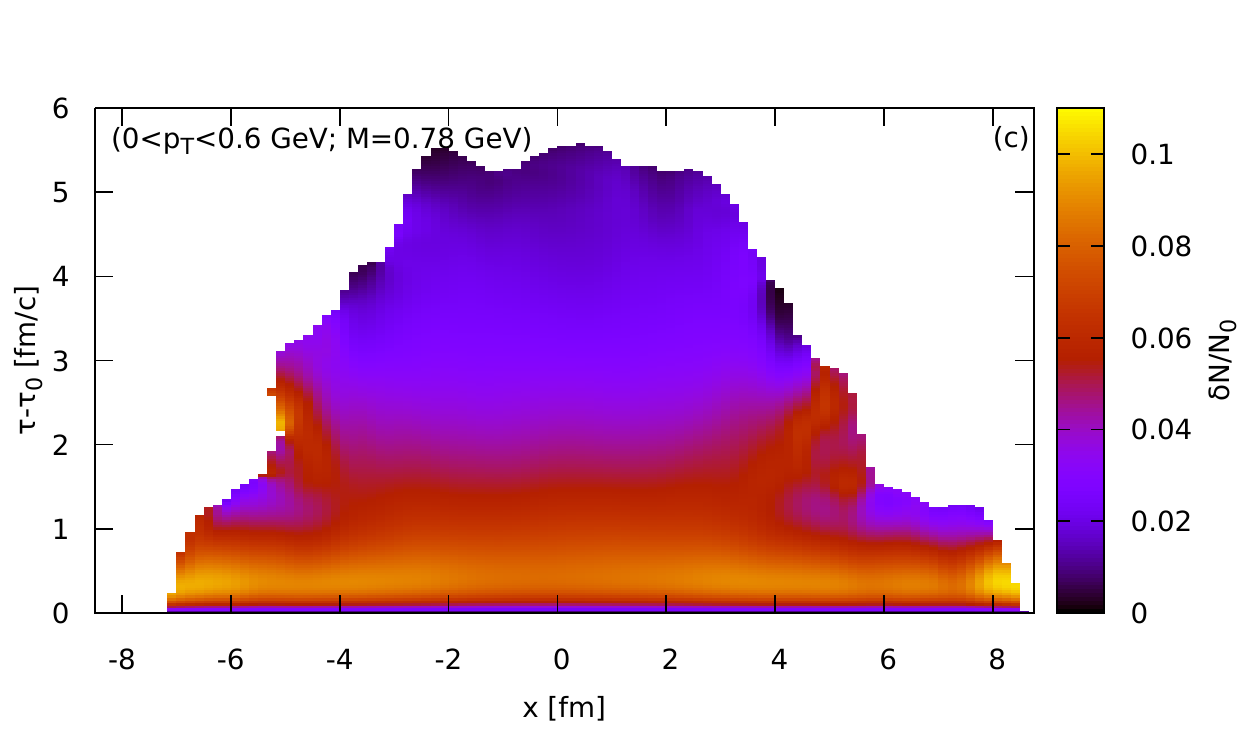} & \includegraphics[width=0.5\textwidth]{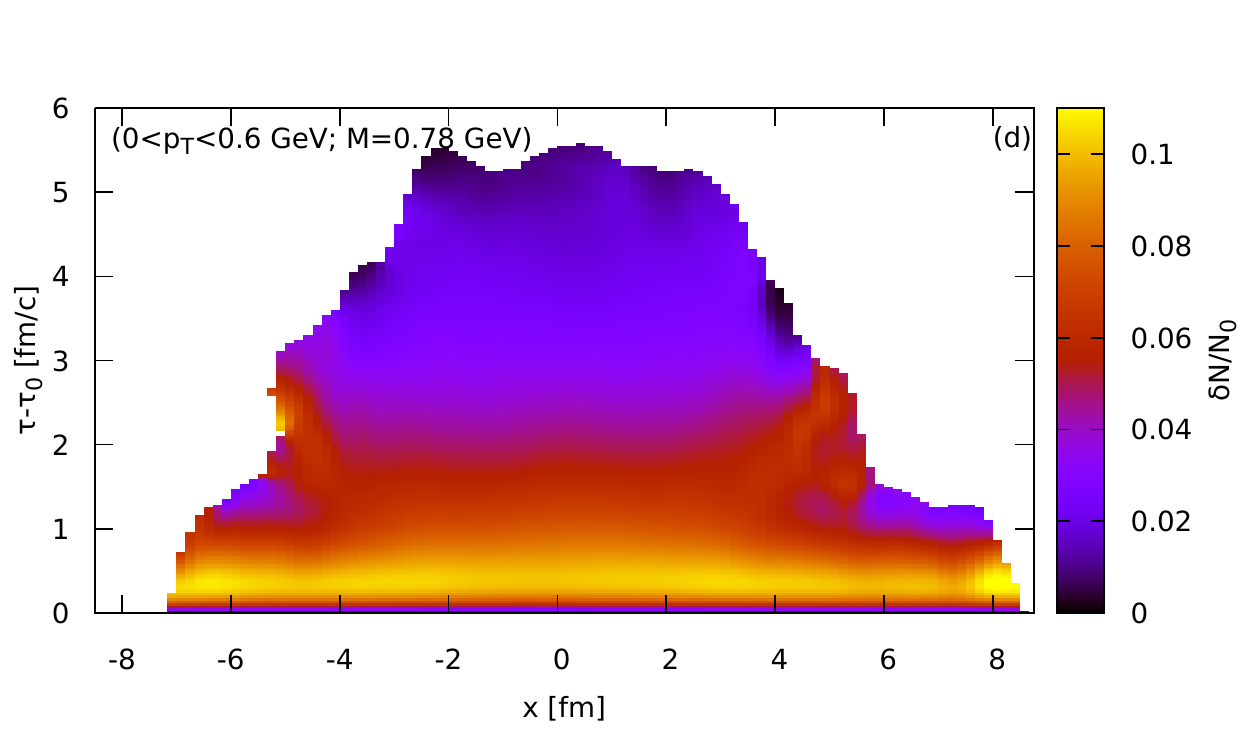}\\
\includegraphics[width=0.5\textwidth]{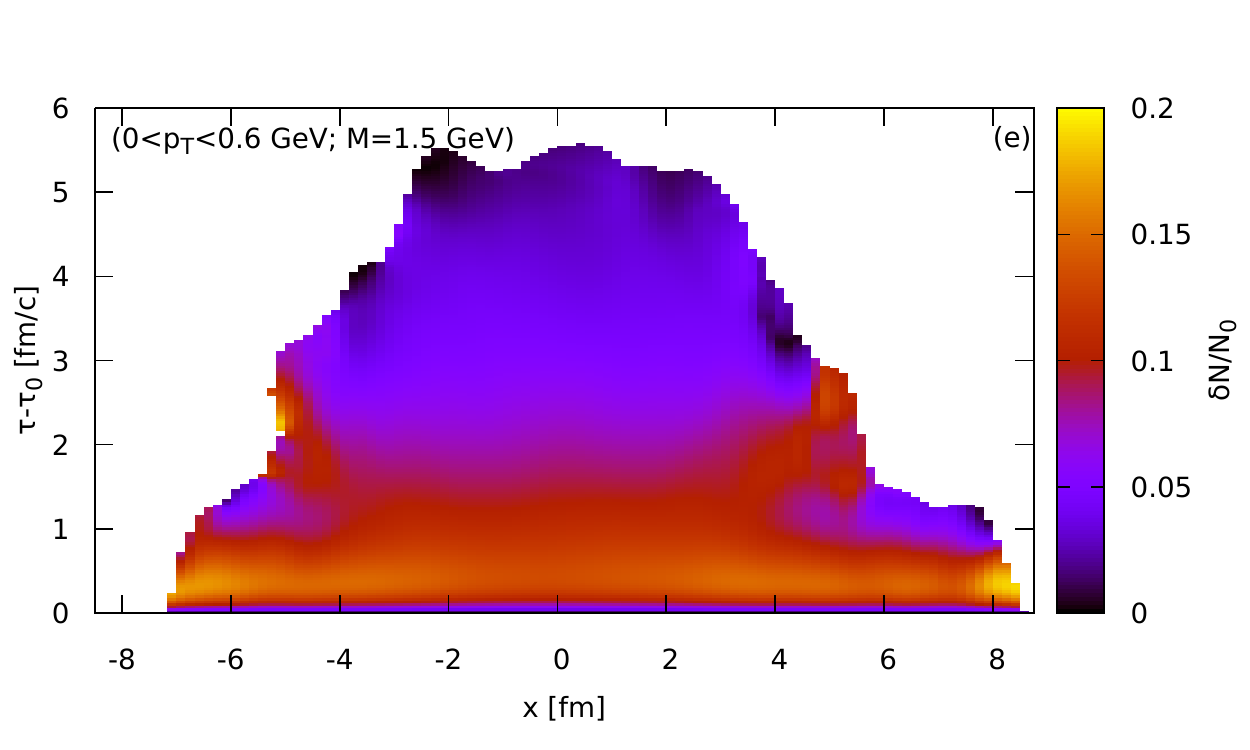}  & \includegraphics[width=0.5\textwidth]{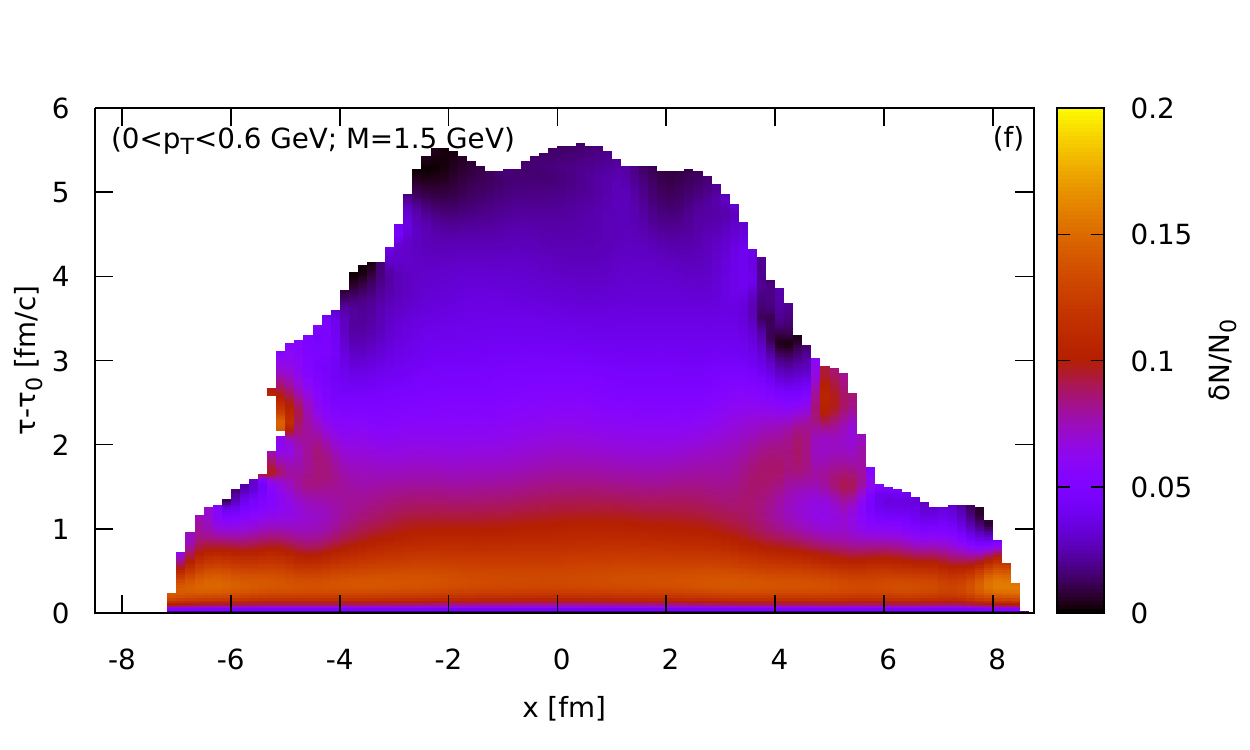}\\
\includegraphics[width=0.5\textwidth]{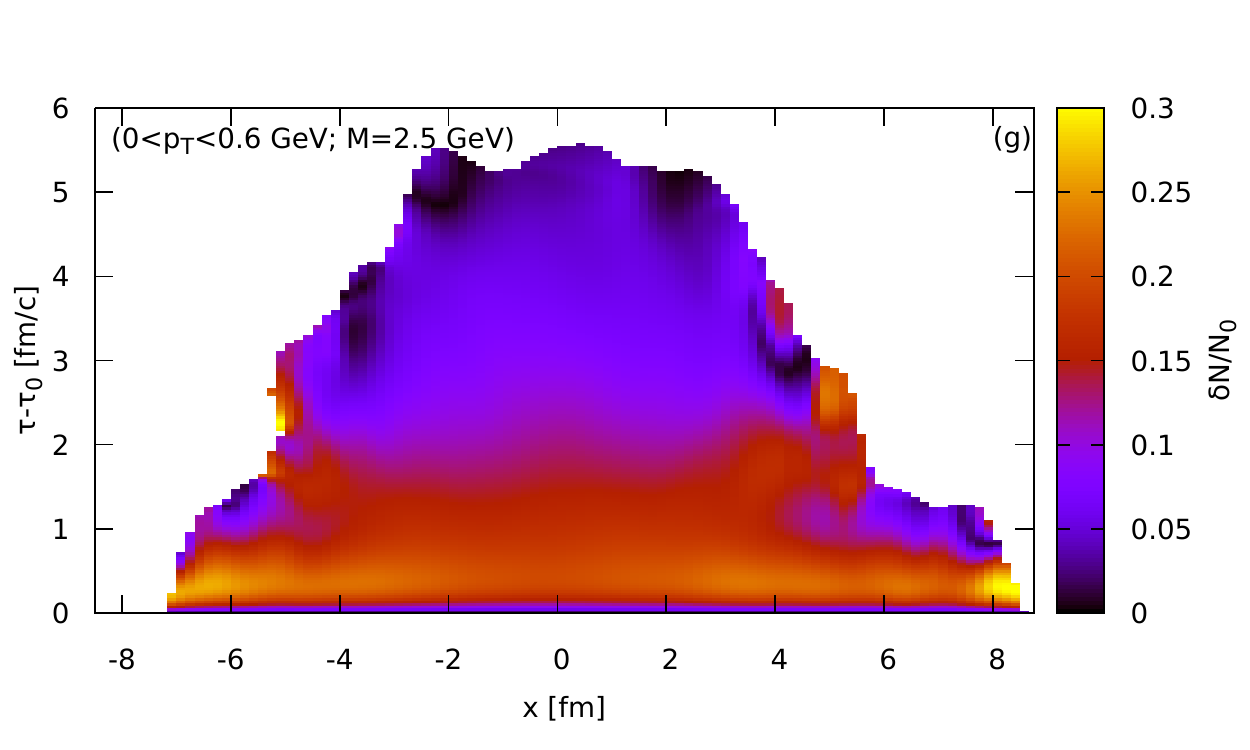}    & \includegraphics[width=0.5\textwidth]{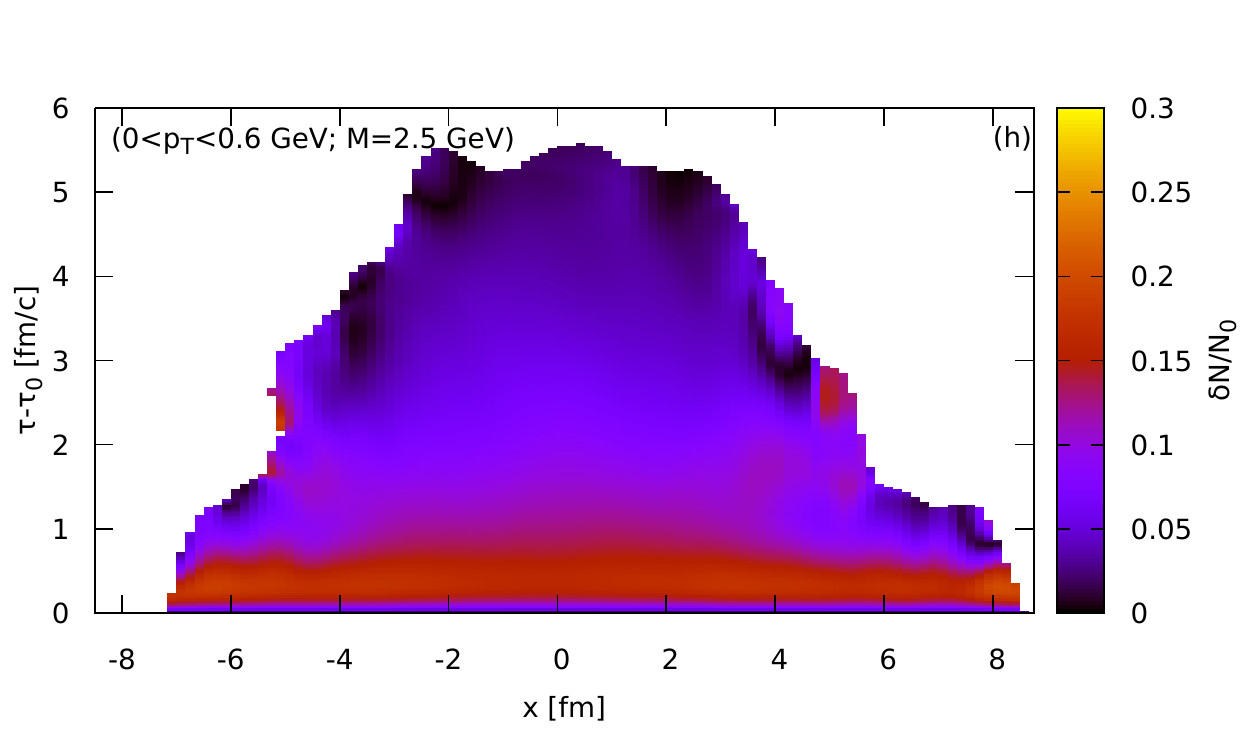}  
\end{tabular}
\caption{(Color online) The size of the viscous contribution to the dilepton yield in the QGP relative to its ideal (inviscid) dilepton production for a medium with constant $\eta/s=1/(4\pi)$. Left column: $\delta N/N_0$ using the IS $\delta R$. Right column: $\delta N/N_0$ using the constant cross section $\delta R$.}
\label{fig:deltaN_over_N0_vs_M}
\end{figure}

\begin{figure}[!h]
\begin{tabular}{cc}
\includegraphics[width=0.5\textwidth]{const_eta_s_deltaN_over_N_old_int_pt_0to0p6_M_0p3.pdf}    & \includegraphics[width=0.5\textwidth]{const_eta_s_deltaN_over_N_new_int_pt_0to0p6_M_0p3.pdf}\\
\includegraphics[width=0.5\textwidth]{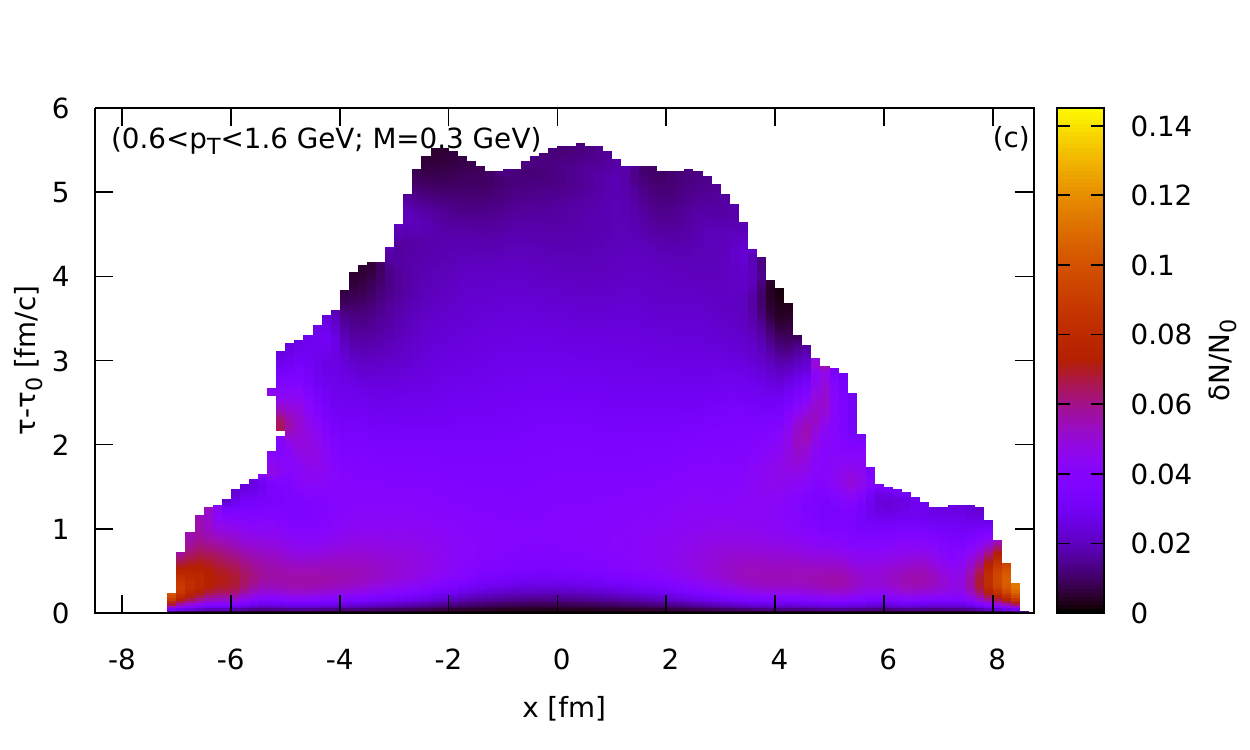} & \includegraphics[width=0.5\textwidth]{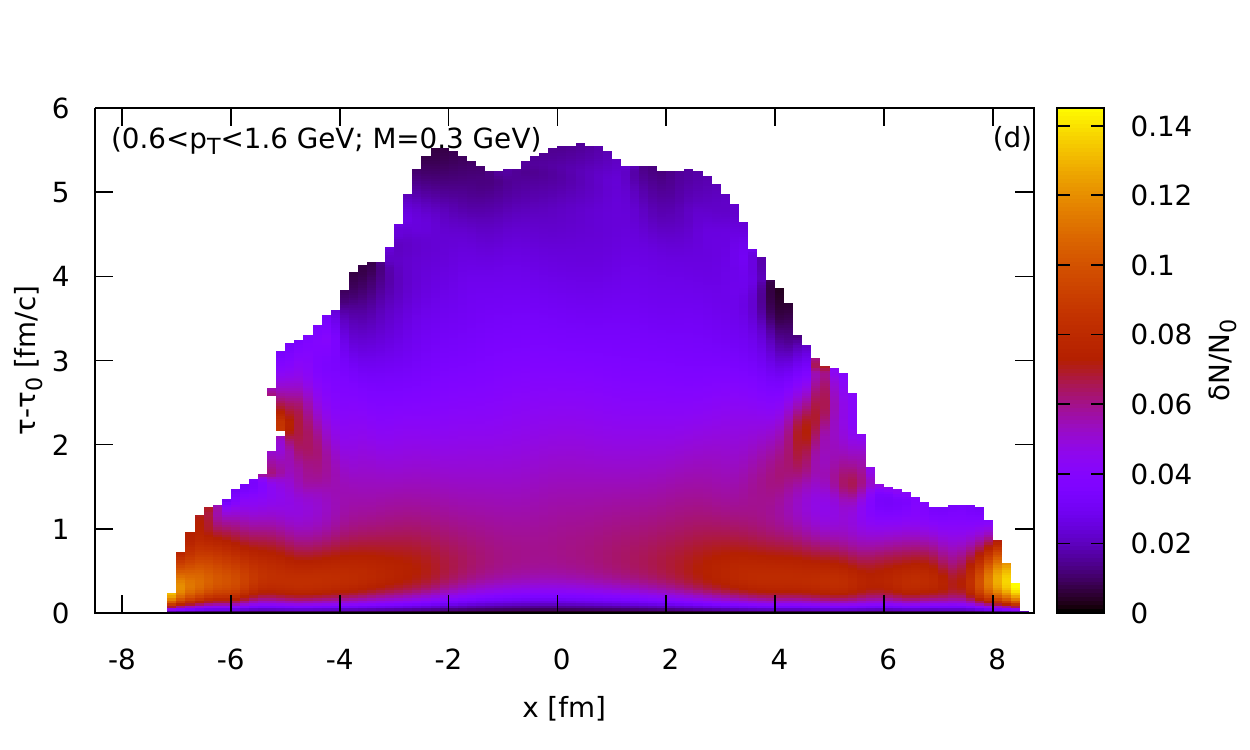}\\
\includegraphics[width=0.5\textwidth]{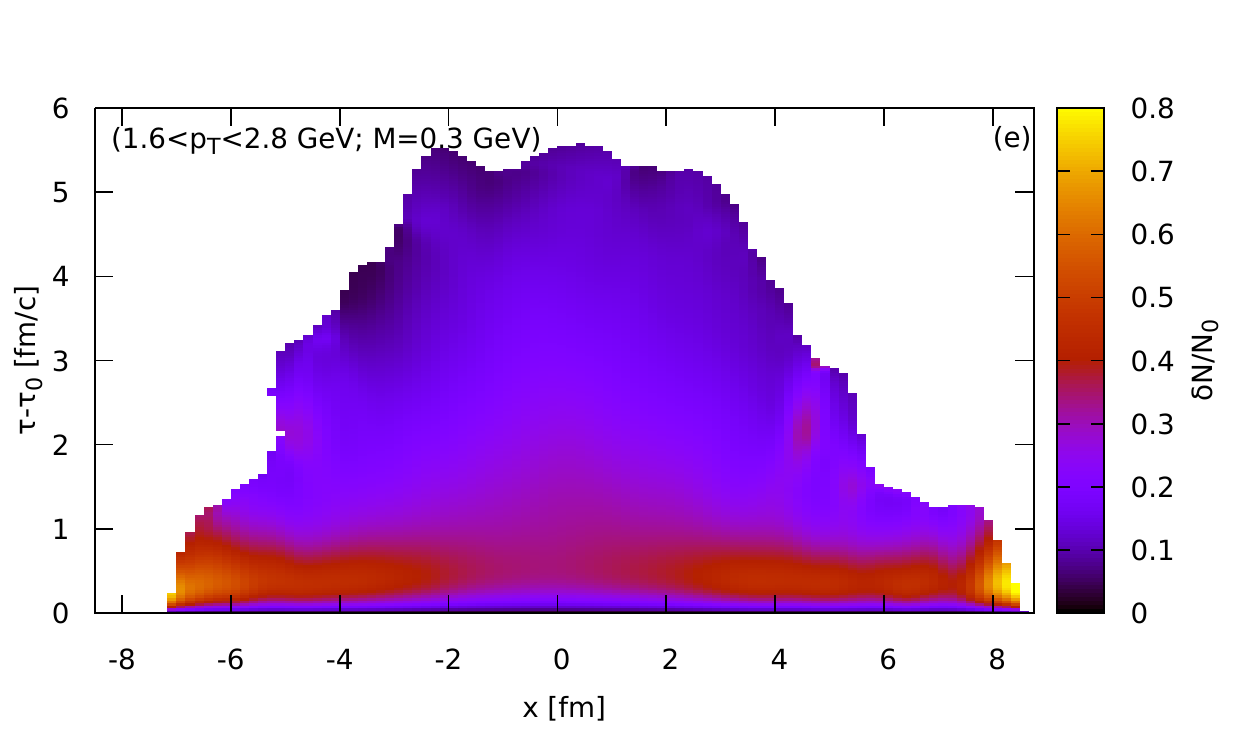}  & \includegraphics[width=0.5\textwidth]{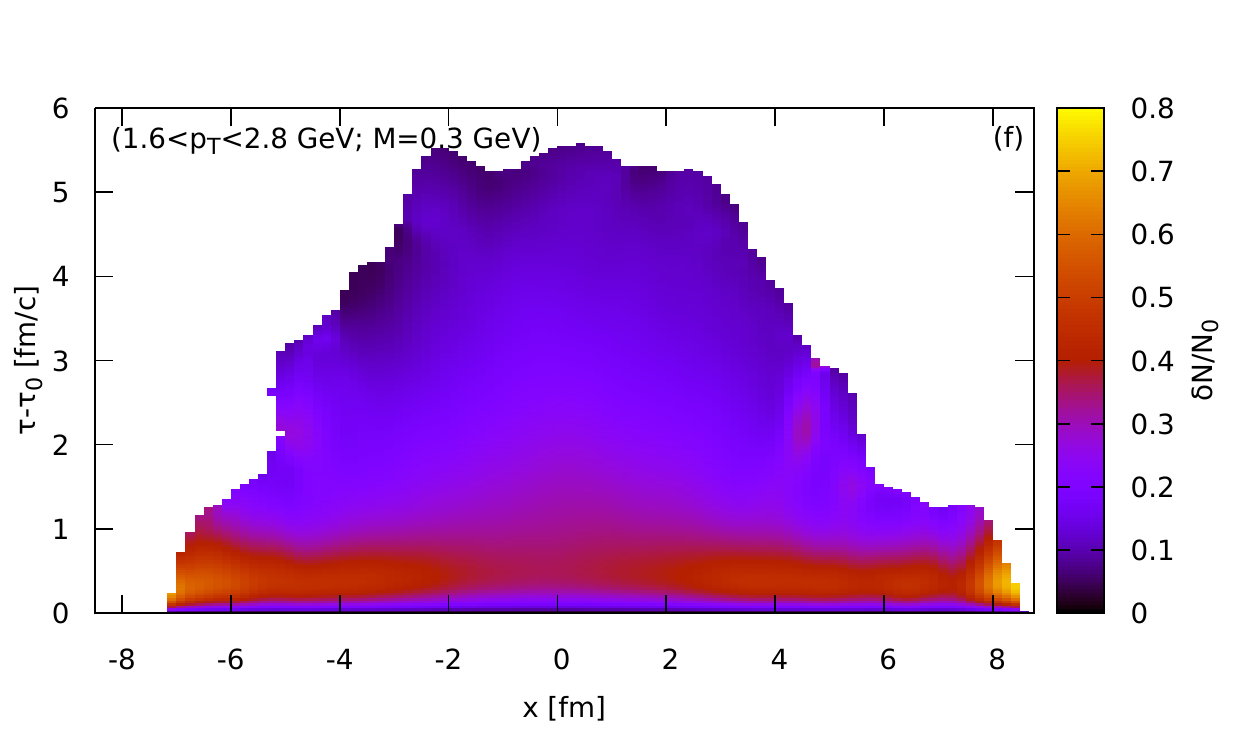}
\end{tabular}
\caption{(Color online) The size of the viscous contribution to the dilepton yield in the QGP relative to its ideal (inviscid) dilepton production for a medium with constant $\eta/s=1/(4\pi)$. Left column: $\delta N/N_0$ using the IS $\delta R$. Right column: $\delta N/N_0$ using the constant cross section $\delta R$.}
\label{fig:deltaN_over_N0_vs_pt}
\end{figure}

The behavior of $\delta N/N_0$ depicted in Figs. \ref{fig:deltaN_over_N0_vs_M} and \ref{fig:deltaN_over_N0_vs_pt} can be understood as an interplay between the decay of the envelope function $b_2$ in Eq. (\ref{eq:dR_born}) and the growth of the term $\frac{p^\alpha p^\beta \pi_{\alpha\beta}}{2T^2(\epsilon + P)}$, which is best described through the ratio $A$ in Eq. (\ref{eq:ratio_A}). For a fixed $M$, the $A$ changes slowly with $p_T$ while the term $\frac{p^\alpha p^\beta \pi_{\alpha\beta}}{2T^2(\epsilon + P)}$ grows quadratically with $p_T$,\footnote{Note that at the highest $p_T$, there is a slight suppression in the case of constant cross section $\delta R$ relative to IS $\delta R$ owing to the envelope $b_2$ in Eq. (\ref{eq:dR_born}).} whereas for a fixed $p_T$ since the envelope decays $1/(p\cdot u)$ the quadratic growth from the term $\frac{p^\alpha p^\beta \pi_{\alpha\beta}}{2T^2(\epsilon + P)}$ is suppressed. The overall result is that $\delta N/N_0$ grows faster in the $p_T$ direction than $M$ direction, as can be seen in Figs. \ref{fig:deltaN_over_N0_vs_M} and \ref{fig:deltaN_over_N0_vs_pt}, and is consistent with the fact that $A$ has a stronger energy dependence than three-momentum dependence.

The behavior of $\delta N/N_0$ as a function of $p_T$ at low invariant mass and low $p_T$, resembles that of photons previously investigated in \cite{Shen:2016zpp}. Like in the photon case, the size of the viscous correction is less than 14\% throughout the entire evolution for $p_T<1.6$ GeV. Viscous corrections only become large at early times with relatively high momenta $1.6<p_T<2.8$ GeV. This is especially true near the edges of the QGP sector in the $x$ direction, as is the case for photons as well \cite{Shen:2016zpp}. Furthermore, the largest contributions from $\delta N/N_0$ for the low invariant mass region are happening at high $p_T$ and at early times. Therefore these contributions will not significantly affect the $p_T$ integrated $v_2(M)$ at low invariant masses, whose biggest contribution comes from the low $p_T$ \cite{Vujanovic:2013jpa} and late times sector (recall Fig. \ref{fig:v2_M_hm_T_munu_hm}). The constant cross section $\delta R$ in the QGP sector will however play a more important role as far as the $p_T$-integrated $v_2(M)$ at higher invariant masses are concerned (and can be seen in the right column of Fig. \ref{fig:deltaN_over_N0_vs_pt}), where an improved description for the $\delta R$ generates a more reliable $v_2(M)$ result.

\begin{figure}[!h]
\begin{tabular}{cc}
\includegraphics[width=0.5\textwidth]{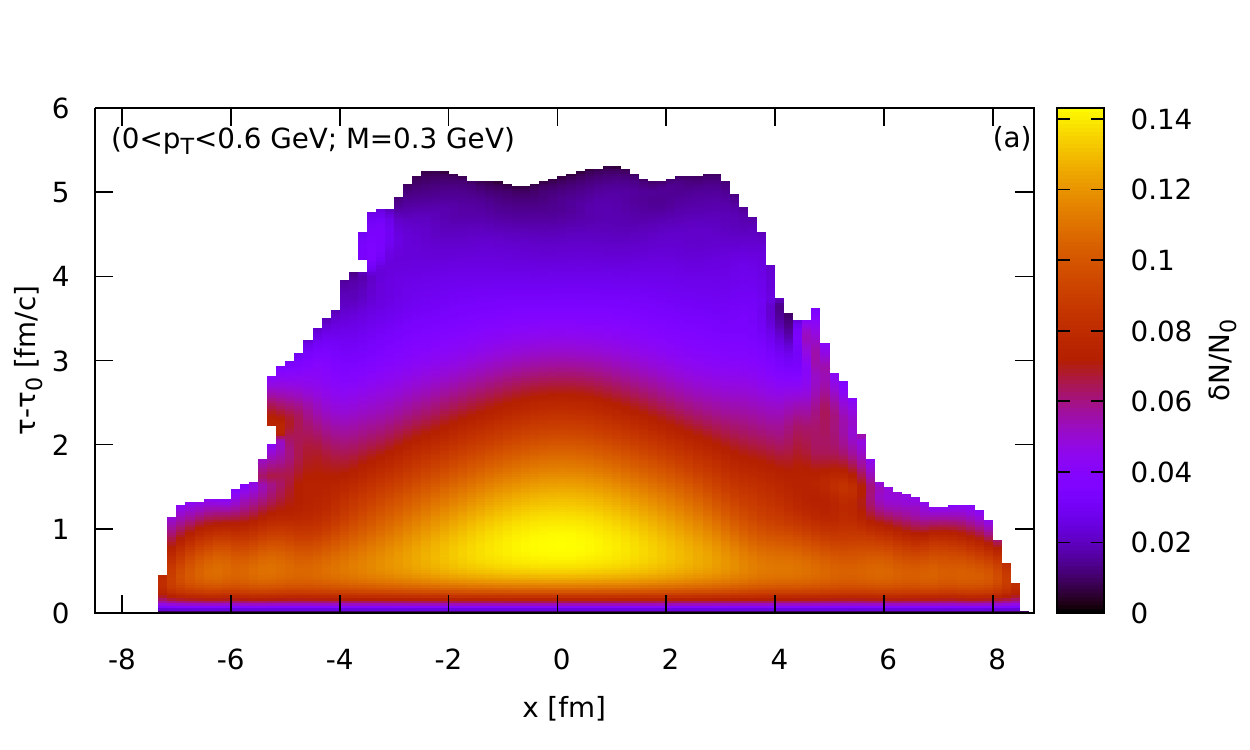} & \includegraphics[width=0.5\textwidth]{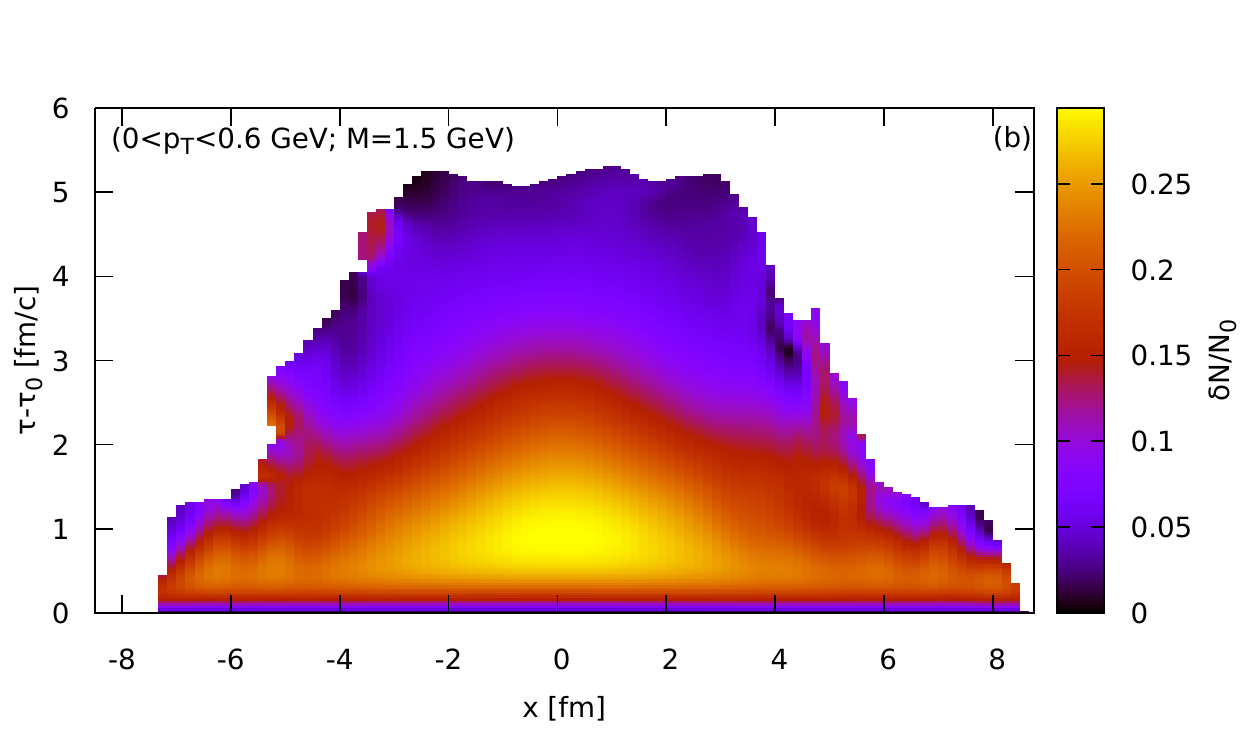} \\
\includegraphics[width=0.5\textwidth]{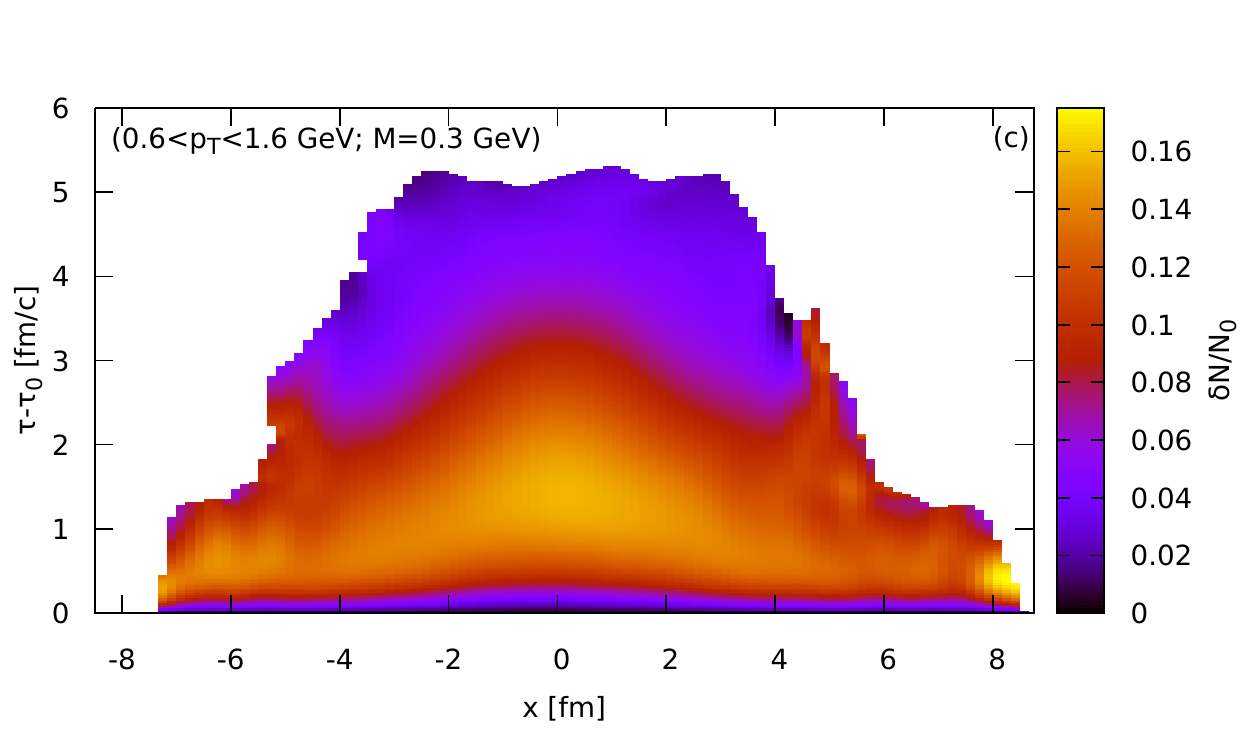} & \includegraphics[width=0.5\textwidth]{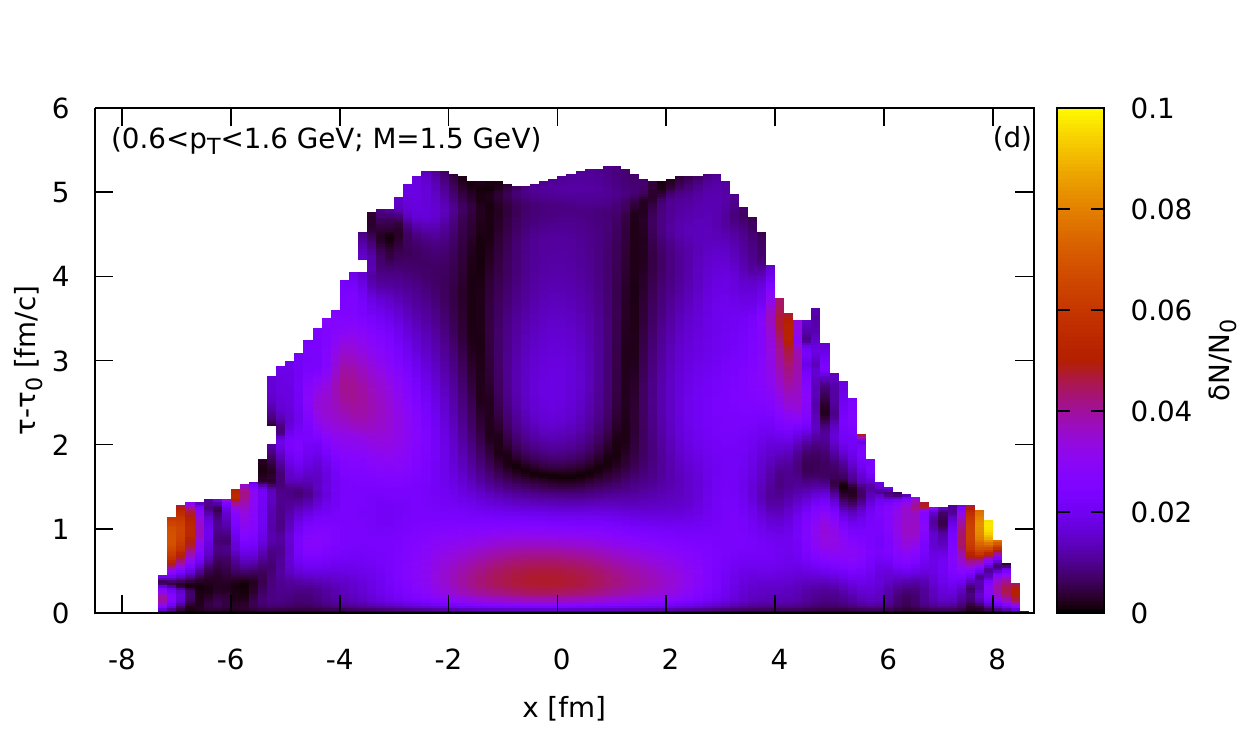} \\
\includegraphics[width=0.5\textwidth]{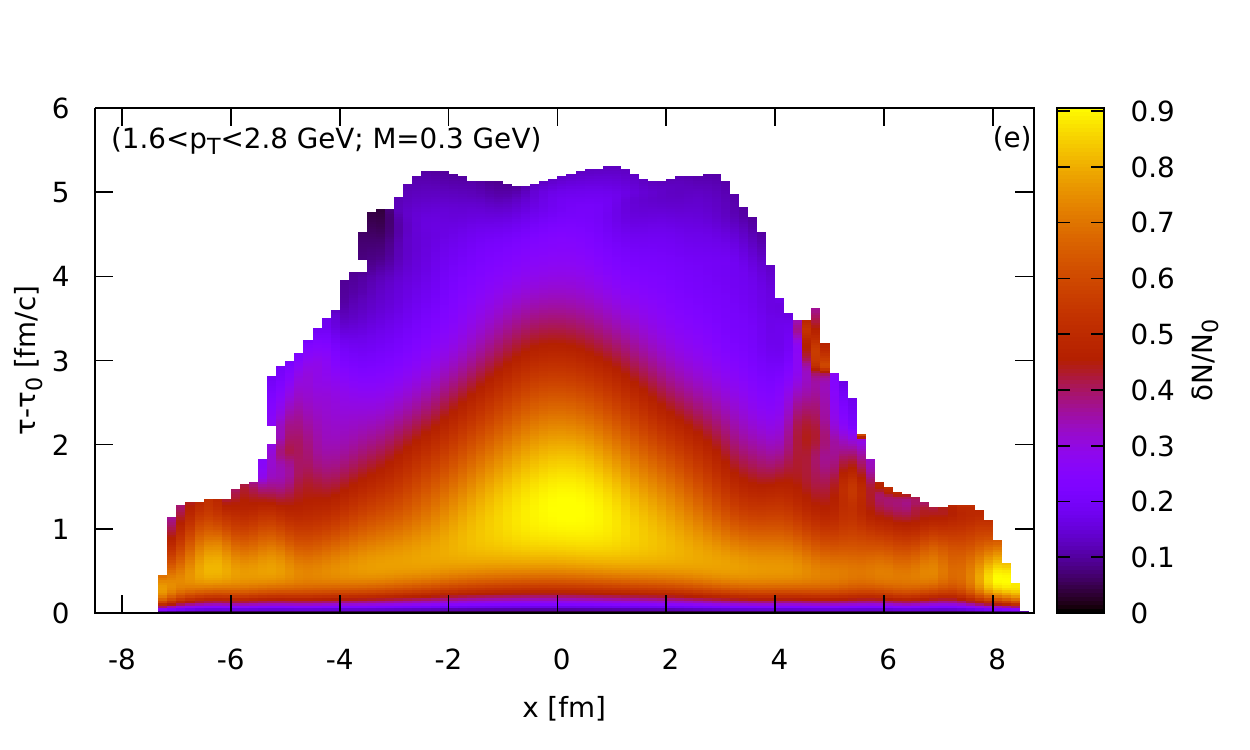} & \includegraphics[width=0.5\textwidth]{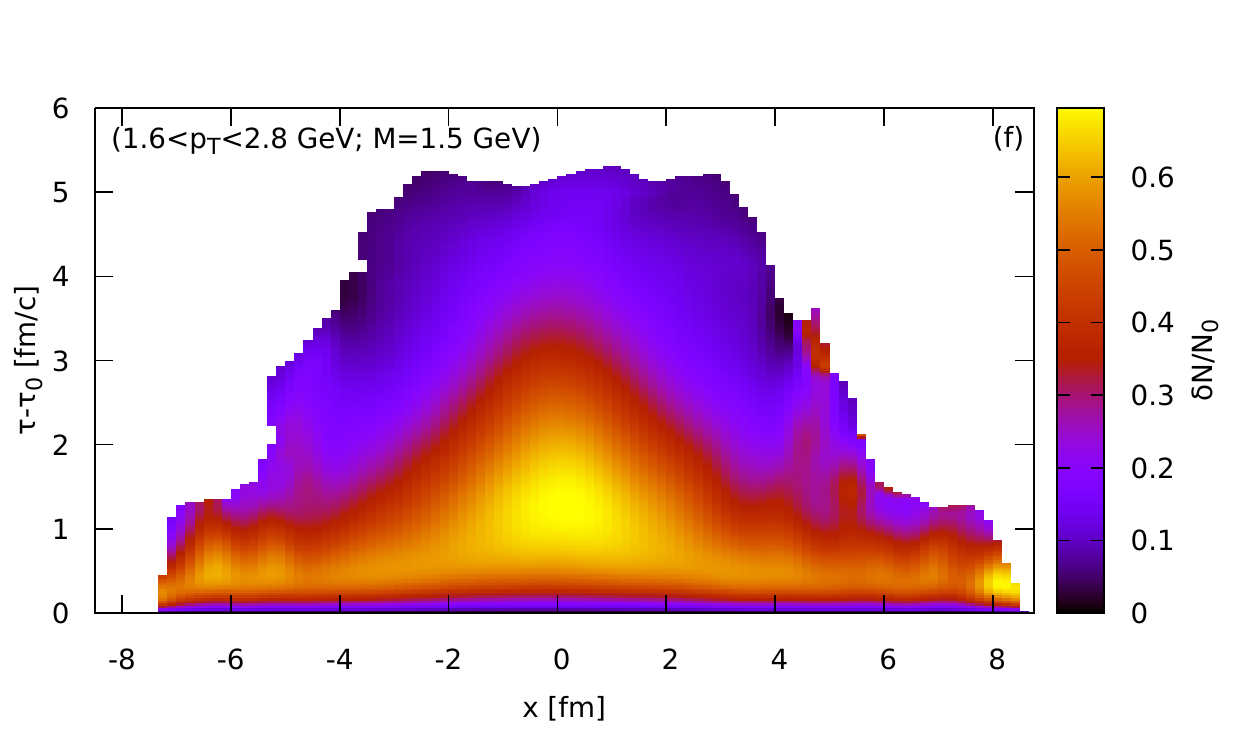} 
\end{tabular}
\caption{(Color online) The size of the viscous contribution to the dilepton yield in the QGP relative to its ideal (inviscid) dilepton production using the constant cross section $\delta R$ for a medium with linear $\eta/s(T)$ and slope $m=0.5516$. The left column plots $\delta N/N_0$ for $M=0.3$ GeV whereas the right column plots the same quantity at $M=1.5$ GeV.}
\label{fig:deltaN_over_N0_vs_pt_eta_s_T}
\end{figure}

Having discussed the effects of constant cross section $\delta R$ in the temperature-independent $\eta/s$, it is instructive to look at how $\delta N/N_0$ behaves once $\eta/s$ is temperature dependent. In particular,  we consider linear $\eta/s(T)$ with the highest slope in Fig. \ref{fig:deltaN_over_N0_vs_pt_eta_s_T}. For a low invariant mass $M=0.3$ GeV and $p_T<1.6$ GeV, the maximum $\delta N$ correction to the differential dilepton yield is $\sim 18$\% whereas at a higher invariant mass $M=1.5$ GeV, the maximum $\delta N$ correction raises to $\sim 29$\%. These percentages are sizable but not alarming. At higher momenta $1.6<p_T<2.8$ GeV, the $\delta N$ correction does significantly increase, however $\delta N/N_0<1$ still holds, which is encouraging.   

It should be emphasized one last time that the effects on the total $v_2(M)$ that were explored Sec. \ref{results} originate from the HM sector of the medium where viscous corrections to the dilepton rate are small and therefore effects seen on the total $v_2(M)$ are mostly independent of said corrections. The QGP only plays an important role once $M\geq1.5$ GeV, and the discussion within this appendix highlights the manner in which QGP dilepton production is modified owing to the constant cross section $\delta R$ relative to the IS one.  

\bibliography{references}
\end{document}